\documentclass[10pt,a5paper]{ETHthesis}

\usepackage{E}
\usepackage{ETHthesis}
\usepackage{makeidx}
\usepackage{enumerate}
\usepackage{epsfig}
\usepackage{amsthm,amssymb,amsmath,latexsym}
\usepackage{palatino}

\theoremstyle{plain}
\newtheorem{theorem}{Theorem}
\newtheorem{corollary}{Corollary}
\newtheorem{lemma}{Lemma}

\theoremstyle{definition}
\newtheorem{definition}{Definition}
\newtheorem{example}{Example}
\newtheorem{protocol}{Protocol}

\numberwithin{equation}{chapter}
\numberwithin{theorem}{chapter}
\numberwithin{lemma}{chapter}
\numberwithin{definition}{chapter}
\numberwithin{corollary}{chapter}

\usepackage{ifpdf}
\ifpdf
	
\else
	
\fi

\newcommand{\cancel}[1]{}

\newcommand{\ol}{\overline}
\newcommand{\ul}{\underline}
\newcommand{\bbR}{\mathbb R}
\newcommand{\bbN}{\mathbb N}
\newcommand{\eps}{\varepsilon}

\newcommand{\mA}{\mathcal A}

\newcommand{\mD}{\mathcal D}
\newcommand{\mE}{\mathcal E}

\newcommand{\mI}{\mathcal I}

\newcommand{\mR}{\mathcal R}
\newcommand{\mS}{\mathcal S}
\newcommand{\mT}{\mathcal T}
\newcommand{\mU}{\mathcal U}
\newcommand{\mV}{\mathcal V}
\newcommand{\mW}{\mathcal W}
\newcommand{\mX}{\mathcal X}
\newcommand{\mY}{\mathcal Y}
\newcommand{\mZ}{\mathcal Z}

\newcommand{\bE}{\mathbf E}
\newcommand{\bF}{\mathbf F}
\newcommand{\bG}{\mathbf G}
\newcommand{\bH}{\mathbf H}
\newcommand{\bI}{\mathbf I}

\newcommand{\bP}{\mathbf P}
\newcommand{\bQ}{\mathbf Q}

\newcommand{\bS}{\mathbf S}
\newcommand{\bT}{\mathbf T}

\DeclareMathOperator{\adv}{Adv}
\DeclareMathOperator{\predadv}{PredAdv}

\DeclareMathOperator{\Hop}{H}
\DeclareMathOperator{\emin}{min}
\newcommand{\Hmin}{\Hop_{\emin}}

\newcommand{\PlayerA}{{\textsf{A}}}
\newcommand{\PlayerB}{{\textsf{B}}}

\newcommand{\Auth}{\textsf{Comm}}

\newcommand{\OT}[3]{{#2 \choose #1}{\textsf{-OT}^{#3}}}
\newcommand{\OTT}{{\textsf{OT}}}
\newcommand{\TO}[3]{{#2 \choose #1}{\textsf{-TO}^{#3}}}

\newcommand{\ROT}[3]{{#2 \choose #1}\textsf{-ROT}^{#3}}
\newcommand{\ROTT}{{\textsf{ROT}}}

\newcommand{\semiROT}[3]{{#2 \choose #1}\textsf{-}{\textsf{ROT}}^{#3}}
\newcommand{\semiROTT}{{{\textsf{ROT}}}}

\newcommand{\EFLOO}{{\textsf{ELFO}}}

\newcommand{\RReduce}{\textsf{R-Reduce}}
\newcommand{\SReduce}{\textsf{S-Reduce}}
\newcommand{\EReduce}{\textsf{E-Reduce}}

\newcommand{\SimWOT}{\textsf{SimWOT}}

\newcommand{\WOT}[3]{(#1,#2,#3)\textsf{-WOT}}
\newcommand{\WOTtwo}[2]{(#1,#2)\textsf{-WOT}}
\newcommand{\WOTT}{\textsf{WOT}}

\newcommand{\compWOT}[3]{(#1,#2,#3)\textsf{-compWOT}}
\newcommand{\compWOTT}{\textsf{compWOT}}

\newcommand{\UOT}[2]{(#1)\textsf{-}{2 \choose 1}\textsf{-UOT}^{#2}}
\newcommand{\UOTT}{\textsf{UOT}}

\newcommand{\ROTfromOT}{\textsf{ROTfromOT}}
\newcommand{\OTfromROT}{\textsf{OTfromROT}}

\newcommand{\ROTfromUOT}{\textsf{ROTfromUOT}}

\newcommand{\ROTOR}{\textsf{ROTOR}}

\newcommand{\TOR}[3]{{#2 \choose #1}\textsf{-TOR}^{#3}}
\newcommand{\TORR}{{\textsf{TOR}}}

\newcommand{\semiTOR}[3]{{#2 \choose #1}\textsf{-}{\textsf{TOR}}^{#3}}

\newcommand{\compIndist}{\stackrel{\rm c}{\equiv}}

\DeclareMathOperator{\maj}{maj}

\DeclareMathOperator{\extr}{Ext}
\DeclareMathOperator{\leak}{Leak}

\DeclareMathOperator{\negl}{negl}

\DeclareMathOperator{\poly}{poly}

\initETHthesis

\makeindex

\begin{document}

\dissnum{17125}
\title{Oblivious-Transfer Amplification}
\degree{Doctor of Sciences}
\author{J\"urg Wullschleger}
\acatitle{Dipl. Inf.-Ing. ETH}
\dateofbirth{July 5, 1975}
\citizen{Vordemwald, AG, Switzerland} 
\examiner{Prof. Dr. Stefan Wolf} 
\coexaminera{Prof. Dr. Ivan Damg{\aa}rd}
\coexaminerb{}

\maketitle
\thispagestyle{empty}
\cleardoublepage

\pagestyle{plain} 

\chapter*{Acknowledgments}

First of all, I would like to thank Stefan Wolf who has been a great
advisor. Many results in this thesis are the outcome of
endless discussions with him.
I also want to thank Ivan Damg{\aa}rd for co-refereeing this thesis.

I would also like to thank all the people I was able to work with or
talk to about my research during the last few years, including
Don Beaver,
Hugue Blier, Gilles Brassard, Anne Broadbent, Daniel Burgarth,
Claude Cr\'epeau, Meriem Debbih, Simon-Pierre Desrosiers,
Thomas D\"ubendor\-fer, Fr\'ed\'eric Dupuis, Serge Fehr, Matthias Fitzi, Viktor Galliard, 
S\'ebastien Gambs, Nicolas Gisin,  Iftach Haitner,  Esther H\"anggi,
Patrick Hayden, Martin Hirt, Thomas Holenstein, Reto Kohlas,
Robert K\"onig, Ueli Maurer, Remo Meier, Andr\'e M\'ethot, Kirill Morozov, 
Yvonne Anne Oswald,
J\"orn M\"uller-Quade, Anderson Nascimento,
Krzysztof Pietrzak, 
Bartosz Przydatek, Melanie Raemy, Dominik Raub, Renato Renner,
Louis Salvail, George Savvides, Valerio Scarani, Christian Schaffner, Jean-Raymond Si\-mard,
Johan Sj\"odin, Christian Sommer, Reto Strobl,
Alain Tapp, Stefano Tessaro, Dominique Unruh, Stephanie Wehner, Douglas Wikstr\"om, Andreas Winter,
Jon Yard, and Vassilis Zikas.

Special thanks to Thomas Holenstein for answering many questions
and for giving me many helpful hints and ideas, to J\"orn M\"uller-Quade, Dominik Raub, Renato Renner and
Dominique Unruh for answering my questions about universal composablility, to Iftach Haitner for helpful comments
on the computational part of this thesis,
and
to Esther H\"anggi, Melanie Raemy and Christian Schaffner for proof-reading this
thesis and pointing out many errors.

This research was supported by the Swiss National Science Foundation (SNF), by the Natural Sciences and Engineering Research Council of Ca\-na\-da (NSERC) and by the Fonds Qu\'eb\'ecois de la Recherche sur la Nature et les Technologies (FQRNT).

\chapter*{Abstract}

In \emph{two-party computation}, two players want to
collaborate in a secure way in order to achieve a common goal, but, they do
not trust each other and do not want the other to learn more than
necessary about their inputs.
Unfortunately, two-party computation is impossible to achieve \emph{unconditionally}
securely, i.e., such that even an adversary with infinite computing power has no chance
in breaking the system. We do have implementations in the \emph{computational} setting,
i.e., where we assume that the computing power of the adversary is bounded,
but the security of these implementations are based on unproven assumptions such
as the assumption that factoring
is hard.

However, if a very simple primitive called \emph{oblivious transfer}
is available, then \emph{any} two party computation can be implemented in an unconditionally
secure way.
In this thesis we investigate what weaker forms of oblivious transfer still allow for
implementing oblivious transfer, and hence any two-party computation.

First of all, we will show that oblivious transfer is equivalent to a randomized
form of oblivious transfer, and that this randomized oblivious transfer is
in fact \emph{symmetric}. It follows that also oblivious transfer is symmetric.

Then, we present a protocol that implements oblivious transfer from a weakened
oblivious transfer called \emph{universal oblivious transfer}, where one of the
two players may get additional information. Our reduction is about twice as efficient
as previous results.

\emph{Weak oblivious transfer} is an even weaker form of oblivious transfer, where
both players may obtain additional information about the other player's input, and
where the output can contain errors. We give a new, weaker definition of weak oblivious transfer,
as well as new reductions with a more detailed analysis.

Finally, we show that any protocol that implements oblivious transfer
from weak oblivious transfer can be used in the computational setting to implement
computationally secure oblivious transfer from \emph{computational weak oblivious transfer},
which is a computational version of weak oblivious transfer, where the additional information
both players may obtain about the other player's input is only \emph{computationally} bounded.

\chapter*{Zusammenfassung}

\emph{Sichere Zweiparteienberechnung} erlaubt es zwei Spielern, 
die einander nicht vertrauen, gemeinsam eine Berechnung durchzuf\"uhren, ohne
dass der jeweils andere Spieler irgend\-welche zus\"atzlichen Informationen \"uber ihre Ein\-gabe
erf\"ahrt.
Leider ist es unm\"oglich eine solche Berechnung so auszu\-f\"uhren, dass sie selbst
gegen einen berechenm\"assig unbe\-schr\"ankten Angreifer sicher ist. Unter der Annahme,
dass der Angreifer berechenm\"assig be\-schr\"ankt ist, existieren sichere Protokolle, jedoch 
basiert die Sicherheit dieser Protokolle auf zus\"atz\-lichen Annahmen, wie zum Beispiel der Annahme,
dass Faktorisieren schwie\-rig ist.

Wenn jedoch eine Primitive mit dem Namen \emph{vergessliche \"Ubertragung}
ge\-ge\-ben
ist, dann kann \emph{jede} Zweiparteienberechnung sicher gegen unbe\-schr\"ankte Angreifer
ausgef\"uhrt werden.
In dieser Arbeit untersuchen wir, welche schw\"acheren Formen von ver\-gesslicher \"Uber\-trag\-ung 
uns immer noch erlauben, eine sichere vergessliche \"Ubertragung auszu\-f\"uhren.

Zuerst zeigen wir, dass vergessliche \"Ubertragung \"aquivalent ist zu einer
randomisierten vergesslichen \"Uber\-tragung,
und dass diese Primitive \emph{symmetrisch} ist. Daraus folgt, dass vergessliche
\"Ubertragung ebenfalls sym\-metrisch ist.

\emph{Universelle vergessliche \"Ubertragung} ist eine schw\"achere Variante von
ver\-gesslicher \"Uber\-tragung, in welcher einer der beiden  Spieler zus\"atzliche Informationen erhalten kann. 
Wir zeigen ein neues, effizienteres Prokoll um daraus vergessliche \"Uber\-tragung herzustellen.

\emph{Schwache vergessliche \"Ubertragung} ist eine noch schw\"achere Form von ver\-gesslicher
\"Uber\-tragung, in welcher beide Spieler zus\"atzliche Information erhalten k\"onnen und die \"Uber\-tragung
falsch sein kann. Wir geben sowohl eine neue, schw\"achere Definition von schwacher ver\-gesslicher \"Uber\-tragung,
als auch neue Protokolle wie man daraus vergessliche \"Uber\-trag\-ung herstellen kann.

Schliesslich zeigen wir, dass jedes Verfahren, welches vergessliche \"Uber\-trag\-ung aus
schwa\-cher ver\-gesslicher \"Uber\-trag\-ung herstellt, auch eingesetzt werden kann, um 
berechenm\"assig sichere vergessliche \"Ubertragung aus \emph{berechenm\"assig
schwacher ver\-gesslicher \"Uber\-trag\-ung} herzu\-stellen.

\tableofcontents

\cleardoublepage

\pagenumbering{arabic}
\pagestyle{headings}

\chapter{Introduction}

On January 16, 1797, Johann Wolfgang von Goethe (1749-1832)
sent a letter
to the publisher Vieweg with the following content (translated to English by \cite{MolTie98}):
\begin{quotation}
''I am inclined to offer Mr.\ Vieweg from Berlin an epic poem, Hermann and Dorothea,
which will have approxima\-tely 2000 hexameters. [\dots] Concerning the royalty we will proceed as follows:
I will hand over to Mr.\ Counsel B\"ottiger a sealed note which contains my demand, and I wait for
what Mr.\ Vieweg will suggest to offer for my work. If his offer is lower than my demand, then
I take my note back, unopened, and the negotiation is broken. If, however, his offer is higher, then
I will not ask for more than what is written in the note to be opened by Mr.\ B\"ottiger.''
\end{quotation}
The reason for Goethe to choose such a complicated scheme was not to maximize his profit --- he would not have earned less
by just selling it to Vieweg --- he wanted to gain information on how much Vieweg was
willing to pay for his work. Indeed, his procedure can be viewed as a second price auction, where
Goethe himself was playing the second bidder \cite{MolTie98}. However, other than in a second price auction,
Goethe would get to know the bid of the highest bidder. To achieve his goal, Goethe needed to
be able to commit to a value that Vieweg would not get to know before placing his bid, but such that
Goethe himself would also not
be able to change it. He did this by giving an envelope to a third, trusted party, Mr.\ B\"ottiger.
Unfortunately, things turned out other than intended by Goethe. 
B\"ottiger opened the envelope
and gave Vieweg a hint, who then bid exactly what Goethe had demanded in his envelope. Vieweg was
therefore able to completely hide the information on how much he was willing to pay.

This is an example of \emph{two-party computation}, where two players want to achieve a common
goal, however they do not trust each other and do not want the other to learn more than
necessary about their inputs. Obviously, such a computation can easily be achieved with the help
of a trusted third party. However, as the example above shows, the two players would rather
not need to trust such a third party. Our goal is therefore to achieve a
two-party computation \emph{without the help of a trusted third party}.

Unfortunately, this task is impossible to achieve \emph{unconditionally}
securely, i.e., such that even an adversary with infinite computing power has no chance
in breaking the system. On the other hand, there exist implementations in the computational setting,
i.e., they are secure against adversaries which only have limited computing power.
However, the security of these implementations are based on unproven assumptions such
as that factoring the product of two large prime numbers is hard.

Needless to say, we would like to base the security of a two-party computation protocol on as few
assumptions as possible.
Surprisingly, it turned out that if a
very simple primitive called \emph{oblivious transfer} is available,
then \emph{any} two party computation can be implemented in an unconditionally secure way.
Oblivious transfer is a primitive that allows a sender to send two bits to a receiver, who
can choose which bits he wants to receive. The receiver will remain completely
ignorant about the other bit, while the sender does not get to
know which bit has been chosen by the receiver.

Even though oblivious transfer is quite simple, it is rather difficult to implement.
For example, in the computational setting quite strong assumptions are needed at the moment.
On the other hand, it is possible to implement oblivious transfer under certain physical
assumptions.  However, such systems generally do not achieve a perfect oblivious transfer,
but one where one or both players may still be able to cheat in some way, and obtain
additional information that he should not be allowed.

The main topic of this thesis is to present different protocols that implement
oblivious transfer from weaker variants. For example, in \emph{weak oblivious
transfer}, there can occur three types of errors: first, even if both
players execute the protocol honestly, the output of the receiver can be wrong with
some probability. Secondly, a dishonest receiver may not remain completely ignorant about
the second input bit. And finally, a dishonest sender may gain partial
information about the receivers choice bit. We show that if these three errors are
not too large, it is possible to implement an almost perfect oblivious transfer.

\section{Background}

\paragraph{Two- and multi-party computation.} The concept of \emph{two- and multi-party computation} was introduced by Yao \cite{Yao82}.
A complete solution of this problem with respect to computational security
was given by Gold\-reich, Micali, and Wigderson \cite{GoMiWi87}, and later but independently, by Chaum, Dam\-g{\aa}rd, and van de Graaf \cite{ChDaGr87}. Later
Ben-Or, Goldwasser, and Wigderson \cite{BeGoWi88} and, independently,
Chaum, Cr\'e\-peau, and Dam\-g{\aa}rd \cite{ChCrDa88} showed that in
a model with only pairwise secure channels, multi-party computation among $n$ players unconditionally secure
against an active adversary is achievable if and only if $t<n/3$
players are corrupted. Beaver \cite{Beaver89b} and independently Rabin and Ben-Or \cite{RabBen89}
showed that this bound can be improved to $t<n/2$,
assuming that global broadcast channels are available.

\paragraph{Security definitions.}
Intuitively, it seems to be very clear what we mean when we say that a two-party protocol
should be \emph{secure}: it should be \emph{correct}, i.e., it should implement
the desired functionality, and it should be \emph{private}, meaning that it should not leak additional
information to any of the players. Unfortunately, these intuitive ad-hoc requirements are hard to 
formalize and often even insufficient.

Inspired by the work of Goldwasser, Micali, and Rackoff
\cite{GoMiRa85} on zero-knowledge proofs of knowledge, Goldreich, Micali and Wigderson
\cite{GoMiWi87} were the first to use the simulation paradigm to define the security of
multi-party computation protocols. Micali and Rogaway \cite{MicRog91} and
Beaver \cite{Beaver91} further formalized this approach.
The idea behind these definitions is very intuitive and goes as follows. We
say that a (real) protocol securely computes a certain functionality if for
any adversary attacking the protocol, there exists a (not much stronger) adversary in an ideal
setting --- where the players only have black-box access to the functionality
they try to implement --- that achieves the same. In other words, a protocol is
secure if any attack in the real model can be simulated in the much more
restrictive ideal model.
As shown by Beaver \cite{Beaver91}, and formally proved by Canetti \cite{Canetti96,Canetti00b},
these security definitions imply that secure protocols are \emph{sequentially composable}: if
in a secure protocol that uses an ideal
functionality, that ideal functionality is replaced by a secure protocol, then the
composed protocol is again a secure protocol. 
Later, Backes, Pfitzmann and Waidner \cite{PfiWai00,BaPfWa03} and independently
Canetti \cite{Canetti00} introduced a stronger security definition called
\emph{universal composability}, which guarantees that protocols can be composed in an arbitrary way.

\paragraph{Oblivious transfer.}
For the special case of two-party computation, there cannot exist
a protocol that is unconditionally
secure against one corrupted player. However, if a primitive called \emph{oblivious transfer (OT)} is
available, then \emph{any} two-party computation can be executed unconditionally secure, which was shown by Goldreich and Vainish \cite{GolVai87} for passive adversaries, and by Kilian \cite{Kilian88} for active adversaries. These results were later improved by Cr\'e\-peau \cite{Crepea89}, Goldwasser and Levin \cite{GolLev90}, and Cr\'e\-peau, van de Graaf, and Tapp \cite{CrvGTa95}. The idea of oblivious transfer goes back to Wiesner~\cite{Wiesner70} in around 1970.
He tried to show that quantum physics allows us to achieve certain (classical)
tasks that otherwise
would not be possible. Since a quantum state can contain more information than what we can get out by measuring
it, he proposed to use quantum communication as
 \emph{``a means for transmitting two messages either but not both of which may be received.''},
which is exactly what OT achieves. More formally, OT is a primitive that receives two bits $x_0$ and $x_1$ from the sender and a bit $c$ from the receiver, and sends $x_c$ to the receiver, while the receiver does not get to know $x_{1-c}$, and the sender does not get to know $c$.
Wiesner proposed a simple protocol that achieves this, but he pointed out that it could be broken
in principle.
Rabin \cite{Rabin81} introduced a similar primitive in 1981, and showed its usefulness to cryptographic
applications. (He also gave oblivious transfer its name.) Even, Goldreich and Lempel \cite{EvGoLe85}
reintroduced Wiesner's version OT.

\paragraph{Computationally secure oblivious transfer.}
There exist different approaches to securely implement OT, with different
degrees of security. If we are only interested in \emph{computational security}, i.e.,
a system that cannot be broken by any adversary limited to polynomial computing time,
then OT can be implemented using noiseless communication only,
given some assumptions are correct. Of course, we would like to make these assumptions as weak
as possible, for example, we would like to have an implementation of OT that
is secure under the assumption that \emph{one-way functions} --- functions that are easy
to evaluate, but hard to invert --- exist. Unfortunately, such an implementation is still not known.
Even worse, Impagliazzo and Rudich \cite{ImpRud89} showed that such an implementation, if it exists, will be very hard to find, because there cannot exist any \emph{black-box reduction} of OT to one-way functions.

Even, Goldreich and Lempel \cite{EvGoLe85} presented an implementation of OT using
trapdoor permutations. 
However, Goldreich \cite{Goldreich04} showed that in fact the
stronger assumption of an \emph{enhanced trapdoor permutations} is needed for the protocol to be secure. This assumption was later weakened by Haitner \cite{Haitne04} to \emph{dense trapdoor permutations}.
Other implementations use more specific assumptions such as the assumption that factoring
a product of two primes is hard, as shown by Rabin \cite{Rabin81},
or the \emph{Diffie-Hellman assumption}, shown by Bellare and Micali, Naor and Pinkas, and Aiello, Ishai and
Reingold \cite{BelMic89,NaoPin01,AiIsRe01}.
Unfortunately, these latter assumptions have turned out to be wrong in the quantum world, as there exists
an efficient algorithm for breaking both assumptions, shown by Shor \cite{Shor94}.

In the universally composable framework, Canetti and Fischlin \cite{CanFis01} showed that there cannot exist an implementation of OT secure against active adversaries\footnote{They showed that bit-commitment is impossible, but since bit-commitment can be implemented from OT,
this implies that also OT is impossible.}.
On the other hand, Canetti, Lindell, Ostrovsky, and Sahai \cite{CLOS02} showed that the protocol
presented in
\cite{GoMiWi87} is secure against passive adversaries in the universally composable framework.
Garay, MacKenzie and Yang \cite{GaMaYa04} proposed an implementation of \emph{enhanced
committed OT} secure against active adversaries under the 
additional assumption of a \emph{common reference string}. Fischlin \cite{Fischl06} proposed a protocol
that does not assume a common reference string, but needs the help of other players.

\paragraph{Unconditionally secure oblivious transfer.}
All known computational implementations of OT --- besides
the assumption that the adversary is computationally
bounded --- are based on quite strong, unproven assumptions about the complexity of certain problems.
\emph{Unconditional security} does not have these shortcomings. It offers a security that
cannot be broken \emph{in principle}, no matter what computing power the adversary has,
and is generally not based on unproven assumptions. Unfortunately, unconditional secure OT is impossible
to achieve if the players only have access to noiseless communication.
In fact, even noiseless \emph{quantum} communication does not help, as has been shown by
Mayers \cite{Mayers97}, and independently by Lo and Chau \cite{LoChau97}\footnote{They showed that bit-commitment is impossible, but since bit-commitment can be implemented from OT,
this implies that also OT is impossible.}.
Therefore, some additional resources must be available in order to achieve unconditionally
secure OT.

\paragraph{Reductions between different variants of OT.}
There exist many different variants of OT, and all of them have been
shown to be equivalent to OT.
Cr\'epeau \cite{Crepea87} showed that OT can be implemented from Rabin's OT, and
Brassard, Cr\'epeau and Robert \cite{BrCrRo86b} showed, among others, that string OT
(where the sender can send strings instead of single bits)
can be implemented from bit OT. More efficient methods to implement string OT from bit OT were presented by 
Brassard, Cr\'epeau and S\'antha \cite{BrCrSa96}, 
by Brassard, Cr\'epeau and Wolf \cite{BraCre97,BrCrWo03}, and by Cr\'epeau
and Savvides \cite{CreSav06}. Imai, Morozov, and Nascimento \cite{ImMoNa06} showed a direct implementation of string
OT from Rabin's OT.
Dodis and Micali \cite{DodMic99} presented a protocol to extend the number of
choices for the receiver. Another interesting property of OT was shown by
Bennett, Brassard, Cr{\'e}peau and Skubiszewska \cite{BBCS92} and Beaver \cite{Beaver95}, namely that
OT can be \emph{precomputed}. This means that OT can be converted into a randomized version of OT,
that can later be converted back into OT. Cr\'epeau and S\'antha \cite{CreSan91},
and independently Ostrovsky, Venkatesan and Yung \cite{OsVeYu91} presented protocols which
implement OT in one direction from OT in the other direction. Wolf and Wullschleger  \cite{WolWul06}
 presented a much simpler and more efficient protocol for this. 

Various weak versions of OT have been proposed where either the sen\-der's or the receiver's security
is weakened.
Cr\'epeau and Kilian \cite{CreKil88} presented an implementation of OT from \emph{$\alpha$-1-2 slightly~OT},
which is a weak version of OT where the sender may get some information about the choice bit of the receiver.
Brassard, Cr\'epeau and Wolf \cite{BraCre97,BrCrWo03} showed that OT can also be implemented from
\emph{XOT}, \emph{GOT} or \emph{UOT with repetitions}, which are weak versions of OT where the receiver
may get information he is not supposed to.
Cachin \cite{Cachin98} proposed a primitive called \emph{Universal OT} (without repetitions), which is
a generalization of XOT, GOT or UOT with repetitions. He proposed a protocol to implement OT, but his proof
turned out to be incorrect. The protocol was finally shown to be secure by Damg{\aa}rd, Fehr, Salvail and Schaffner \cite{DFSS06}. The bound for the protocol were later improved by Wullschleger \cite{Wullsc07}.
Damg{\aa}rd, Kilian and Salvail \cite{DaKiSa99} presented an even weaker form of OT called \emph{weak OT} (WOT),
where the security for \emph{both} players is weakened and the output to the receiver may be faulty.
They presented some bounds for which OT can be implemented from WOT. Later Wullschleger \cite{Wullsc07} 
showed that their definition of WOT implicitly uses quite strong assumptions, and proposed a new,
weaker definition together with new reductions.

\paragraph{OT from physical assumptions.}
Cr\'epeau and Kilian \cite{CreKil88} were the first to present
protocols for OT using \emph{noise} as additional resource
in form of an \emph{erasure channel}.
Cr\'epeau \cite{Crepea97} presented a protocol for the \emph{binary-symmetric noisy channel}, which was later
generalized by Korjik and Morozov \cite{KorMor01}. Cr\'epeau, Morozov and Wolf \cite{CrMoWo04} finally presented a protocol for \emph{any non-trivial channel}.
As shown by Imai, M\"uller-Quade, Nascimento and Winter, \cite{IMNW04}, Wolf and Wullschleger
\cite{WolWul04}, and Nascimento and Winter \cite{NaWi06},
these results also translate to the model where the players receive distributed
randomness\footnote{A similar model has already been studied in
the context of \emph{key agreement} by Ahlswede and Csisz\'ar \cite{AhlCsi93}
and Maurer \cite{Maurer93}.}.

Damg{\aa}rd, Kilian and Salvail \cite{DaKiSa99} introduced a more realistic, \emph{unfair} model
in which the adversary is given more information than the honest players. 
For example, if a noisy channel is implemented using a transmitter
and an antenna, an adversary may be able to replace the official antenna
by a larger one, and may, therefore, receive the transmitted signal with less noise than
an honest receiver would.
They presented explicit bounds for the \emph{unfair binary noisy channel}, which were later improved
by Damg{\aa}rd, Fehr, Morozov and Salvail \cite{DFMS04,Morozo05}. 
A central part of these results was the algorithm that implements OT from WOT. However, for the reduction to
work, the definition of \cite{Wullsc07} must be used.

\section{Outline of the Thesis}

\paragraph{Preliminaries.}
In Chapter~\ref{chap:pre}, we introduce the three distance measures that we will be using in this thesis. We will present some of the properties they have and how they are related. The
\emph{distinguishing advantage} and the \emph{statistical distance} are standard measures for
the distance between two distributions. On the other hand, the \emph{maximal bit-prediction advantage} is a special measure that we will use in Chapters~\ref{chap:wot} and \ref{chap:compWOT}.

\paragraph{Definition of secure two-party computation.}
In Chapter~\ref{chap:secTPC}, we give a simplified, formal framework for
two-party computation that is universally composable.
We will define two different models: the malicious model, where the corrupted players may behave arbitrarily, and the semi-honest model, where the corrupted players follow the protocol, but may try to obtain as much information as they can during the protocol. We will also show
that these definitions allow protocols to be composed. Finally, we show that
security in the malicious model does not imply security in the semi-honest model, and give a
weaker security definition for the semi-honest model for which this implication holds.

\paragraph{Oblivious transfer.}
In Chapter~\ref{chap:ot}, we will introduce the main topic of this thesis: oblivious transfer (OT).
We will also define a randomized version of OT, called \emph{randomized OT} (ROT), and show that OT and ROT are equivalent if communication is free.
We will then give a very simple protocol which shows that ROT is symmetric. In connection with
the other protocols, this gives us a simple way to reverse the direction of OT. 
Finally, we will present \emph{information-theoretic conditions}
that imply that a protocol
securely implements ROT.

\emph{Contribution.} 
Our reduction that reverses ROT and hence also OT is joint work with Stefan Wolf \cite{WolWul06}, and
is much simpler and more efficient
than previous reductions presented in \cite{CreSan91,OsVeYu91}.
The information-theoretic conditions for the security of ROT presented here build on prior joint work with Claude Cr\'epeau, George Savvides and Christian Schaff\-ner \cite{CSSW06}. There, we presented information-theoretic conditions that imply that a protocol securely implements secure function evaluation in a  sequentially composable model.
These conditions replace many ad-hoc definitions for the security of protocols
which often have been faulty. Here, we only present conditions for ROT, however we show a stronger statement about ROT,
 as our conditions imply that a protocol is \emph{universally composable}, and not only sequentially. Also, our conditions
have explicit error terms, which makes them easier to use.

\paragraph{Universal oblivious transfer.}
In Chapter~\ref{chap:uot}, we will present a protocol that implements ROT from a weak variant of ROT
called \emph{universal OT} (UOT). In contrast to ROT, UOT allows a corrupted receiver to receive
\emph{any} information he wants about the input, as long as he does not receive too much information. For example, he could be allowed to receive a bit string of a certain size that is an arbitrary function of his choice of
the sender's inputs.

\emph{Contribution.}
Our proof, which is also presented in \cite{Wullsc07}, shows that in the reduction of OT to
UOT, the string length of the resulting OT can be about twice as long as
for the bound presented in
\cite{DFSS06}, which is optimal for that protocol.
(The same bound that we present here has already been claimed in \cite{Cachin98},
but the proof presented there was incorrect, which was discovered by \cite{DFSS06}.)
Our proof makes use of a novel \emph{distributed leftover hash lemma},
which is a generalization of the well-known leftover hash lemma \cite{BeBrRo88,ILL89},
and of independent interest.

\paragraph{Weak oblivious transfer.}
In Chapter~\ref{chap:wot}, we introduce \emph{weak oblivious transfer} (WOT), a weak
variant of ROT where the security for \emph{both} players is weak, and where the output may be incorrect.
We give formal definitions of WOT in both the semi-honest and the malicious model. We show that
for certain parameters (when the instances of WOT are too weak), it is impossible to implement ROT from WOT.
Then we present several protocols that implement ROT from WOT, and give
upper bounds on how many instances of WOT are needed.
Unfortunately, these reductions do not meet the impossibility bound.

\emph{Contribution.} We give several improvements over the results presented in
\cite{DaKiSa99}, most of which are also presented in \cite{Wullsc07}.
First of all, we give new, weaker definitions of WOT
that replaces the definition presented in \cite{DaKiSa99,DFMS04}, which was too strong
and had only a very limited range of applications.
Also, our definitions make the need
for the more general notion of \emph{generalized weak oblivious transfer} of \cite{DFMS04} unnecessary.
For the special case where the WOT does not make any
error, we present a more detailed proof and a better upper bound on the amount of instances
used than in \cite{DaKiSa99}. Then, using a different error-reduction protocol
that also works with our weaker definitions, we give bounds for the special case where information is leaked
only to one of the two players, as well as several new bounds for the general case.

\paragraph{Computational weak oblivious transfer.}
In Chapter~\ref{chap:compWOT} we transfer the results from Chapter~\ref{chap:wot} to the computational setting. We define \emph{computational weak oblivious transfer} (compWOT), which is a computational version of WOT, where the adversary may get some additional
\emph{computational} knowledge about the value he is not supposed to. Using Holenstein's hard-core lemma \cite{Holens05,Holens06},  we show that any protocol that is secure in the information-theoretic setting can also be used in the computational setting. Hence, the reductions presented in Chapter~\ref{chap:wot} can be used to amplify compWOT to
a computationally secure OT.

\emph{Contribution.} We give a simplified but slightly stronger version of the
\emph{pseudo-randomness extraction theorem} from \cite{Holens06}, and fix the proof given in \cite{Holens06},
where a step was missing.
Then, we show that computationally secure OT can
be implemented from a large set of compWOT. This improves
 the results presented in \cite{Haitne04}, where only one special case was solved.

\chapter{Preliminaries} \label{chap:pre}

\section{Notation}

We will use the following convention: lower case letters will denote fixed values and 
upper case letters will denote random variables and algorithms. Calligraphic letters
will denote sets and domains of random variables.
For a random variable
$X$ over $\mX$, we denote its distribution by $P_X: \mX \rightarrow [0,1]$ with
$\sum_{x\in \mX} P_X(x) = 1$. For a given distribution
$P_{XY}: \mX \times \mY \rightarrow [0,1]$, we write for the marginal distribution
$P_{X}(x) := \sum_{y \in \mY} P_{XY}(x,y)$ and, if $P_Y(y) \neq 0$, 
$P_{X \mid Y}(x \mid y) := P_{XY}(x,y) / P_{Y}(y)$ for the conditional distribution. 
By $x^n$ we denote the list $(x_0,\dots,x_{n-1})$.

We use the function $\exp(x) := e^x$. $\ln(x)$ denotes the natural logarithm, and $\log(x)$ denotes the logarithm to the base 2.

\section{Distances between Distributions} \label{sec:statDist}

In this section, we will introduce two measures for the distance between two distributions:
the \emph{distinguishing advantage} and the \emph{statistical distance}.

\begin{definition} \label{def:adv-4-RV}
The \emph{distinguishing advantage} of an algorithm $A: \mU \rightarrow \{0,1\}$ (called the \emph{distinguisher})
to
distinguish $X$ from $Y$, which are random variables over the domain $\mU$, is
\[ \adv^A(X,Y) := \big| \Pr[A(X) = 1] - \Pr[A(Y)=1] \big |\;.\]
The distinguishing advantage of a class $\mD$ of distinguishers in distinguishing $X$ from $Y$ is
\[ \adv^{\mD}(X,Y) := \max_{A \in \mD} \adv^{A}(X,Y)\;.\]
\end{definition}

We have $\adv^\mD(X,X) = 0$ and $\adv^\mD(X,Y) = \adv^\mD(Y,X)$ for all $X$ and $Y$.
It is also easy to see that probabilistic distinguishers do not perform better than deterministic ones: let $A_R$ be a probabilistic distinguisher that takes additionally some randomness $R$ as input. We have
\[  \adv^{A_R}(X,Y) := \sum_r P_R(r) \cdot \big| \Pr[A_r(X) = 1] - \Pr[A_r(Y)=1] \big |\;.\]
Now let $r \in \mR$ be the value that maximizes the expression 
\[\big| \Pr[A_r(X) = 1] - \Pr[A_r(Y)=1] \big |\;.\]
Then $A_r$ is a deterministic distinguisher with
\[\adv^{A_r}(X,Y) \geq \adv^{A_R}(X,Y)\;.\]

In the following, we will therefore only consider deterministic distinguishers.
Lemma~\ref{lem:advTri} shows that the \emph{triangle inequality} holds for the
distinguishing advantage.

\begin{lemma}[Triangle inequality] \label{lem:advTri}
For any $X$, $Y$, and $Z$ over $\mU$, we have
\[ \adv^A(X,Z) \leq \adv^A(X,Y) + \adv^A(Y,Z)\;.\]
\end{lemma}

\begin{proof} We have
\begin{align*}
\adv^A(X,Z)
 & = \big| \Pr[A(X) = 1] - \Pr[A(Z)=1] \big | \\
 & = \big| \Pr[A(X) = 1] - \Pr[A(Y) = 1]\\
 & \qquad \qquad +  \Pr[A(Y) = 1] - \Pr[A(Z)=1] \big | \\
 & \leq \big| \Pr[A(X) = 1] - \Pr[A(Y) = 1] \big|\\
 & \qquad \qquad + \big| \Pr[A(Y) = 1] - \Pr[A(Z)=1] \big | \\
 & = \adv^A(X,Y) + \adv^A(Y,Z)\;.
\end{align*}
\end{proof}

It is easy to see that the same also holds for \emph{classes} of distinguishers, i.e., for any
$\mD$, we have
$\adv^\mD(X,Z) \leq \adv^\mD(X,Y) + \adv^\mD(Y,Z)$.

\begin{definition}
The \emph{statistical distance} of two random variables $X$ and $Y$ (or two distributions $P_{X}$ and $P_{Y}$) over the same domain $\mU$
is defined as
\[
\Delta(X,Y) = \Delta(P_X,P_Y)
:= \frac 1 2 \sum_{u \in \mU} \Big | P_X(u)  - P_Y(u) \Big |\;.
\]
\end{definition}

We say that $P_X$ is $\eps$-close to $P_Y$, denoted by $P_X \equiv_\eps P_Y$,
if $\Delta(P_X,P_Y) \leq \eps$.
We say that a random variable \emph{$X$ is $\eps$-close to uniform with respect to $Y$}, if $P_{XY} \equiv_\eps P_U P_Y$, where $P_U$ is the uniform
distribution over $\mX$.

\begin{lemma} \label{lem:statDistT}
For all $X$ and $Y$, we have
\[
\Delta(X,Y)
= \Pr[X \in \mT]- \Pr[Y \in \mT] = \sum_{u \in \mT} \Big ( P_X(u)  - P_Y(u) \Big )
\]
for $\mT := \{u \in \mU \mid P_X(u) > P_Y(u)\}$\;.
\end{lemma}

\begin{proof}
We have
\begin{align*}
\Delta(X,Y)
&= \frac 1 2 \sum_{u \in \mT} \Big ( P_X(u)  - P_Y(u) \Big ) + \frac 1 2 \sum_{u \not \in \mT} \Big ( P_Y(u)  - P_X(u) \Big ) \\[.1cm]
&= \frac {\Pr[X \in \mT]} 2 + \frac {\Pr[Y \not \in \mT]} 2
- \frac {\Pr[X \not \in \mT]} 2 - \frac {\Pr[Y \in \mT]} 2 \\[.1cm]
&= \Pr[X \in \mT]- \Pr[Y \in \mT]\;.
\end{align*}
\end{proof}

\begin{lemma} \label{lem:statDistMaxSet}
For all $X$ and $Y$, we have
\[
\Delta(X,Y) = \max_{\mS \subseteq \mU} \Big ( \Pr[X \in \mS] - \Pr[Y \in \mS] \Big )\;.
\]
\end{lemma}

\begin{proof}
Follows directly from Lemma~\ref{lem:statDistT}, since 
\[\Pr[X \in \mS] - \Pr[Y \in \mS]\]
is maximal for $\mS = \mT$.
\end{proof}

From  Lemma~\ref{lem:statDistMaxSet} follows now that
\[\adv^\mD(X,Y) = \Delta(X,Y)\;,\] 
where $\mD$ is the class of all (also inefficient) distinguishers.

\begin{lemma}\label{lem:statDist-dataprocessing}
For any $X$ and $Y$  over $\mU$ and $f: \mU \rightarrow \mV$, we have
\[ \Delta(f(X),f(Y)) \leq \Delta(X,Y)\;. \]
\end{lemma}

\begin{proof}
Let $\mD$ be the class of all (also inefficient) distinguishers, and let $D(v)$ be a distinguisher such that
\[\adv^D(f(X),f(Y)) = \adv^\mD(f(X),f(Y))\;.\]
Then, for $D'(u) := D(f(u))$, we have
\[\adv^{D'}(X,Y) = \adv^\mD(f(X),f(Y))\;.\]
Since $D' \in \mD$, we have
\begin{align*}
\Delta(f(X),f(Y))
& = \adv^\mD(f(X),f(Y)) = \adv^{D'}(X,Y) \\
& \leq \adv^\mD(X,Y) = \Delta(X,Y)\;.
\end{align*}
\end{proof}

\begin{lemma} \label{lem:statDistEvent2}
Let $P_{BX}$ and $P_{CY}$ be distributions over $\{0,1\} \times \mU$ such that
\[\Pr[B=1] = \Pr[C=1] = \eps\;.\]
Then
\[ \Delta( P_{X}, P_{Y}) \leq \eps + \Delta( P_{X \mid B=0}, P_{Y \mid C=0})\;.\]
\end{lemma}

\begin{proof}
For any set $\mS \subseteq \mU$, we have
\begin{align*}
& \Pr[X \in \mS] - \Pr[Y \in \mS] \\
& \qquad = \eps \cdot \big ( \Pr[X \in \mS \mid B=1] - \Pr[Y \in \mS \mid C=1] \big ) \\
& \qquad \qquad + (1-\eps) \cdot \big ( \Pr[X \in \mS \mid B=0] - \Pr[Y \in \mS \mid C=0] \big )\\[0.1cm]
& \qquad \leq \eps + \big ( \Pr[X \in \mS \mid B=0] - \Pr[Y \in \mS \mid C=0] \big )\\
& \qquad \leq \eps + \max_{\mS' \subseteq \mU} \Big ( \Pr[X \in \mS' \mid B=0] - \Pr[Y \in \mS' \mid C=0] \Big ) \\
& \qquad = \eps + \Delta( P_{X \mid B=0}, P_{Y \mid C=0})\;,
\end{align*}
and therefore
\begin{align*}
\Delta(X,Y)
 = \max_{\mS \subseteq \mU} \Big ( \Pr[X \in \mS] - \Pr[Y \in \mS] \Big ) \leq \eps + \Delta( P_{X \mid B=0}, P_{Y \mid C=0})\;.
\end{align*}
\end{proof}

\section{Prediction of Random Variables}

For the case where $X \in \{0,1\}$, we will also use another measure of its closeness
to uniform with respect to a random variable $Y$, the \emph{maximal bit-prediction advantage},
which measures how well
$X$ can be predicted from $Y$. See also Section 2.1 in \cite{Holens06}.

\begin{definition}
Let $P_{XY}$ be a distribution over $\{0,1\} \times \mY$.
The \emph{maximal bit-prediction advantage} of $X$ from $Y$ is
\[ \predadv(X \mid Y) := 2 \cdot \max_{f} \Pr[f(Y) = X] - 1\;.\]
\end{definition}

In other words, if $\predadv(X \mid Y) = \delta$, then we have for all functions $f: \mY \rightarrow \{0,1\}$
\[ \Pr[f(Y) = X] \leq \frac{1 + \delta}{2}\;.\]
First, we show that $\predadv(X \mid Y) \leq 2\eps$, if and only if $X$ is
$\eps$-close to uniform with respect to $Y$.

\begin{lemma} \label{lem:PredAdvStadDist}
Let $P_{XY}$ be a distribution over $\{0,1\} \times \mY$. Then
\[\predadv(X \mid Y)
= 2 \cdot \Delta( P_{XY},P_U P_Y)\;,\]
where $P_U$ is the uniform distribution over $\{0,1\}$.
\end{lemma}

\begin{proof}
Obviously, the best function $f: \mY \rightarrow \{0,1\}$ for guessing $X$ is
\[ f(y) := \left \{
\begin{array}{ll}
0 & \textrm{if $P_{XY}(0,y) \geq P_{XY}(1,y)$,} \\
1 & \textrm{otherwise.}
\end{array}
\right.\]
We have
\begin{align*}
2 \Pr[f(Y) = X] - 1
&= 2 \sum_y P_Y(y) P_{X \mid Y=y}(f(y)) - \sum_y P_Y(y) \\
&= \sum_y P_Y(y) \left ( P_{X \mid Y=y}(f(y)) - (1 - P_{X \mid Y=y}(f(y))) \right )\\
&= \sum_y P_Y(y) \left ( P_{X \mid Y=y}(f(y)) - P_{X \mid Y=y}(1 - f(y)) \right )\\
&= \sum_y P_Y(y) \big |P_{X \mid Y=y}(0) - P_{X \mid Y=y}(1) \big| \\
&= \sum_y P_Y(y) \sum_x \Big |P_{X \mid Y=y}(x) - \frac 1 2 \Big | \\
&= 2 \cdot \Delta( P_{XY},P_U P_Y) \;.
\end{align*}
\end{proof}

Lemma~\ref{lem:predAdv-dataprocessing} follows immediately from Lemmas~\ref{lem:statDist-dataprocessing} and \ref{lem:PredAdvStadDist}.

\begin{lemma} \label{lem:predAdv-dataprocessing}
Let $P_{XY}$ be a distribution over $\{0,1\} \times \mY$, and let $f: \mY \rightarrow \mY'$.
Then
\[\predadv(X \mid f(Y))
\leq \predadv(X \mid Y)\;.\]
\end{lemma}

The following lemma shows that for any distribution $P_{XY}$ over $\{0,1\} \times \mY$,
we can define an event that has probability $1-\predadv(X \mid Y)$, such that conditioned on that
event,
$X$ is uniformly distributed given $Y$, and therefore no function $f(Y)$ can predict $X$.

\begin{lemma} \label{lem:Hol22}
Let $P_{XY}$ be any distribution over $\{0,1\} \times \mY$. There exists a conditional distribution
$P_{B\mid XY}$
over $\{0,1\} \times \{0,1\} \times \mY$ such that
\[\Pr[B=1] \leq \predadv(X \mid Y)\]
and such that for all functions $f: \mY \rightarrow \{0,1\}$,
\[ \Pr[f(Y) = X \mid B=0] = 1/2\;.\]
\end{lemma}

\begin{proof}
We define
\[ P_{B \mid X,Y}(0 \mid x,y) := \frac {\min(P_{XY}(0,y),P_{XY}(1,y))}{P_{XY}(x,y)}\;.\]
Using Lemma~\ref{lem:PredAdvStadDist}, we get
\begin{align*}
\Pr[B=1]
& = \sum_{x,y} P_{XY}(x,y) P_{B \mid X,Y}(1 \mid x,y) \\
& = \sum_{x,y} P_{XY}(x,y) \left ( 1 - \frac {\min(P_{XY}(0,y),P_{XY}(1,y))}{P_{XY}(x,y)}\right ) \\
& = \sum_{x,y} \left (P_{XY}(x,y) - \min(P_{XY}(0,y),P_{XY}(1,y))\right ) \\
& = \sum_{y} \big | P_{XY}(0,y) - P_{XY}(1,y) \big | \\
& = \sum_y P_Y(y) \sum_x \Big |P_{X \mid Y=y}(x) - \frac 1 2 \Big | \\
& = 2 \cdot \Delta( P_{XY},P_U P_Y) 
 = \predadv(X \mid Y)\;.
\end{align*}
For $x \in \{0,1\}$, we have
\begin{align*}
P_{X \mid BY}(x \mid 0,y)
& = \frac{P_{X \mid Y}(x \mid y) \cdot P_{B \mid XY}(0 \mid x,y)}{P_{B \mid Y}(0 \mid y)} \\[0.3cm]
& = \frac{P_{X \mid Y}(x \mid y) \cdot \min(P_{XY}(0,y),P_{XY}(1,y))}{P_{B \mid Y}(0 \mid y) \cdot
P_{Y}(y) \cdot P_{X \mid Y}(x \mid y)} \\[0.3cm]
& = \frac{\min(P_{XY}(0,y),P_{XY}(1,y))}{P_{BY}(0,y)}\;.
\end{align*}
Since $P_{X \mid BY}(x \mid 0,y)$ does not depend on $x$, it must be equal to $1/2$, and, therefore, we have, for
 all functions $f$ and for all values $y$, 
 \[\Pr[f(Y)=X \mid B=0, Y=y] = 1/2\;.\]
\end{proof}

Lemma~\ref{lem:Hol22-converse} shows that the statement of
Lemma~\ref{lem:Hol22} also works in the other
direction. If there exists an event with probability $1-\delta$ under which $X$ cannot
be guessed from $Y$ with any advantage, then $\predadv(X \mid Y) \leq \delta$.

\begin{lemma} \label{lem:Hol22-converse}
Let $P_{XY}$ be any distribution over $\{0,1\} \times \mY$. If there exists a conditional distribution 
$P_{B\mid XY}$
over $\{0,1\} \times \{0,1\} \times \mY$ such that
for all functions $f: \mY \rightarrow \{0,1\}$ we have
\[ \Pr[f(Y) = X \mid B=0] = 1/2\;,\]
then 
\[\predadv(X \mid Y) \leq \Pr[B=1]\;.\]
\end{lemma}

\begin{proof}
For any function $f$, we have
\begin{align*}
\Pr[f(Y) = X]
& = \Pr[B=0] \cdot \Pr[f(Y) = X \mid B=0] \\
& \qquad + \Pr[B=1] \cdot \Pr[f(Y) = X \mid B=1] \\[0.1cm]
& \leq 1/2 \cdot \Pr[B=0] + \Pr[B=1]\;,
\end{align*}
and, therefore,
\begin{align*}
\predadv(X \mid Y)
& = 2 \cdot \max_{f} \Pr[f(Y) = X] - 1 \\
& \leq \Pr[B=0] + 2 \cdot \Pr[B=1] - 1
= \Pr[B=1]\;.
\end{align*}
\end{proof}

The following lemmas show some rules for $\predadv(X \mid Y)$.

\begin{lemma} \label{lem:predXOR}
Let $P_{X_0Y_0},\dots,P_{X_{n-1}Y_{n-1}}$ be distributions over $\{0,1\} \times \mY_i$. Then
\[\predadv(X_0 \oplus \cdots \oplus X_{n-1} \mid Y^n) \leq \prod_{i=0}^{n-1} \predadv(X_i \mid Y_i)\;.\]
\end{lemma}

\begin{proof}
For $i \in \{0,\dots,n-1 \}$, let $B_i$ be the random variable defined by Lemma~\ref{lem:Hol22}.
Let $B = \min_i(B_i)$. If $B = 0$ then for a $j \in \{0,\dots,n-1\}$ we have $B_j = 0$. Therefore,
$X_j$ is uniformly at random given $Y_j$, and any $f: \mY^n \rightarrow \{0,1\}$ will output
$X_0 \oplus \cdots \oplus X_{n-1}$ with probability $1/2$. 
The statement now follows from Lemma~\ref{lem:Hol22-converse}, and from the fact that
\[\Pr[B=1] = \prod_{i=0}^{n-1} \Pr[B_i=1]\;.\]
\end{proof}

\begin{lemma} \label{lem:predComm}
Let $P_{X_0Y_0},\dots,P_{X_{n-1}Y_{n-1}}$ be distributions over $\{0,1\} \times \mY_i$,
 and let $D_i := X_{i} \oplus X_{n-1}$.
Then
\[\predadv(X_{n-1} \mid  Y^n,D^{n-1}) \leq 1 - \prod_{i=0}^{n-1} \left( 1 - \predadv(X_i \mid Y_i) \right)\;.\]
\end{lemma}

\begin{proof}
For $i \in \{0,\dots,n-1 \}$, let $B_i$ be the random variable defined by Lemma~\ref{lem:Hol22},
and let $B = \max_i(B_i)$. 
If $B=0$ then for all $0 \leq i < n$ we have $B_i = 0$, and therefore $X_i$ will be
uniformly at random given $Y_i$. It follows that $X_{n-1}$ is independent from $(Y^n,D^{n-1})$
and any $f: \mY^n \times \mD \rightarrow \{0,1\}$ will output
$X_{n-1}$ with probability $1/2$. 
The statement now follows from Lemma~\ref{lem:Hol22-converse}, and from the fact that
\[\Pr[B=1] = 1 - \prod_{i=0}^{n-1} (1- \Pr[B_i=1])\;.\] 
\end{proof}

\begin{lemma} \label{lem:predXOR2}
For all $X,Y \in \{0,1\}$ and $Z \in \mZ$, we have
\[\predadv(X \oplus Y \mid YZ) = \predadv(X \mid YZ)\;.\]
\end{lemma}

\begin{proof}
If a function $f(y,z)$ can predict $X$
with advantage $a$, then the function $f'(y,z) := f(y,z) \oplus y$ can predict
$X \oplus Y$ with advantage $a$, and if $g(y,z)$ can predict $X \oplus Y$ with advantage $a$, then the function $g'(y,z) := g(y,z) \oplus y$ can predict $X$ with advantage $a$.
\end{proof}

\chapter{Secure Two-Party Computation} \label{chap:secTPC}

In this chapter we give an introduction to a simplified version of
\emph{universally composable two-party computation}. We define security in the malicious
and the semi-honest models, and show that these definitions allow protocols to be composed.
Finally, we show that
security in the malicious model does not imply security in the semi-honest model, and give a
weaker security definition for the semi-honest model for which this implication holds.

\section{Two-Party Computation}

We start with some basic definitions.
Our definitions are based on the formalism by Maurer \cite{Maurer06}, as well
as the formalisms of Backes, Pfitzmann and Waidner
\cite{PfiWai00,BaPfWa03} and Canetti \cite{Canetti00}, but simplified and adapted for our needs.
Since we will only consider two players interacting with each other, we can simplify the notation.
For example, we will not use any identification tags.

We will model everything in terms of \emph{systems} which may interact with other systems or the environment
via \emph{interfaces}. We say that system $\bF$ \emph{implements} a set $\mI$ of interfaces.
There are two players present, which we will call $\PlayerA$ and $\PlayerB$.
The set of interfaces $\mI$ can be divided into two sets: the set $\mI_\PlayerA$
of the interfaces belonging to player $\PlayerA$, and the set $\mI_\PlayerB$ of the interfaces belonging to
player $\PlayerB$.

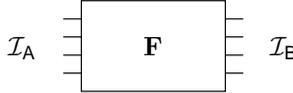
\begin{center} { \input{ps/system.pstex_t} } \end{center}

A system has an internal, possibly infinite supply of randomness. Every output of the system is a function of the received messages so far, and the internal
randomness. The system is efficient if these functions can be evaluated efficiently,
i.e., using a polynomial time turing machine.
The whole interaction between systems is \emph{asynchronous}, i.e.,
there is no global time.

Two systems $\bF$ and $\bG$ can be \emph{composed in parallel} to a new system, denoted by $\bF \| \bG$. The two sub-systems $\bF$ and $\bG$ do not interact with each other, and the resulting system
has all the interfaces of the two subsystems.

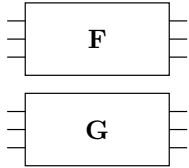
\begin{center} { \begin{picture}(0,0)%
\includegraphics{ps/parallel.pstex}%
\end{picture}%
\setlength{\unitlength}{3947sp}%
\begingroup\makeatletter\ifx\SetFigFont\undefined%
\gdef\SetFigFont#1#2#3#4#5{%
  \reset@font\fontsize{#1}{#2pt}%
  \fontfamily{#3}\fontseries{#4}\fontshape{#5}%
  \selectfont}%
\fi\endgroup%
\begin{picture}(1149,1036)(4151,-3616)
\put(4726,-2873){\makebox(0,0)[b]{\smash{{\SetFigFont{10}{12.0}{\sfdefault}{\mddefault}{\updefault}$\bF$}}}}
\put(4726,-3436){\makebox(0,0)[b]{\smash{{\SetFigFont{10}{12.0}{\sfdefault}{\mddefault}{\updefault}$\bG$}}}}
\end{picture}%
 } \end{center}

We denote the parallel composition of $n$ times the same system $\bF$ by $\bF^{\|n}$.

A system  $\bG$ may use another system $\bF$ as a subsystem, which we denote by
$\bG(\bF)$. $\bG$ may have some interfaces that are connected to
some interfaces of $\bF$. We use this notation because $\bG$ can be viewed as a function that
transforms a system $\bF$ into a system $\bG(\bF)$. $\bF \| \bG$ is a special case of this composition.

\begin{center} { \begin{picture}(0,0)%
\includegraphics{ps/subsystem.pstex}%
\end{picture}%
\setlength{\unitlength}{3947sp}%
\begingroup\makeatletter\ifx\SetFigFont\undefined%
\gdef\SetFigFont#1#2#3#4#5{%
  \reset@font\fontsize{#1}{#2pt}%
  \fontfamily{#3}\fontseries{#4}\fontshape{#5}%
  \selectfont}%
\fi\endgroup%
\begin{picture}(1825,1037)(3813,-3504)
\put(4726,-2817){\makebox(0,0)[b]{\smash{{\SetFigFont{10}{12.0}{\sfdefault}{\mddefault}{\updefault}$\bF$}}}}
\put(4726,-3379){\makebox(0,0)[b]{\smash{{\SetFigFont{10}{12.0}{\sfdefault}{\mddefault}{\updefault}$\bG$}}}}
\end{picture}%
 } \end{center}

\section{Distinguishing Systems}

Definition~\ref{def:adv-4-RV} in Section~\ref{sec:statDist}, which defines the distinguishing advantage for random variables, can be generalized to systems in a straightforward way.
A distinguisher is now an algorithm $D$ that interacts with a system $\bF$
and outputs $0$ or $1$.

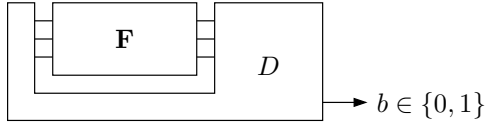
\begin{center} { \input{ps/distinguisher.pstex_t} } \end{center}

\begin{definition} \label{def:adv}
For two systems $\bF$ and $\bF'$, the \emph{distinguishing advantage} of a distinguisher $D$ in distinguishing $\bF$ from $\bF'$ is
\[ \adv^{D}(\bF,\bF') := \big |\Pr[D(\bF)=1] - \Pr[D(\bF')=1] \big |\;.\]
The distinguishing advantage of a class $\mD$ of distinguishers in distinguishing $\bF$ from $\bF'$ is
\[ \adv^{\mD}(\bF,\bF') := \max_{D \in \mD} \adv^{D}(\bF,\bF')\;.\]
\end{definition}

The distinguishing advantage of systems still has the same important properties as the distinguishing advantage for random variables. Obviously, we have $\adv^\mD(\bF,\bF) = 0$ and $\adv^\mD(\bF',\bF) = \adv^\mD(\bF,\bF')$, for all $\bF$ and $\bF'$. Furthermore, it also satisfies the triangle inequality:
\[ \adv^\mD(\bF,\bF'') \leq \adv^\mD(\bF,\bF') + \adv^\mD(\bF',\bF'')\;,\]
for all $\mD$, $\bF$, $\bF'$, and $\bF''$.

Except in Chapter~\ref{chap:compWOT}, $\mD$ will be the set of all possible (also inefficient) distinguishers. In this case, we will omit the $\mD$ and only write $\adv(\bF,\bF')$. We also write $\bF \equiv_{\eps} \bF'$ for $\adv(\bF,\bF') \leq \eps$, and $\bF \equiv \bF'$ for $\adv(\bF,\bF') = 0$.

Similar to Lemma~\ref{lem:statDist-dataprocessing}, we have for all systems $\bG$, $\bF$, and $\bF'$
\[ \adv(\bG(\bF),\bG(\bF')) \leq \adv(\bF,\bF') \;,\]
since any distinguisher $D$ that distinguishes $\bG(\bF)$ from $\bG(\bF')$ with an
advantage of $\eps$ can be used to distinguish $\bF$ from $\bF'$, by first applying
$\bG$. If $\mD$ is the class of all \emph{efficient} distinguishers, then
\[ \adv^{\mD}(\bG(\bF),\bG(\bF')) \leq \adv^{\mD}(\bF,\bF') \;,\]
if $\bG$ is efficient.

Note that for the case where $\bF$ and $\bF'$ have no inputs and output random variables $X$ and $X'$, respectively, this definition is equivalent to Definition~\ref{def:adv-4-RV}, and we have
\[ \adv(\bF,\bF') = \Delta(X,X')\;.\]

\section{Adversaries and Secure Protocols}

In this section we define protocols and their security.
In the following, we will often use special systems that only have interfaces for one player $p \in \{\PlayerA,\PlayerB\}$.
We denote such systems by $\bF_p$. For any systems $\bF_\PlayerA$, $\bF_\PlayerB$, and $\bG$,
we have
\[ \bF_\PlayerA(\bF_\PlayerB(\bG)) = \bF_\PlayerB(\bF_\PlayerA(\bG))\;.\]

A system of the form $\bP(\bF) = (\bP_\PlayerA \| \bP_\PlayerB)(\bF) = \bP_\PlayerA(\bP_\PlayerB(\bF))$ is called a \emph{(two-party) protocol}.

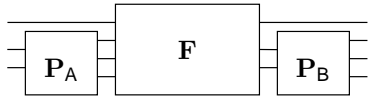
\begin{center} { \input{ps/protocol.pstex_t} } \end{center}

Players may be \emph{honest}, which means that they follow the protocol, or they may be \emph{corrupted} in two different ways. If a player is \emph{actively corrupted}, he may behave in an arbitrary way. If a player
is \emph{passively corrupted}, then he follows the protocol, but forwards everything he sends or receives
immediately over an additional interface that we will call \emph{auxiliary interface}. Such players are also called \emph{honest, but curious}.

The set of all corrupted players are called the \emph{adversary}.
Let \[\mA \subset \{\PlayerA,\PlayerB, \widehat \PlayerA, \widehat \PlayerB \}\] be the set of corrupted players, where $\PlayerA$ and $\PlayerB$ are actively and $\widehat \PlayerA$ and $\widehat \PlayerB$ passively corrupted players. We will assume that this set is \emph{static}, i.e., it is already
determined before the protocol starts.
We will not mix actively and passively corrupted players, and consider two different models.
In the \emph{malicious model}, the players may be actively corrupted, and in the \emph{semi-honest model}
the players may be passively corrupted.
Furthermore, we can ignore the case where $\mA = \{\PlayerA,\PlayerB\}$ or $\mA = \{\widehat \PlayerA,\widehat \PlayerB\}$,
as we never have any requirement for these cases.
Therefore, we only have to consider the case $|\mA| \leq 1$.

Because an adversary may be able to use a system in a different way than the honest players, we will
use the following generalized notion of a system. A \emph{collection of systems} 
\[ \bF = (\bF_{\emptyset}, \bF_{\{\PlayerA\}}, \bF_{\{\PlayerB\}})\;, \qquad
 \bF = (\bF_{ \emptyset}, \bF_{ \{\widehat \PlayerA\}}, \bF_{\{\widehat \PlayerB\}}) \]
(in the malicious or the semi-honest model) defines a different system $\bF_\mA$ for every possible set of corrupted players $\mA$, where the honest players always
have the same interfaces as in $\bF_{\emptyset}$. This means that in $\bF_{\{\PlayerA\}}$ and $\bF_{\{\widehat \PlayerA\}}$, $\PlayerB$ must have the same interfaces as in $\bF_{\emptyset}$, and in $\bF_{\{\PlayerB\}}$ and $\bF_{\{\widehat \PlayerB\}}$, $\PlayerA$ must have the same interfaces as in
$\bF_{\emptyset}$. Furthermore, the system $\bF_\mA$ 
should be at least as good for the adversary as the system $\bF_{\emptyset}$, i.e., the adversary
should always be able to behave honestly.
$\bF_\mA$ can be interpreted as a model of a system where the adversary $\mA$ can corrupt a part of the system $\bF$.

In the following, we will abuse the term ``system'', and also use it for collections of systems. 

\subsection{The Malicious Model}

In the \emph{malicious model}, the adversary is allowed to cheat actively, in an arbitrary way. Therefore,
we do not have any restrictions on how the interface to the adversary may look like,
as long as it allows him to behave honestly, if he wants.

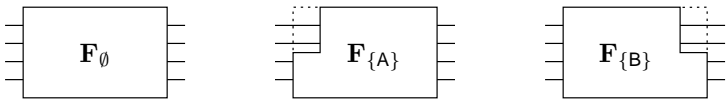
\begin{center} { \input{ps/collection.pstex_t} } \end{center}

We will now define the security of protocols.
We say that a protocol $\bP$ having access to the system $\bF$ securely implements a system $\bG$, if, first of all, $\bG_{\emptyset} \equiv \bP(\bF_{\emptyset})$, i.e., the protocol implements the system $\bG$ correctly, given that both players are honest. Additionally, for $\mA = \{p\}$, we require that the adversary attacking
the protocol has no advantage over another adversary that attacks
$\bG$ directly. We therefore require that there exists a \emph{simulator} $\bS_p$ that simulates
exactly what the adversary would get in the execution of the protocol $\bP(\bF)$. Since the adversary may not follow the protocol, his view of the protocol is in fact the ``raw'' interface of
$\bF$, without his part of the protocol.

\begin{definition} \label{def:sec}
A protocol $\bP(\bF) = (\bP_\PlayerA \| \bP_\PlayerB)(\bF)$ \emph{securely implements a system $\bG$ in the malicious model with an error of at most $\eps$}, if
\begin{itemize}
\item(Correctness) $\bP(\bF_{\emptyset}) \equiv_\eps \bG_{\emptyset}$\;.
\item(Security for $\PlayerA$) There exists a system $\bS_\PlayerB$ (called \emph{the simulator for $\PlayerB$}), such that
\[\bP_\PlayerA(\bF_{\{\PlayerB\}}) \equiv_\eps \bS_\PlayerB(\bG_{\{\PlayerB\}})\;.\]
\item(Security for $\PlayerB$) There exists a system $\bS_\PlayerA$ (called \emph{the simulator for $\PlayerA$}), such that
\[\bP_\PlayerB(\bF_{\{\PlayerA\}}) \equiv_\eps \bS_\PlayerA(\bG_{\{\PlayerA\}})\;.\]
\end{itemize}
\end{definition}

Note that the protocol $\bP(\bF)$ can also be viewed as
a new system $\bE$, defined by $\bE_{\emptyset} := \bP(\bF_{\emptyset})$,
$\bE_{\{\PlayerB\}} := \bP_\PlayerA(\bF_{\{\PlayerB\}})$, and
$\bE_{\{\PlayerA\}} := \bP_\PlayerB(\bF_{\{\PlayerA\}})$. Definition~\ref{def:sec} could then be stated by
comparing the systems $\bE$ and $\bG$.

\begin{figure}
\begin{center} { \input{ps/security.pstex_t} } \end{center}
\caption{The three conditions for the security in the malicious model of a protocol $\bP = (\bP_\PlayerA \| \bP_\PlayerB)$  that uses
$\bF = (\bF_{\emptyset}, \bF_{\{\PlayerA\}}, \bF_{\{\PlayerB\}})$
and implements $\bG = (\bG_{\emptyset}, \bG_{\{\PlayerA\}}, \bG_{\{\PlayerB\}})$.}
\end{figure}

We do generally not require the simulation to be efficient. Therefore, an attack that is efficient
in $\bP(\bF)$ may be mapped to a very inefficient attack in $\bG$. This means that if
 the system $\bG$ is replaced by the protocol $\bP(\bF)$, the adversary
may gain extra possibilities because he may be able to execute some attacks more efficiently
in the new setting. More precisely, he gains the extra possibility of executing
the simulator for free. Depending on the setting, this may be a problem. For example,
if the simulator allows him to invert a one-way function, a system that relies on
the assumption that inverting this one-way function is hard may not be secure anymore. On the other hand, if $\bP(\bF)$ is used in a protocol that is information-theoretically secure, the additional, virtual computing power of the adversary
will be of little use to him.
Therefore, an efficient simulation is preferable, even in the model where the adversary is (potentially) unbounded, because it allows the protocol to be used also in the computational setting.
A very important property of this security definition is that it allows protocols to be \emph{composed}.

\begin{theorem}[Composition theorem, malicious model]
If $\bP(\bF)$ securely implements $\bG$ in the malicious model with an error of at most $\eps_1$, and $\bQ(\bH)$ securely implements $\bF$ in the malicious model with an error of at most $\eps_2$, then
$\bP(\bQ(\bH))$ securely implements $\bG$ in the malicious model with an error of at most $\eps_1 + \eps_2$.
\end{theorem}

\begin{proof}
From $\bQ(\bH_\emptyset) \equiv_{\eps_2} \bF_\emptyset$ follows that
$\bP(\bQ(\bH_\emptyset)) \equiv_{\eps_2} \bP(\bF_\emptyset)$. Since $\bP(\bF_\emptyset) \equiv_{\eps_1} \bG_\emptyset$,
it follows from the triangle inequality that
\[\bP(\bQ(\bH_\emptyset)) \equiv_{\eps_1 + \eps_2} \bG_\emptyset\;.\]

There exists a simulator $\bS_\PlayerB$, such that
$\bQ_\PlayerA(\bH_{\{\PlayerB\}}) \equiv_{\eps_2} \bS_\PlayerB(\bF_{\{\PlayerB\}})$. It follows that
\[\bP_\PlayerA(\bQ_\PlayerA(\bH_{\{\PlayerB\}})) \equiv_{\eps_2}
\bP_\PlayerA(\bS_\PlayerB(\bF_{\{\PlayerB\}})) = \bS_\PlayerB(\bP_\PlayerA(\bF_{\{\PlayerB\}}))\;.\]
Since there exists a simulator $\bT_\PlayerB$ such that
$\bP_\PlayerA(\bF_{\{\PlayerB\}}) \equiv_{\eps_1} \bT_\PlayerB(\bG_{\{\PlayerB\}})$, we have
\[\bS_\PlayerB(\bP_\PlayerA(\bF_{\{\PlayerB\}})) \equiv_{\eps_1} \bS_\PlayerB( \bT_\PlayerB(\bG_{\{\PlayerB\}}))\;.\]
It follows from the triangle inequality that
\[\bP_\PlayerA(\bQ_\PlayerA(\bH_{\{\PlayerB\}})) \equiv_{\eps_1 + \eps_2} \bS_\PlayerB( \bT_\PlayerB(\bG_{\{\PlayerB\}}))\;,\]
and hence the protocol is secure for $\PlayerA$, with an error of at most $\eps_1 + \eps_2$. The security for $\PlayerB$ can be shown in the same way.
\end{proof}

\subsection{The Semi-Honest Model} \label{sec:semi-honest}

In the \emph{semi-honest model}, the adversary is \emph{passive}. Instead of executing $\bP_p$, 
a passively corrupted player $p$ executes $\underline \bP_p$, which is equal to $\bP_p$, but forwards
everything it sends or receives immediately over an auxiliary interface. Note that
the output of the auxiliary interface contains the entire \emph{view} of the corrupted player, and therefore also the output of the honest interface.

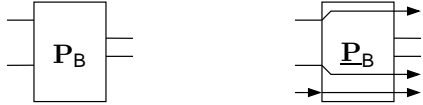
\begin{center} { \input{ps/passiveAdversary.pstex_t} } \end{center}

We require that every system in a collection must also have the same interfaces for the adversary as
for the honest player, because the adversary executes the protocol honestly and can only connect to these interfaces. However, the system has auxiliary output interfaces for the adversary, that provide him with some extra information.

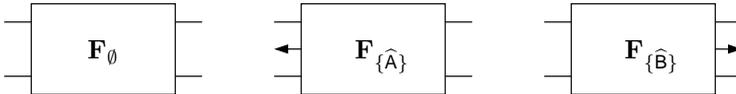
\begin{center} { \input{ps/passiveCollection.pstex_t} } \end{center}

Let $\mA = \{\widehat p \}$. A protocol $\bP(\bF)$ securely implements a system $\bG$ in the
semi-honest model if there exists a simulator $\bS_p$ that accesses the interaction of the system $\bG_{\{\widehat p\}}$ with player $p$ and produces the same output as $\underline \bP_p$. Furthermore, the simulator $\bS_p$ is not allowed to modify the inputs and outputs on the interfaces of the honest player, because we require that the simulated adversary attacking $\bG$ is also only passively, and not actively corrupted. Otherwise, the protocol could not be composed. We get the following definition.

\begin{definition} \label{def:passiveSec}
A protocol $\bP(\bF) = (\bP_\PlayerA \| \bP_\PlayerB)(\bF)$ \emph{securely implements $\bG$ in the
semi-honest model with an error of at most $\eps$}, if
\begin{itemize}
\item(Correctness) $\bP(\bF_{\emptyset}) \equiv_\eps \bG_{\emptyset}$\;.
\item(Security for \PlayerA) There exists a system $\bS_\PlayerB$ (called \emph{the simulator for $\PlayerB$}), that only modifies the auxiliary interfaces, such that
\[(\bP_\PlayerA \| \underline \bP_\PlayerB)(\bF_{\{\widehat \PlayerB\}}) \equiv_\eps \bS_\PlayerB(\bG_{\{\widehat \PlayerB\}})\;.\]
\item(Security for \PlayerB) There exists a system $\bS_\PlayerA$ (called \emph{the simulator for $\PlayerA$}), that only modifies the auxiliary interfaces, such that
\[(\underline \bP_\PlayerA \| \bP_\PlayerB)(\bF_{\{\widehat \PlayerA\}}) \equiv_\eps \bS_\PlayerA(\bG_{\{\widehat \PlayerA\}})\;.\]
\end{itemize}
\end{definition}

\begin{figure}
\begin{center} { \input{ps/passiveSecurity.pstex_t} } \end{center}
\caption{The three conditions for the security in the semi-honest model of a two-party protocol $\bP = (\bP_\PlayerA \| \bP_\PlayerB)$ that uses
$\bF = (\bF_{\emptyset}, \bF_{\{\widehat \PlayerA\}}, \bF_{\{\widehat \PlayerB\}})$
and implements $\bG = (\bG_{\emptyset}, \bG_{\{\widehat \PlayerA\}}, \bG_{\{\widehat \PlayerB\}})$.}
\end{figure}

As in the malicious model, we can show that protocols in the semi-honest model compose.

\begin{theorem}[Composition theorem, semi-honest model] \label{thm:comSemiHonest}
If $\bP(\bF)$ securely implements $\bG$ in the semi-honest model with an error of at most $\eps_1$, and $\bQ(\bH)$ securely implements $\bF$ in the semi-honest model with an error of at most $\eps_2$, then
$\bP(\bQ(\bH))$ securely implements $\bG$ in the semi-honest model with an error of at most $\eps_1 + \eps_2$.
\end{theorem}

\begin{proof}[Proof sketch]
From $\bQ(\bH_\emptyset) \equiv_{\eps_2} \bF_\emptyset$ follows that
$\bP(\bQ(\bH_\emptyset)) \equiv_{\eps_2} \bP(\bF_\emptyset)$. Since $\bP(\bF_\emptyset) \equiv_{\eps_1} \bG_\emptyset$,
it follows from the triangle inequality that 
\[\bP(\bQ(\bH_\emptyset)) \equiv_{\eps_1 + \eps_2} \bG_\emptyset\;.\]

There exists a simulator $\bS_\PlayerB$, such that
$(\bQ_\PlayerA \| \underline \bQ_\PlayerB)(\bH_{\{\widehat \PlayerB\}}) \equiv_{\eps_2} \bS_\PlayerB(\bF_{\{\widehat \PlayerB\}})$. It follows that
\[(\bP_\PlayerA \| \underline \bP_\PlayerB)((\bQ_\PlayerA \| \underline \bQ_\PlayerB)(\bH_{\{\widehat \PlayerB\}})) \equiv_{\eps_2}
(\bP_\PlayerA \| \underline \bP_\PlayerB)(\bS_\PlayerB(\bF_{\{\widehat \PlayerB\}}))\;.\]
Note that $\underline \bP_\PlayerB$ passes all its communication to $\PlayerB$, and $\bS_\PlayerB$ only modifies the additional output, but leaves
the messages of the honest player unchanged. Furthermore, all messages that
$\bS_\PlayerB$ sees will be passed along by the protocol $\underline \bP_\PlayerB$. Hence, we can
move $\bS_\PlayerB$ to the outside, i.e., 
\[(\bP_\PlayerA \| \underline \bP_\PlayerB)(\bS_\PlayerB(\bF_{\{\widehat \PlayerB\}})) =
\bS_\PlayerB((\bP_\PlayerA \| \underline \bP_\PlayerB)(\bF_{\{\widehat \PlayerB\}}))\;.\]
Since there exists a simulator $\bT_\PlayerB$ such that
$(\bP_\PlayerA \| \underline \bP_\PlayerB)(\bF_{\{\widehat \PlayerB\}})
\equiv_{\eps_1} \bT_\PlayerB(\bG_{\{\widehat \PlayerB\}})$, we have
\[ \bS_\PlayerB((\bP_\PlayerA \| \underline \bP_\PlayerB)(\bF_{\{\widehat \PlayerB\}}))
\equiv_{\eps_1}
\bS_\PlayerB(\bT_\PlayerB(\bG_{\{\widehat \PlayerB\}}))\;.
\]
It follows from the triangle inequality that
\[ (\bP_\PlayerA \| \underline \bP_\PlayerB)((\bQ_\PlayerA \| \underline \bQ_\PlayerB)(\bH_{\{\widehat \PlayerB\}}))
\equiv_{\eps_1 + \eps_2} \bS_\PlayerB(\bT_\PlayerB(\bG_{\{\widehat \PlayerB\}}))
= (\bS_\PlayerB(\bT_\PlayerB))(\bG_{\{\widehat \PlayerB\}}) \;.\]
Since $\bS_\PlayerB(\bT_\PlayerB)$ only modifies the auxiliary output, it is a valid simulator,
and hence the protocol is secure for $\PlayerA$ with an error of at most $\eps_1 + \eps_2$. The security for $\PlayerB$ can be shown in the same way.
\end{proof}

\paragraph{From passive to active security.}
Since security against passively corrupted players is quite weak in practice,
it is preferable to have a protocol that is secure against active adversaries.
\cite{GoMiWi87} showed that it is possible to convert any protocols that is secure
in the semi-honest model into a protocol that is secure in the malicious model,
by forcing all players to follow the protocol. To achieve this, every player
must commit himself to all the values he has, and in every step of the protocols,
he must proof in zero-knowledge that he has executed the computation correctly.
We will not further comment on this method, and refer to
\cite{GoMiWi87,Crepea89,CrvGTa95,DaKiSa99,CLOS02,DFMS04} for any details.

\subsection{The Weak Semi-Honest Model} \label{subsec:weakSemi}

We would expect that every protocol that is secure in the malicious model is also secure in the semi-honest model,
since the adversary is restricted in the latter case. Unfortunately, this is not always true. The security condition
in the malicious model only tells us that for any (also semi-honest) adversary, there exists a \emph{malicious} adversary for the ideal system. On the other hand, the security condition in the semi-honest model
requires the adversary for the ideal system to be \emph{semi-honest}. The following example, which we call the \emph{asymmetric dating problem}, illustrates the difference.

\begin{example}[The asymmetric dating problem]
Let the system $\bF$ be defined as follows. It receives  a value $x \in \{0,1\}$ from $\PlayerA$, 
and a value $y \in \{0,1\}$ from $\PlayerB$. Then, it outputs $z := x \cdot y$ to $\PlayerB$.
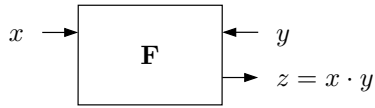
\begin{center} { \input{ps/weakSemiEx.pstex_t} } \end{center}
Let $\Auth$ be a communication channel, and let the protocol $\bP(\Auth)$ be defined as follows. $\bP_\PlayerA$ receives input $x \in \{0,1\}$ and sends $x$ over $\Auth$ to $\PlayerB$.
$\bP_\PlayerB$ receives input $y \in \{0,1\}$ from $\PlayerB$ and $x$ over $\Auth$ and outputs $z := x \cdot y$.
Let us look at the security for $\PlayerA$. It is easy to see that $\bP(\Auth)$ securely implements $\bF$
in the malicious model, since the simulator $\bS_\PlayerB$ can
always input $y=1$ to $\bF$ and obtain the same information as in $\bP(\bF)$. However, the
protocol $\bP(\Auth)$ is \emph{not secure in the semi-honest model}. Since the simulator $\bS_\PlayerB$ is not allowed to
change the value $y$, $\bS_\PlayerB$ cannot simulate $x$ if $y=0$. 
\end{example}

We will now present a weaker security definition for the semi-honest model that is also strictly weaker than 
 the security definition of the malicious model. The only difference to Definition~\ref{def:passiveSec} is that we allow arbitrary simulators, i.e., the simulator may modify the inputs as it likes.

\begin{definition} \label{def:passiveSec2}
A protocol $\bP(\bF) = (\bP_\PlayerA \| \bP_\PlayerB)(\bF)$ \emph{securely implements $\bG$ in the weak semi-honest model with an error of at most $\eps$}, if
\begin{itemize}
\item(Correctness) $\bP(\bF_{\emptyset}) \equiv_\eps \bG_{\emptyset}$\;.
\item(Security for \PlayerA) There exists a system $\bS_\PlayerB$ (called \emph{the simulator for $\PlayerB$}), such that
\[(\bP_\PlayerA \| \underline \bP_\PlayerB)(\bF_{\{\widehat \PlayerB\}}) \equiv_\eps \bS_\PlayerB(\bG_{\{ \PlayerB\}})\;.\]
\item(Security for \PlayerB) There exists a system $\bS_\PlayerA$ (called \emph{the simulator for $\PlayerA$}), such that
\[(\underline \bP_\PlayerA \| \bP_\PlayerB)(\bF_{\{\widehat \PlayerA\}}) \equiv_\eps \bS_\PlayerA(\bG_{\{ \PlayerA\}})\;.\]
\end{itemize}
\end{definition}

\begin{lemma} \label{lem:malIsWeakSemiHonest}
If a protocol $\bP(\bF) = (\bP_\PlayerA \| \bP_\PlayerB)(\bF)$ securely implements $\bG$ in the semi-honest model or in the malicious model with an error of at most $\eps$, then it also securely implements $\bG$ in the weak semi-honest model with an error of at most $\eps$.
\end{lemma}

\begin{proof}
It is obvious that security in the semi-honest model implies security
in the weak semi-honest model.

Let us assume that $\bP(\bF) = (\bP_\PlayerA \| \bP_\PlayerB)(\bF)$ securely implements $\bG$ in the malicious model. The correctness conditions in the weak semi-honest model is the same as in the malicious model.

From the security for $\PlayerA$ follows that there exists a simulator $\bS_\PlayerB$, such that
\[\bP_\PlayerA(\bF_{\{\PlayerB\}}) \equiv_\eps \bS_\PlayerB(\bG_{\{\PlayerB\}})\;.\]
Therefore, we have
\[(\bP_\PlayerA\|\ul \bP_\PlayerB)(\bF_{\{\PlayerB\}})
=
\ul \bP_\PlayerB(\bP_\PlayerA(\bF_{\{\PlayerB\}}))
\equiv_\eps
\ul \bP_\PlayerB (\bS_\PlayerB(\bG_{\{\PlayerB\}}))
= (\ul \bP_\PlayerB (\bS_\PlayerB))(\bG_{\{\PlayerB\}})\;.\]
The system $\bT_{\PlayerB} := \ul \bP_\PlayerB (\bS_\PlayerB)$ is a simulator, which implies security
for $\PlayerA$ in the weak semi-honest model. The security for $\PlayerB$ can be shown in the same way.
\end{proof}

Unfortunately, Definition~\ref{def:passiveSec2} is too weak to allow for composition,
and is therefore not a very useful definition for the security of protocols. The only composition
that is possible is the following, where the outer protocol is secure in
the weak semi-honest model, and the inner protocol is secure in the semi-honest model.

\begin{theorem}[Simple composition theorem, weak semi-honest model] \label{thm:compWeakSemiHonest}
If $\bP(\bF)$ securely implements $\bG$ in the weak semi-honest model with an error of at most $\eps_1$, and $\bQ(\bH)$ securely implements $\bF$ in the semi-honest model with an error of at most $\eps_2$, then
$\bP(\bQ(\bH))$ securely implements $\bG$ in the weak semi-honest model with an error of at most $\eps_1 + \eps_2$.
\end{theorem}

\begin{proof}[Proof sketch]
The proof can be done in the same way as the proof of Theorem~\ref{thm:comSemiHonest}.
The only difference is that now, the simulator $\bT_\PlayerB$ is not restricted in any way.
The argument works in the same way, except that the resulting simulator $\bS_\PlayerB(\bT_\PlayerB)$
will not be restricted either. Hence, the protocol is secure in the weak semi-honest model.
\end{proof}

The weak semi-honest model is useful to prove impossibilities, since it is weaker than the definitions in both the malicious and the semi-honest models. If we
can show that there cannot exist a protocol in the weak semi-honest model, then there can neither exist
a protocol secure in the malicious, nor in the semi-honest model.

\section{Discussion}

In this chapter we presented a simplified universally composable framework for two-party computation.
We did not use the frameworks presented in \cite{PfiWai00,BaPfWa03} or \cite{Canetti00}
because they are far too complex and too general for what we will need them. Our simplified framework
 will make the results in the following chapters easier to state, 
and hopefully also easier to understand. However, this also means that in order to fit our results into
more general frameworks such as \cite{PfiWai00,BaPfWa03} or \cite{Canetti00}, additional work will be needed.

If our protocols are to be executed in an environment where more players are present, we have to make sure that all the other players do not get any information over the inputs or the outputs of $\PlayerA$ and $\PlayerB$.
This can be achieved by requiring that all our 
two-party systems are completely independent of the other players. This means for example that all
channels must be secure and authentic.

\chapter{Oblivious Transfer} \label{chap:ot}

In this chapter we introduce the primitives oblivious transfer (OT) and randomized oblivious transfer (ROT), which is
a variant of OT where the inputs of the honest players are chosen at random.

We start by showing that OT and ROT are equivalent if noiseless communication
is available for free.
Then, we show that ROT is symmetric by presenting a protocol that converts an instance of ROT into an instance of ROT in the opposite direction.
This implies that also the
direction of OT can be reversed in a very simple way (Theorem~\ref{thm:otto}).

In Theorems~\ref{thm:SecCondforROT} and \ref{thm:passiveSecCondforSROT2} we give information-theoretic conditions for the security of ROT.
These conditions are similar to the ones presented in \cite{CSSW06},
however we are able to show a stronger result, as our
conditions imply that a protocol which satisfies them
is \emph{universally composable}, and not only sequentially. Also, our conditions
have explicit error terms, which makes them easier to use.

All the results will be stated in the malicious and the semi-honest model.

\section{(Randomized) Oblivious Transfer}

In this section we will introduce \emph{oblivious transfer (OT)}, and a randomized version of OT called
\emph{randomized OT (ROT)}.

\begin{definition}[Oblivious transfer]
The system $\OT{1}{n}{\ell}$ (or, if the values of $n$ and $\ell$ are clear from the context, $\OTT$) is defined as follows. First, it waits for $\PlayerB$ to send his input $c \in \{0,\dots,n-1\}$, and sends $\PlayerA$
$\bot$\footnote{This is a message without any content, which notifies
$\PlayerA$ about the fact that $\PlayerB$ has sent his input $c$.}
. After having received input $x^n = (x_0,\dots,x_{n-1}) \in \{0,1\}^{\ell \cdot n}$ from $\PlayerA$, it sends 
$y := x_c$ to $\PlayerB$. (Notice that $\OTT = \OTT_{\emptyset} = \OTT_{\{\PlayerA\}} = \OTT_{\{\PlayerB\}}$.)
\end{definition}

\begin{center} { \input{ps/ot.pstex_t} } \end{center}

(Note that from now on, the drawings will also include timing aspects. The time flows
from the top to the bottom.  The dotted lines indicate waiting points, where the system waits
to receive all messages above the line before it continues.)

We use the same version of OT as \cite{CLOS02}, where the sender is notified about the fact
that the receiver has made his choice. Notice that in \cite{Canetti00, Fischl06}, OT has been defined differently.
There, the honest sender does not get this notification.
We do not know how to securely implement OT if the malicious sender does
not get to know the fact that the receiver has made his
choice. Therefore, it is preferable to also give this information to the honest sender. For example,
this allows us to easily implement a bit-commitment protocol from the receiver to the sender.
Also, only this definition allows us
to show that OT and ROT are equivalent if noiseless communication is available for free.

Often, it is much easier to implement a randomized version of OT, called \emph{randomized oblivious transfer} (ROT), first. One way of defining ROT would be to make it equivalent to OT, but
where all the inputs are chosen uniformly at random by the system.
This definition would, however, not be very
useful, because it is too strong: any secure implementation would
have to make sure that all values are
indeed chosen uniformly at random, which can be very difficult.
Furthermore, it turns out that in most applications
this is not needed. We will, therefore, define ROT as a collection of
systems, where the adversary can choose
her own output.

\begin{definition}[Randomized oblivious transfer, malicious model] \label{def:rot}
The system $\ROT{1}{n}{\ell}$ (or, if the values of $n$ and $\ell$ are clear from the context, $\ROTT$) is defined as a collection of
systems 
\[\ROTT = (\ROTT_\emptyset, \ROTT_{\{\PlayerA\}}, \ROTT_{\{\PlayerB\}})\;,\]
where
\begin{itemize}
\item $\ROTT_\emptyset$: The system chooses uniformly at random the value
$x^n \in \{0,1\}^{\ell \cdot n}$ and $c \in \{0,\dots,n-1\}$. It sends
$x^n$ to $\PlayerA$ and $(c,y)$
to $\PlayerB$ where $y = x_c$.
\item $\ROTT_{\{\PlayerA\}}$: The system waits for $\PlayerA$ to send the value
$x^n \in \{0,1\}^{\ell \cdot n}$. Then, it  chooses the value $c \in \{0,\dots,n-1\}$ uniformly at random and sends $(c,y)$
to $\PlayerB$, where $y = x_c$.
\item $\ROTT_{\{\PlayerB\}}$:  The system waits for $\PlayerB$ to send the value
$(c,y) \in \{0,\dots,n-1\} \times \{0,1\}^{\ell}$. Then, it sets $x_c = y$, chooses the values $x_i \in \{0,1\}^{\ell}$ uniformly at random for $i \neq c$, 
 and sends $x^n \in \{0,1\}^{\ell \cdot n}$ to $\PlayerA$.
\end{itemize}
\end{definition}

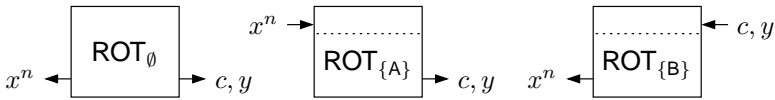
\begin{center} { \input{ps/rot.pstex_t} } \end{center}

We will now show that $\OTT$ and $\ROTT$ are equivalent if communication is given for free, by presenting two protocols that securely implement
one system using one instance of the other and a communication channel.

Protocol $\ROTfromOT = \ROTfromOT_\PlayerA \| \ROTfromOT_\PlayerB$ securely implements $\ROTT$ from one instance of $\OTT$, and is defined as follows.

\begin{protocol} $\ROTfromOT_\PlayerA$:
\begin{enumerate}
\item Choose $x^n \in \{0,1\}^{\ell \cdot n}$ uniformly at random.
\item Send $x^n$ to $\OTT$.
\item Receive $\bot$ from $\OTT$.
\item Output $x^n$. 
\end{enumerate}
$\ROTfromOT_\PlayerB$:
\begin{enumerate}
\item Choose $c \in \{0,\dots,n-1\}$ uniformly at random.
\item Send $c$ to $\OTT$.
\item Receive $y \in \{0,1\}^{\ell}$ from $\OTT$.
\item Output $(c,y)$.
\end{enumerate}
\end{protocol}

\begin{center} { \input{ps/ROTfromOT.pstex_t} } \end{center}

\begin{lemma}
$\ROTfromOT(\OT{1}{n}{\ell})$ securely implements $\ROT{1}{n}{\ell}$ in the malicious model.
\end{lemma}

\begin{proof}
Obviously, we have $\ROTT_{\emptyset} \equiv \ROTfromOT(\OTT)$. 

$\ROTfromOT_\PlayerA(\OTT)$ waits for input $c$ from $\PlayerB$, and then
outputs $x^n$ to $\PlayerA$, where all $x_i$ are chosen uniformly at random and independently of the rest,
and $y := x_c$ to $\PlayerB$.
We define $\bS_\PlayerB$ as follows.
It waits for input $c$ from $\PlayerB$. Then it chooses $y \in \{0,1\}^\ell$ uniformly
at random, sends $(c,y)$ to $\ROTT_{\{\PlayerB\}}$, and outputs $y$.
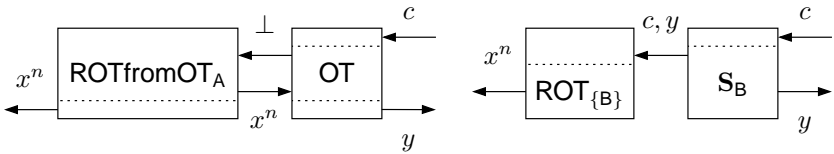
\begin{center} { \input{ps/ROTfromOT-SB.pstex_t} } \end{center}
It is easy to verify that
$\ROTfromOT_\PlayerA(\OTT) \equiv \bS_\PlayerB(\ROTT_{\{\PlayerB\}})$.

$\ROTfromOT_\PlayerB(\OTT)$ outputs $\bot$ to $\PlayerA$. It waits for input $x^n$ from $\PlayerA$, chooses a value $c  \in \{0,\dots,n-1\}$ uniformly at random, and sends $c$ and $y := x_c$ to $\PlayerB$.
We define $\bS_\PlayerA$ as follows.
It outputs $\bot$ to $\PlayerA$. It waits for input $x^n$ from $\PlayerA$ and sends it to $\ROTT_{\{\PlayerA\}}$.
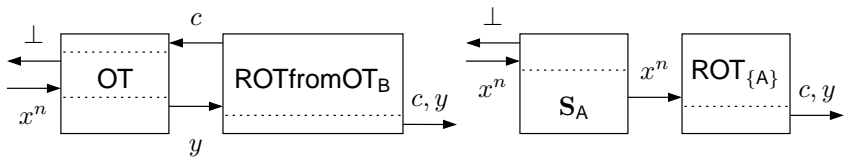
\begin{center} { \input{ps/ROTfromOT-SA.pstex_t} } \end{center}
It is easy to verify that
$\ROTfromOT_\PlayerB(\OTT) \equiv \bS_\PlayerA(\ROTT_{\{\PlayerA\}})$.
\end{proof}

To implement $\OTT$ from $\ROTT$, $\PlayerA$ and $\PlayerB$ need to be able to communicate.
We will therefore additionally need the system $\Auth$, which implements a communication channel from $\PlayerA$ to $\PlayerB$ and from $\PlayerB$ to $\PlayerA$.
Note that, in contrast to $\OTT$ or $\ROTT$, $\Auth$ can be used many
times.

\begin{definition}[Channel]
The system $\Auth$ is defined as follows. Every time it receives a message 
$m \in \{0,1\}^*$ from $p \in \{\PlayerA,\PlayerB\}$, it sends it to the other player in $\{\PlayerA,\PlayerB\}$.
\end{definition}

We can now state the protocol $\OTfromROT$, which was first proposed in
\cite{BBCS92} to securely implements $\OTT$ using $\ROTT$ and $\Auth$.
The protocol is defined as follows.

\begin{protocol} $\OTfromROT_\PlayerA$:
\begin{enumerate}
\item Receive $d \in \{0,\dots,n-1\}$ from
$\Auth$ and $(x')^n \in \{0,1\}^{\ell \cdot n}$ from $\ROTT$.
\item Output $\bot$ to $\PlayerA$.
\item Receive $x^n \in \{0,1\}^{\ell \cdot n}$ from $\PlayerA$.
\item Send $m^n \in \{0,1\}^{\ell \cdot n}$
to $\Auth$, where $m_i := x_i \oplus x'_{i + d \pmod n}$.
\end{enumerate}
$\OTfromROT_\PlayerB$:
\begin{enumerate}
\item Receive $c \in \{0,\dots,n-1\}$ from $\PlayerB$ and $(c',y') \in
 \{0,\dots,n-1\} \times \{0,1\}^{\ell}$ from $\ROTT$.
\item Send $d := c' - c \pmod n$ to $\Auth$.
\item Receive $m^n \in \{0,1\}^{\ell \cdot n}$ from $\Auth$.
\item Output $y := m_c \oplus y'$ to $\PlayerB$.
\end{enumerate}
\end{protocol}

\begin{center} { \input{ps/OTfromROT.pstex_t} } \end{center}

\begin{lemma} \label{lem:ROTfromOTactive}
$\ROTfromOT(\ROT{1}{n}{\ell}\|\Auth)$ securely implements $\OT{1}{n}{\ell}$ in the malicious model.
\end{lemma}

\begin{proof}
$\OTfromROT(\ROTT_{\emptyset} \| \Auth)$ waits for input $c$ from $\PlayerB$, and sends
$\bot$ to $\PlayerA$. After receiving $x^n$ from $\PlayerA$,
it sends 
\begin{align*}
y &= m_c \oplus y'
= x_c \oplus x'_{c + d \!\pmod n} \oplus y'
= x_c \oplus x'_{c + c' - c \!\pmod n} \oplus y' \\
&= x_c \oplus x'_{c} \oplus y'
= x_c
\end{align*}
to $\PlayerB$. (We used the fact that $y' = x_c'$.) Hence, we have
\[\OTT \equiv \OTfromROT(\ROTT_{\emptyset} \| \Auth)\;.\]

$\OTfromROT_\PlayerA(\ROTT_{\{\PlayerB\}} \| \Auth)$ waits for $(c',y')$ and $d$ from
$\PlayerB$, and then outputs
$\bot$ to $\PlayerA$. It then waits for its input $x^n$ from $\PlayerA$ and outputs $m^n$ to $\PlayerB$, where $m_{c' - d} = x_{c'-d} \oplus y'$, and all the other values $m_i$ are uniformly distributed and independent of the rest.
We define $\bS_\PlayerB$ as follows. It waits for input $(c',y')$ on the $\ROTT_{\{\PlayerB\}}$ interface, and $d$ on the $\Auth$ interface. Then it sends $c := c' - d$ to $\OTT$.
It receives $y = x_{c'-d}$ from $\OTT$, sets $m_{c' - d} := y \oplus y'$ and chooses all other
$m_i$ uniformly at random. Finally, it outputs $m^n$ on the $\Auth$ interface.
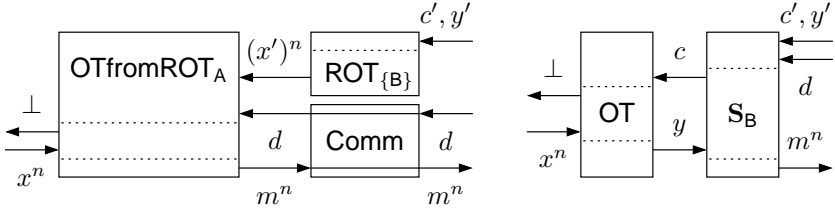
\begin{center} { \input{ps/OTfromROT-SB.pstex_t} } \end{center}
It is easy to verify that
$\OTfromROT_\PlayerA(\ROTT_{\{\PlayerB\}} \| \Auth) \equiv \bS_\PlayerB(\OTT)$.

$\OTfromROT_\PlayerB(\ROTT_{\{\PlayerA\}} \| \Auth)$ waits for $(x')^n$ from the $\ROTT$ interface, and the input $c$ from $\PlayerB$. It chooses $d$ uniformly at random and sends it to $\PlayerA$. After receiving also
$m^n$ from the $\Auth$ interface from $\PlayerA$, it outputs $y = m_c \oplus y' = m_c \oplus x'_{c+d \pmod n}$ to $\PlayerB$.
We define $\bS_\PlayerA$ as follows. It waits for $x'^n$ on the $\ROTT$ interface,
and $\bot$ from $\OTT$. Then, it chooses $d$ uniformly at random and sends it to $\PlayerA$ on the $\Auth$ interface. After receiving $m^n$ on the $\Auth$ interface, it sends
the inputs $x_i := m_i + x'_{i+d}$ for $i \in \{0,\dots,n-1\}$ to $\OTT$.
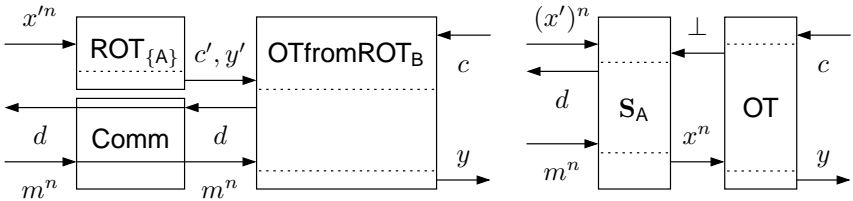
\begin{center} { \input{ps/OTfromROT-SA.pstex_t} } \end{center}
It is easy to verify that $\OTfromROT_\PlayerB(\ROTT_{\{\PlayerA\}} \| \Auth) \equiv \bS_\PlayerA(\OTT)$.
\end{proof}

\section{Oblivious Transfer is Symmetric}

Even though $\ROT{1}{2}{1}$ does not look very symmetric, it is \emph{almost} symmetric,
as we will show in this section.
In particular, we will show that $\ROT{1}{2}{1}$ can be \emph{reversed}, using a very simple
transformation that we will call $\ROTOR$. Let $\TOR{1}{2}{1}$ be $\ROT{1}{2}{1}$ in the opposite direction.

The protocol $\ROTOR$
implements $\TOR{1}{2}{1}$ using $\ROT{1}{2}{1}$ and is defined as follows.

\begin{protocol}
$\ROTOR_\PlayerA$:
\begin{enumerate}
\item Receive $(x'_0,x'_1)$ from $\ROTT$.
\item Output $(c,y)$ to $\PlayerA$, where $y = x'_0$ and $c = x'_0 \oplus x'_1$.
\end{enumerate}
$\ROTOR_\PlayerB$:
\begin{enumerate}
\item Receive $(c',y')$ from $\ROTT$.
\item Output $(x_0,x_1)$, where $x_0 = y'$ and $x_1 = c' \oplus y'$.
\end{enumerate}
\end{protocol}

\begin{center} { \input{ps/ROTOR.pstex_t} } \end{center}

\begin{lemma} \label{lem:ROTOR}
$\ROTOR(\ROT{1}{2}{1})$ securely implements $\TOR{1}{2}{1}$ in the malicious model.
\end{lemma}

\begin{proof} From
\begin{align*}
x_c
& = x_0 \oplus (x_0 \oplus x_1) \cdot c
 = y' \oplus (y' \oplus c' \oplus y') \cdot ( x'_0 \oplus x'_1) \\
& = y' \oplus c' \cdot ( x'_0 \oplus x'_1)
 = y' \oplus x'_{c'} \oplus x'_0
 = x'_0
 = y
\end{align*}
follows that $\TORR_{\emptyset} \equiv \ROTOR(\ROTT_{\emptyset})$. We choose
$\bS_{\PlayerB} := \ROTOR_\PlayerB$ and $\bS_{\PlayerA} := \ROTOR_\PlayerA$. It is easy
to verify that $\ROTOR_\PlayerA(\ROTT_{\{\PlayerB\}}) \equiv \bS_{\PlayerA}(\TORR_{\{\PlayerB\}})$
and $\ROTOR_\PlayerB(\ROTT_{\{\PlayerA\}}) \equiv \bS_{\PlayerB}(\TORR_{\{\PlayerA\}})$.
\end{proof}

Let $\TO{1}{2}{1}$ be $\OT{1}{2}{1}$ in the opposite direction.
Using the protocols
$\ROTfromOT$, $\ROTOR$ and $\OTfromROT$, we can implement $\OT{1}{2}{1}$ using one instance of $\TO{1}{2}{1}$, and get the following theorem.

\begin{theorem} \label{thm:otto}
$\OT{1}{2}{1}$ can be securely implemented in the malicious model using $\Auth$ and one instance of $\TO{1}{2}{1}$.
\end{theorem}

Protocols that implement $\OT{1}{2}{1}$ from $\TO{1}{2}{1}$ have previously been 
presented in \cite{CreSan91}, and independently in \cite{OsVeYu91}. However, Theorem~\ref{thm:otto} leads to a much simpler and more
efficient protocol. The protocol of Theorem~\ref{thm:otto} has been proposed in \cite{WolWul06},
together with an even more efficient protocol, that only used one bit of communication. Unfortunately,
that protocol does not work here. The problem is that we are not able to send the
value $\bot$ to $\PlayerA$ as soon as $\PlayerB$ has made his choice, if $\PlayerB$ makes his choice before $\PlayerA$
has given her input.

\section {In the Semi-Honest Model}

In Section~\ref{subsec:weakSemi} we have seen that
security in the malicious model does not always imply security in the semi-honest model.
We will therefore show that the protocols $\ROTfromOT$, $\OTfromROT$ and $\ROTOR$
are also secure in the semi-honest model.

First of all, we have to adjust the definition of $\ROTT$.
Since a semi-honest adversary will always choose its random inputs truly random, we have
$ \semiROTT_{\{\widehat \PlayerA\}} = \semiROTT_{\{\widehat \PlayerB\}} = \semiROTT_{\emptyset}$.

\begin{lemma}
Protocol $\ROTfromOT(\OT{1}{n}{\ell})$ securely implements $\semiROT{1}{n}{\ell}$ in the semi-honest model.
\end{lemma}

\begin{proof}
Obviously, we have $\semiROTT_{\emptyset} \equiv \ROTfromOT(\OTT)$. 

$(\ROTfromOT_\PlayerA \| \ul \ROTfromOT_\PlayerB)(\OTT)$ outputs $x^n$ to $\PlayerA$
and $c$ (on the auxiliary interface) and $(c,y)$ to $\PlayerB$.
$\bS_\PlayerB$ receives $(c,y)$, outputs $c$ on the auxiliary interface, and passes $(c,y)$ along to $\PlayerB$.
We have
\[(\ROTfromOT_\PlayerA \| \ul  \ROTfromOT_\PlayerB)(\OTT) \equiv \bS_\PlayerB(\semiROTT_{\{\widehat \PlayerB\}})\;.\]

$(\ul \ROTfromOT_\PlayerA \| \ROTfromOT_\PlayerB)(\OTT)$ outputs $x^n$ and $\bot$ (on the auxiliary interface) and $x^n$ to $\PlayerA$, and $(c,y)$ to $\PlayerB$.
$\bS_\PlayerA$ receives $x^n$, outputs $x^n$ and $\bot$ on the 
auxiliary interface and then passes $x^n$ along to $\PlayerA$. We have
\[(\ul \ROTfromOT_\PlayerA \| \ROTfromOT_\PlayerB)(\OTT) \equiv \bS_\PlayerA(\semiROTT_{\{\widehat \PlayerA\}})\;.\]
Hence, the protocol is secure  in the semi-honest model.
\end{proof}

\begin{lemma}
$\OTfromROT(\semiROT{1}{n}{\ell}\|\Auth)$ securely implements $\OT{1}{n}{\ell}$ in the semi-honest model.
\end{lemma}

\begin{proof}
We have seen in Lemma~\ref{lem:ROTfromOTactive} that $\OTT \equiv \OTfromROT(\ROTT_{\emptyset}\| \Auth)$.

$(\OTfromROT_\PlayerA\|\ul \OTfromROT_\PlayerB)(\semiROTT_{\{\widehat \PlayerB\}} \| \Auth)$ chooses $(c',y')$ uniformly at random, outputs it on the auxiliary interface to $\PlayerB$, and waits for input $c$ from $\PlayerB$. Then it outputs $d = c'-c$ on the auxiliary interface to $\PlayerB$, and $\bot$ to $\PlayerA$. After receiving $x^n$ from $\PlayerA$, it outputs
$m^n$ on the auxiliary interface and $y = x_c$ on the normal interface to $\PlayerB$, where $m_c = y' \oplus y$ and
all the other values $m_i$ are chosen uniformly at random.

$\bS_{\PlayerB}$ chooses $(c',y')$ uniformly at random and
outputs it on the auxiliary interface. It waits for input $c$, passes it along to $\OTT$,
and outputs $d := c' - c \pmod n$ on the auxiliary interface. After receiving $y = x_c$ from $\OTT$, it outputs
$m^n$ to $\PlayerB$, where $m_c = y' \oplus y$ and the remaining values are chosen uniformly at random.
Finally, it outputs $y$. It is easy to verify that
\[(\OTfromROT_\PlayerA\|\ul \OTfromROT_\PlayerB)(\semiROTT_{\{\widehat \PlayerB\}} \| \Auth) \equiv \bS_\PlayerB(\OTT)\;.\]

$(\ul \OTfromROT_\PlayerA\|\OTfromROT_\PlayerB)(\ROTT_{\{\widehat \PlayerA\}} \| \Auth)$ chooses $(x')^n$ uniformly at random and
outputs it on the auxiliary interface to $\PlayerA$. After receiving $c$ from $\PlayerB$, it chooses $d$ uniformly at random and outputs $d$ and $\bot$ to
$\PlayerA$ on the auxiliary interface. After receiving $x^n$ from $\PlayerA$, it outputs
$m^n$ to $\PlayerA$, where $m_i := x_i \oplus x'_{i + d \pmod n}$, and $y=x_c$ to $\PlayerB$.

$\bS_{\PlayerA}$ chooses $(x')^n$ at random and outputs it on the auxiliary interface to
$\PlayerA$. After receiving $\bot$ from $\OTT$, it outputs $d$ chosen uniformly at random 
on the auxiliary interface and passes $\bot$ along to $\PlayerA$. After receiving
$x^n$, it outputs $m^n$ to $\PlayerA$, where $m_i := x_i \oplus x'_{i + d \pmod n}$, and passes $x^n$ along
to $\OTT$.
It is easy to verify that
\[\OTfromROT_\PlayerB(\semiROTT_{\{\widehat \PlayerA\}} \| \Auth) \equiv \bS_\PlayerA(\OTT)\;.\]
Hence, the protocol is secure.
\end{proof}

Protocol $\ROTOR$ applies a bijective function on the output of $\semiROTT$. Hence, all
the auxiliary output can be simulated from the output of $\semiTOR{1}{2}{1}$, and we get the following lemma.

\begin{lemma} \label{lem:SROTfromROT}
$\ROTOR(\semiROT{1}{2}{1})$ securely implements $\semiTOR{1}{2}{1}$ in the semi-honest model.
\end{lemma}

\section{Information-Theoretic Security Conditions} \label{sec:infoSecCond}

We will now present information-theoretic conditions, which imply that a protocol securely implements
$\ROTT$ either in the malicious or the semi-honest models.
 
\subsection{In the Malicious Model}

The following information-theoretic conditions are similar to the conditions presented in \cite{CSSW06}, and to the definitions of randomized oblivious transfer used in \cite{DFSS06} and \cite{Wullsc07}. However, our correctness
condition is stron\-ger, because we require the outputs to be random, if the players are honest.

\begin{theorem} \label{thm:SecCondforROT} 
  A protocol $\bP(\bF) = (\bP_\PlayerA \| \bP_\PlayerB)(\bF)$ securely implements $\ROT{1}{n}{\ell}$ with
  an error of at most $\eps$ in the malicious model, if
\begin{itemize}
\item(Correctness) $\bP(\bF_{\emptyset}) \equiv_\eps \ROTT_{\emptyset}$.
\item(Security for \PlayerA) $\bP_\PlayerA(\bF_{\{\PlayerB\}})$ interacts over the interfaces
belonging to $\PlayerB$ (which produces a transcript $V$), and after the last
input is received, it outputs $X^n \in \{0,1\}^{\ell \cdot n}$ to $\PlayerA$.
There exists a conditional probability distribution $P_{C \mid X^n V}$ that
produces a random variable $C \in \{0,\dots,n-1\}$ such that
$(X_0,\dots,X_{C-1},X_{C+1},\dots,X_{n-1})$ is $\eps$-close to uniform with respect to
$(C,X_{C},V)$.
\item(Security for \PlayerB) $\bP_\PlayerB(\bF_{\{\PlayerA\}})$ interacts over the interfaces belonging to $\PlayerA$ (which produces a transcript $U$), and after the last
input is received, it outputs $(C,Y) \in \{0,\dots,n-1\} \times  \{0,1\}^\ell$ to $\PlayerB$ where $C$ is
$\eps$-close to uniform with respect to $U$.
\end{itemize}
\end{theorem}

\begin{proof}
Let $\bP(\bF)$ satisfy these conditions.
The correctness condition is the same as in Definition \ref{def:sec}.

Let $\bS_\PlayerB$ first simulate $\bP_\PlayerA(\bF_{\{\PlayerB\}})$ which interacts with $\PlayerB$ and outputs $(X')^n$ and the transcript $V$ of the interaction with $\PlayerB$. Then, it 
samples $C$ according to $P_{C \mid X^n=(x')^n,V=v}$
 and sends $(C,Y)$ to $\ROTT_{\{\PlayerB\}}$, where $Y := X'_{C}$. $\ROTT_{\{\PlayerB\}}$ will output $X^n$ to $\PlayerA$, where $X_{C} = X'_{C}$ and 
 \[(X_0,\dots,X_{C-1},X_{C+1},\dots,X_{n-1})\] is chosen uniformly at random and independent from the rest. Since 
 \[(X'_0,\dots,X'_{C-1},X'_{C+1},\dots,X'_{n-1})\] is $\eps$-close to uniform with respect to $(C,X_{C},V)$, we have
 \[ (X^n,C,V) \equiv_\eps ((X')^n,C,V) \;,\]
from which follows that
\[\bS_\PlayerB(\ROTT_{\{\PlayerB\}}) \equiv_{\eps} \bP_\PlayerA(\bF_{\{\PlayerB\}})\;.\]

$\bS_\PlayerA$ is defined as follows. First, it simulates $\bP_\PlayerB(\bF_{\{\PlayerA\}})$,
which interacts with $\PlayerA$ and outputs
$(C',Y')$ and the transcript $U$ of the interaction with $\PlayerA$.  Since $C'$ is $\eps$-close to uniform with
respect to $U$, we have
\[ P_{C'Y'U} = P_{C'U}P_{Y'\mid UC'} \equiv_{\eps} P_{\overline C} P_{U} P_{Y' \mid UC'}\;,\]
where $P_{\overline C}$ is the uniform distribution over $\{0,1\}$. 
$\bS_\PlayerA$ now calculates $X'^n$, where $X'_i$ is sampled according to
the probability distributions $P_{Y' \mid U,C'=i}$, and sends them to $\ROTT_{\{\PlayerA\}}$.
Note that the behavior of
the system $\bP_\PlayerB(\bF_{\{\PlayerA\}})$ is known, and therefore also the probability
distribution $P_{Y' \mid U,C'=i}$.
$\PlayerB$ receives a value $C$ chosen uniformly at random, and $Y=X'_C$ distributed according to
$P_{Y' \mid U,C'=c}$. We have
\[P_{CYU} = P_{U} P_{C \mid U} P_{Y' \mid UC'} = P_{U} P_{\ol C} P_{Y' \mid UC'} \equiv_\eps P_{C'Y'U}\;,\] and, 
therefore,
\[\bS_\PlayerA(\ROTT_{\{\PlayerA\}}) \equiv_{\eps}  \bP_\PlayerB(\bF_{\{\PlayerA\}}) \;.\]
\end{proof}

Note that the simulation given in Theorem~\ref{thm:SecCondforROT} is not necessarily efficient.

\subsection{In the Semi-Honest Model}

\begin{theorem} \label{thm:passiveSecCondforSROT2}
Let $\eps \geq 0$. Let $\bP(\bF) = (\bP_\PlayerA \| \bP_\PlayerB)(\bF)$ 
 be a protocol that outputs $X^n$ to $\PlayerA$ and $(C,Y)$ to $\PlayerB$, and let $U$ be the auxiliary output to $\PlayerA$ given  by $\ul \bP_\PlayerA$, and $V$ be the auxiliary output to $\PlayerB$ given by $\ul \bP_\PlayerB$.
$\bP(\bF)$ securely implements $\semiROT{1}{n}{\ell}$ with
  an error of at most $3\eps$ in the semi-honest model, if

\begin{itemize}
\item(Correctness) $\bP(\bF_{\emptyset}) \equiv_\eps \ROTT_{\emptyset}$.
\item(Security for $\PlayerA$)
$(X_0,\dots,X_{C-1},X_{C+1},\dots,X_{n-1})$ is $\eps$-close to uniform with respect to $(C,Y,V)$.
\item(Security for $\PlayerB$)
$C$ is $\eps$-close to uniform with respect to $(X^n,U)$.
\end{itemize}
\end{theorem}

\begin{proof}
Let $\bP(\bF)$ satisfy these conditions and
let $P_{\ol {X}^n \ol{C Y}}$ be the output distribution of $\semiROT{1}{n}{\ell}$.
We have
$P_{X^n C Y} \equiv_{\eps} P_{\ol {X}^n \ol{C Y}}$.
Obviously, the correctness condition is satisfied with an error of at most $\eps$.

We define $\bS_\PlayerB$ as follows. After receiving $(C,Y)$, it samples a value
$V'$ distributed according to $P_{V \mid CY}$ and outputs $(C,Y,V')$.
We get
\begin{align*}
P_{X_0X_1CYV}
&=\; P_{X_0\dots X_{C-1}X_{C+1}\dots X_{n-1}CYV} P_{X_C \mid X_0\dots X_{C-1}X_{C+1}\dots X_{n-1}CYV}\\
&\equiv_{\eps} P_{X_0\dots X_{C-1}X_{C+1}\dots X_{n-1}CYV} P_{\ol{X}_{\ol{C}} \mid \ol{CY}}  \\
&\equiv_{\eps} P_{CYV} P_{\ol X_0\dots \ol X_{C-1}\ol X_{C+1}\dots \ol X_{n-1}} P_{\ol X_{\ol C} \mid \ol C\ol Y} \\
&=\; P_{CY} P_{V \mid CY} P_{\ol X_0\dots \ol X_{C-1}\ol X_{C+1}\dots \ol X_{n-1}} P_{\ol X_{\ol C} \mid \ol C\ol Y} \\
&\equiv_{\eps}
P_{\ol{CY}} P_{V \mid CY} P_{\ol X_0\dots \ol X_{\ol C-1}\ol X_{\ol C+1}\dots \ol X_{n-1}} P_{\ol X_{\ol C} \mid \ol C\ol Y} \\
&=\; P_{\ol {X}^n \ol{C Y}} P_{V \mid CY}
= P_{\ol {X}^n \ol{C Y}} P_{V' \mid CY}
\end{align*}
and, therefore,
\[(\bP_\PlayerA \| \ul  \bP_\PlayerB)(\bF_{\{\widehat \PlayerB\}}) \equiv_{3\eps} \bS_\PlayerB(\ROTT_{\{\widehat \PlayerB\}})\;.\]

We define $\bS_\PlayerA$ as follows. After receiving $X^n$, it samples a value
$U'$ distributed according to $P_{U \mid X^n}$ and outputs $(X^n,U')$.
We get
\begin{align*}
P_{X^n CYU}
&=\; P_{X^n CU} P_{Y \mid X^n CU} \equiv_\eps P_{X^n CU} P_{\ol{Y} \mid \ol{X}^n \ol{C}}\\
&\equiv_{\eps} P_{X^n U} P_{\ol C} P_{\ol Y \mid \ol {X}^n \ol{C}} 
= P_{X^n} P_{U \mid X^n} P_{\ol C} P_{\ol Y \mid \ol {X}^n \ol{C}} \\
&\equiv_{\eps} P_{\ol{X}^n}  P_{\ol C} P_{\ol Y \mid \ol {X}^n \ol{C}} P_{U \mid X^n}
= P_{\ol {X}^n \ol{CY}} P_{U' \mid X^n}
\end{align*}
and, therefore,
\[(\ul \bP_\PlayerA \|  \bP_\PlayerB)(\bF_{\{\widehat \PlayerA\}}) \equiv_{3\eps} \bS_\PlayerA(\ROTT_{\{\widehat \PlayerA\}})\;.\]
\end{proof}

One way to sample $V'$ according to $P_{V \mid C,Y}$ is to simulate the protocol $(\bP_\PlayerA \| \ul \bP_\PlayerB)(\bF)$ until $(V',C',Y')$ is received where $C'=C$ and $Y'=Y$. This simulation needs
exponential time in the parameter $\ell$ and $n$, but is efficient if $\ell$ and $n$ are small and
$(\bP_\PlayerA \| \ul \bP_\PlayerB)(\bF)$ is efficient.
Similarly, we can sample $U'$ by simulating the protocol $(\ul \bP_\PlayerA \| \bP_\PlayerB)(\bF)$ until $(U',(X')^n)$ is received where $(X')^n = X^n$.

\chapter{Universal Oblivious Transfer} \label{chap:uot}

\emph{Universal oblivious transfer} (UOT) is a variant of ROT where the security of the sender
is weakened. A malicious receiver is allowed to receive
\emph{any} information he wants about the sender's input, as long as he does not receive too much information. A parameter $\alpha$ specifies a lower bound on the amount of uncertainty the receiver must have over
the sender's input, measured in terms of min-entropy.
UOT was introduced in \cite{Cachin98}, together with a protocol that implements ROT from UOT.
However, the security proof contained an error which was discovered
in \cite{DFSS06}.
It was showed that
ROT with a string length of $\ell$ can be implemented from one instance of UOT with
an error of at most $\eps$ if $\ell \leq \alpha/4 - \frac34 \log(1/\eps) - 1$, which is only about
half as much as originally claimed in \cite{Cachin98}.

In Theorem~\ref{thm:universal-tight} we give a new proof for the same protocol that was also used in \cite{Cachin98,DFSS06},
and show that the protocol is also secure for
\[ \ell \leq \alpha/2  - 3\log(1/\eps)\]
with an error of at most $2\eps$. This improves the bound of \cite{DFSS06} by a factor of 2 (at the cost of a larger error term) and achieves the bound that has been originally claimed in \cite{Cachin98}, which is asymptotically optimal for this protocol. 

Our proof makes use of a new \emph{distributed leftover hash lemma} (Lemma~\ref{lem:distRandExt}) which is of independent interest.

\section{Min-Entropy and Randomness Extraction} \label{sec:randExt}

In this section we show how almost uniform randomness can be extracted out
of non-uniform randomness.
We use the \emph{min-entropy} to
measure the amount of randomness a random variable has.
 
\begin{definition} [Conditional Min-entropy]
Let $X$ and $Y$ be random variables. The \emph{min-entropy of $X$ given $Y$} is defined as
\begin{align*}
\Hmin(X \mid Y) &:= \min_{xy: P_{XY}(x,y)>0} \log \frac 1 {P_{X \mid Y}(x \mid y)}\;.
\end{align*}
\end{definition}

We will need the following lemma.

\begin{lemma} \label{lem:entropy-cond}
For all $X$, $Y$, and $Z$, we have $\Hmin(X \mid Z) \geq \Hmin(X \mid YZ)$.
\end{lemma}

\begin{proof} This inequality follows from
\begin{align*}
 \max_{x,z} P_{X \mid Z}(x \mid z)
& =  \max_{x,z} \sum_{y} P_{Y}(y) P_{X \mid YZ}(x \mid y,z) \\
& \leq  \max_{x,z} \sum_{y} P_{Y}(y) \max_{x,y,z} P_{X \mid YZ}(x \mid y,z) \\
& = \max_{x,y,z} P_{X \mid YZ}(x \mid y,z)\;.
\end{align*}
\end{proof}

We will use \emph{$2$-universal hash functions} to extract randomness.

\begin{definition}[\cite{CarWeg79}]
A function $h:\mX \times \mS \rightarrow \mY$ is called a \emph{$2$-universal hash function}, if
for all $x_0 \neq x_1 \in \mX$, we have 
\[ \Pr [h(x_0,S) = h(x_1,S)] \leq \frac 1{|\mY|}\;,\]
if $S$ is uniform over $\mS$.
\end{definition}

The \emph{leftover hash lemma} \cite{ILL89}
shows that a $2$-universal hash function is able to extract almost all randomness,
if some additional uniform randomness $S$ is provided as a catalyst. Notice that the
extracted randomness is independent from $S$.
A slightly less general form of this lemma has been proved before in \cite{BeBrRo88}, where
it was called \emph{privacy amplification}. \cite{BBCM95} generalized the notion of
privacy amplification to basically the same statement as \cite{ILL89}, in a slightly different notion.

\begin{lemma}[Leftover hash lemma \cite{BeBrRo88,ILL89}] \label{lem:randExt}
Let $X$ be a random variable over $\mX$ and let $m > 0$. Let
$h:\mS \times \mX \rightarrow \{0,1\}^m$ be a $2$-universal hash function.
If 
\[m \leq \Hmin(X) - 2 \log(1/\eps)\;,\]
then for $S$ uniform over $\mS$,
$h(S,X)$ is $\eps$-close to uniform with respect to $S$.
\end{lemma}

We will now give a distributed version of the leftover hash lemma, where
two players independently extract randomness from two dependent random variables $X$ and $Y$.
The (normal) leftover hash lemma tells us that if the extracted randomness of $X$ and $Y$, respectively, is smaller than
the min-entropy of $X$ and $Y$, respectively, then the extracted strings are close to uniform. However, the
two extracted strings might depend on each other. Lemma \ref{lem:distRandExt} now states that
if the total length of the extracted randomness is smaller than the min-entropy of $(X,Y)$,
then the two strings are also almost independent. Clearly, this bound is optimal.

\begin{lemma}[Distributed leftover hash lemma] \label{lem:distRandExt}
Let $X$ and $Y$ be random variables over $\mX$ and $\mY$, and let $m, n > 0$. Let $g:\mS \times \mX \rightarrow \{0,1\}^m$ and $h:\mR \times \mY \rightarrow \{0,1\}^n$ be $2$-universal hash functions.
If 
\begin{align*}
m &\leq \Hmin(X) - 2\log(1/\eps)\;, \\
n &\leq \Hmin(Y) - 2\log(1/\eps)\;, \quad \textrm{and}  \\
m + n &\leq \Hmin(XY) - 2\log(1/\eps)\;,
\end{align*}
then, for $(S,R)$ uniform over $\mS \times \mR$,
$(g(S,X),h(R,Y))$ is $\eps$-close to uniform with respect to $(S,R)$.
\end{lemma}

\begin{proof}
For any $W$ having distribution $P_W$ over $\mW$, and $W'$ uniformly distributed over $\mW$, we have
\begin{align*}
\Delta(W,W')
 = & \frac 1 2 \sum_{w} \left | P_{W}(w) - \frac{1}{|\mW|} \right|
=  \frac 1 2 \sqrt{\left ( \sum_{w} \left | P_{W}(w) - \frac{1}{|\mW|} \right| \right )^2}\\
\leq & \frac 1 2 \sqrt{|\mW|}
\sqrt{\sum_{w}\left ( P_{W}(w) - \frac{1}{|\mW|} \right)^2} \\
= & \frac 1 2 \sqrt{|\mW|}
\sqrt{\sum_{w}P^2_{W}(w) - \frac{1}{|\mW|}}\;.
\end{align*}
Here we used Lemma~\ref{lem:cauchySchwartz2}.
 
Let $V = g(S,X)$, $V' = h(R,Y)$ and $U,U'$ be two
uniform random variables over $\{0,1\}^m$ and $\{0,1\}^n$. Choosing $W := (V,V',S,R)$ and $W' := (U,U',S,R)$
in the above inequality, we get
\begin{align*}
& \Delta((V,V',S,R),(U,U',S,R)) \\
& \qquad \leq  \frac 1 2 \sqrt{|\mS||\mR| 2^{m+n}}
\sqrt{\sum_{vv'sr}P^2_{VV'SR}(v,v',s,r) - \frac{1}{|\mS||\mR| 2^{m+n}}}\;.
\end{align*}
Since $\sum_{x} P^2_{X}(x)$ is the \emph{collision probability}\footnote{
Let $X_0$ and $X_1$ be distributed according to $P_X$. 
The \emph{collision probability} is $\Pr[X_0 = X_1] = \sum P_X(x)^2$.}
 of a random variable $X$, we have
for $(X_0,Y_0)$ and $(X_1,Y_1)$ independently distributed according to $P_{XY}$
and for uniformly random $S_0$, $S_1$, $R_0$, and $R_1$ that
\begin{align*}
& \sum_{vv'sr} P^2_{VV'SR}(v,v',s,r) = \Pr[S_0 = S_1 \wedge R_0=R_1] \\
& \qquad \cdot \Pr[g(X_0,S_0) = g(X_1,S_0) \wedge h(Y_0,R_0) = h(Y_1,R_0)]\;.
\end{align*}

Because $g$ and $h$ are 2-universal hash functions, we have
\begin{align*}
& \Pr[g(X_0,S_0) = g(X_1,S_0) \wedge h(Y_0,R_0) = h(Y_1,R_0)] \\
& \qquad \leq  \Pr[X_0 = X_1 \wedge Y_0 = Y_1]
+ 2^{-m} \Pr[X_0 \neq X_1 \wedge Y_0 = Y_1] \\
& \qquad \qquad + 2^{-n} \Pr[X_0 = X_1 \wedge Y_0 \neq Y_1] +  2^{-m-n} \\
& \qquad \leq  2^{-m-n} \cdot \eps^2 +  2^{-m} 2^{-n} \cdot \eps^2 +  2^{-n}  2^{-m} \cdot \eps^2 +  2^{-m-n} \\
& \qquad =  (1 + 3 \eps^2)2^{-m-n}\;,
\end{align*}
which implies that
\begin{align*}
& \Delta((V,V',S,R),(U,U',S,R)) \\
& \qquad \leq \frac 1 2
\sqrt{|\mS||\mR| 2^{m+n}}
\sqrt{\frac{1}{|\mS||\mR|}\frac{1 + 3\eps^2}{2^{m+n}} - \frac{1}{|\mS||\mR| 2^{m+n}}}
\leq \frac {\sqrt{3}} 2 \eps\;.
\end{align*}
\end{proof}

Notice that Lemma~\ref{lem:distRandExt} implies Lemma~\ref{lem:randExt}.

\section{Definition of Universal Oblivious Transfer} \label{sec:uotDef}

We now define \emph{universal oblivious transfer}, or $\UOT{\alpha}{n}$, which is a variant of $\ROT{1}{2}{n}$ that provides weaker security for $\PlayerA$. For $\mA = \emptyset$ or $\mA= \{\PlayerA\}$, $\UOTT$ is equal to $\ROTT$. But for $\mA= \{\PlayerB\}$,
instead of requiring that $\PlayerB$ does not know anything about
one of the two strings, we only require that he does not entirely know both of them, i.e., the a min-entropy of sender's input is at least $\alpha$.
Note that from Lemma~2 in \cite{RenWol05}, it follows that there
is no need to use different kinds of R\'enyi-entropies \cite{Renyi61} as done in \cite{Cachin98} or \cite{DFSS06}, as they are basically all equivalent to the min-entropy.

\begin{definition}[Universal oblivious transfer] \label{def:uot}
The system $\UOT{\alpha}{n}$ (or, if $\alpha$ and $n$ are clear from the context, $\UOTT$) is defined as a collection of
systems
\[\UOTT = (\UOTT_\emptyset, \UOTT_{\{\PlayerA\}}, \UOTT_{\{\PlayerB\}})\;,\]
where
$\UOTT_\emptyset = \ROT{1}{2}{n}_\emptyset$ and $\UOTT_{\{\PlayerA\}} = \ROT{1}{2}{n}_{\{\PlayerA\}}$. $\UOTT_{\{\PlayerB\}}$ is defined
as follows. The system waits for $\PlayerB$ to input a distribution 
\[p \in \{ P_{X_0X_1} \mid \Hmin(X_0,X_1) \geq \alpha\}\;,\]
where $(X_0,X_1) \in \{0,1\}^n \times \{0,1\}^n$.
After receiving $p$, it chooses $(x_0,x_1)$ according to $p$ and outputs $(x_0,x_1)$ to $\PlayerA$.
\end{definition}

Notice that our definition UOT is slightly weaker than the
definitions used in \cite{Cachin98,DFSS06}. Because our UOT is a weak version of ROT, we do not
only allow the malicious receiver to receive arbitrary information about his input, but we also allow
him to freely choose his output. For example, we allow him to select $2n-\alpha$ bit and freely fix their values. UOT will then choose the remaining $\alpha$ bit randomly.

\section{Universal Oblivious Transfer Amplification} \label{sec:uotProtocol}

Our protocol $\ROTfromUOT$ is basically the same as the protocols used in
\cite{BraCre97,Cachin98,BrCrWo03,DFSS06}. It securely implements $\ROT{2}{1}{\ell}$ using one instance of
$\UOT{\alpha}{n}$ and $\Auth$ in the malicious model.
Let $h: \{0,1\}^n \times \mR \rightarrow \{0,1\}^\ell$ be a $2$-universal hash function.
The protocol is defined as follows.

\begin{protocol} 
$\ROTfromUOT_\PlayerA$:
\begin{enumerate}
\item Receive $(x_0, x_1) \in \{0,1\}^n \times \{0,1\}^n$ from $\UOTT$.
\item Choose $(r_0, r_1) \in \mR^2$ uniformly at random.
\item Send $(r_0, r_1)$ to $\Auth$.
\item Output $(u_0,u_1) \in \{0,1\}^\ell \times \{0,1\}^\ell$ to $\PlayerA$, where $u_0  := h(x_0,r_0)$ and $u_1 := h(x_1,r_1)$.
\end{enumerate}
$\ROTfromUOT_\PlayerB$:
\begin{enumerate}
\item
Receive $(c,w) \in \{0,1\} \times \{0,1\}^n$ from $\UOTT$ and $(r_0, r_1) \in \mR^2$ from $\Auth$.
\item
Output $(c,y) \in \{0,1\} \times \{0,1\}^\ell$ to $\PlayerB$, where $y := h(r_c,w)$.
\end{enumerate}
\end{protocol}

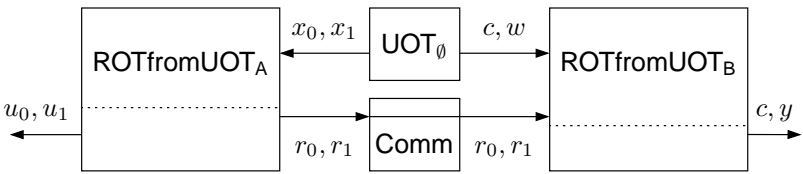
\begin{center} { \input{ps/ROTfromUOT.pstex_t} } \end{center}

We will now show that this protocol indeed achieves the optimal bound
of $\ell \approx \alpha/2$.
The proof works roughly as follows. We define
an additional random variable $A \in \{0,1,2\}$ that distinguishes between three different cases, and show that in each case there
exists a random variable $C$ such that $U_{1-C}$ is almost uniform and independent of the rest. If $A \in \{0,1\}$,
we can lower-bound the min-entropy of $X_{1-A}$ conditioned on
$X_{A}$, and are therefore able to apply Lemma~\ref{lem:randExt} for $C=A$.
If $A=2$ we have lower bounds for the min-entropy of $X_{0}$, $X_{1}$,
and $(X_0,X_1)$, which allow us to
apply Lemma~\ref{lem:distRandExt}.
We need that $\Pr[A=2] \geq \eps$.
If this is not
the case, we ignore the events $A=2$ at the cost of an additional error of at most $\eps$.

\begin{theorem} \label{thm:universal-tight}
Let $\alpha, n, \ell, \eps > 0$.
Protocol $\ROTfromUOT(\UOT{\alpha}{n})$ securely implements $\ROT{1}{2}{\ell}$
in the malicious model with an error of at most $2\eps$, if $\ell \leq \alpha/2 - 3 \log(1/\eps)$.
\end{theorem}

\begin{proof}
Obviously, for $\mA = \emptyset$, we have $\ROTfromUOT(\UOTT_{\{\emptyset\}}) \equiv \ROTT_{\{\emptyset\}}$.

Let $\mA = \{\PlayerA\}$. $\ROTfromUOT_{\PlayerB}(\UOTT_{\{\PlayerA\}})$ waits for receiving $(x_0,x_1)$ and $(r_0,r_1)$ from $\PlayerA$ and then outputs $(c,y)$ to $\PlayerB$, where $c$ is chosen uniformly at random and $y = h(x_c,r_c)$.
We define $\bS_{\PlayerA}$ as follows. It waits for receiving $(x_0,x_1)$ and $(r_0,r_1)$ from $\PlayerA$ and sends
$(h(x_0,r_0),h(x_1,r_1))$ to $\ROTT$. It is easy to see that
$\ROTfromUOT_{\PlayerB}(\UOTT_{\{\PlayerA\}}) = \bS_{\PlayerA}(\ROTT)$.

Let $\mA = \{\PlayerB\}$. The system $\ROTfromUOT_\PlayerA(\UOTT_{\{\PlayerB\}})$ receives the value $p$
from $\PlayerB$, and then outputs $(U_0,U_1)$ to $\PlayerA$ and $(R_0,R_1)$ to $\PlayerB$. In the following,
we will implicitly condition on the values $P=p$.
Let 
\[S_i := \left \{x_i \in \mX_i: \Pr[X_i=x_i] \leq 2^{-\alpha/2} \right \}\;,\]
for $i \in \{0,1\}$. Let
\begin{eqnarray} \label{eq:f2}
g(x_0,x_1) := \left \{
\begin{array}{ll}
2 & \textrm{if }(x_0 \in S_0) \wedge (x_1 \in S_1) \\
0 & \textrm{if }(x_{0} \not \in S_{0})\wedge(x_1 \in S_1) \\
1 & \textrm{if }(x_{0} \in S_{0})\wedge(x_1 \not \in S_1) \\
u & \textrm{if }(x_0 \not \in S_0) \wedge (x_1 \not \in S_1)\;,
\end{array}
\right. 
\end{eqnarray}
and $A := g(X_0,X_1)$,
for $u$ chosen uniformly at random from $\{0,1\}$.
If $\Pr[A=2] < \eps$, let $\mE$ be the event that $A < 2$, and let $\mE$ be the event with probability $1$ otherwise.
We have $\Pr[\mE] \geq 1 - \eps$, and the event $(A=2) \cap \mE$
either has probability $0$ or at least $\eps$. Let $C = \min(A,1)$.
\begin{itemize}
\item For $A=a \in \{0,1\}$ and $\Pr[A=a \wedge \mE] > 0$, we have $C=a$. 
All $x_a \in S_a$ have $\Pr[X_a = x_a \mid A=a \wedge \mE] =0$. For all $x_a \not \in S_a$ we have
\begin{align*}
 & \Pr[X_a = x_a \wedge A=a \wedge \mE] \\
 & \qquad = \Pr[X_a = x_a \wedge X_{1-a} \in S_{1-a}]
    + \frac {\Pr[X_a = x_a \wedge X_{1-a} \not \in S_{1-a}]} 2 \\
 & \qquad \geq \frac {\Pr[X_a = x_a]} 2 \geq 2^{-\alpha/2-1}\;.
\end{align*}
It follows that
\begin{align*}
 &\Pr[X_{1-a} = x_{1-a} \mid  X_a = x_a \wedge A=a \wedge \mE] \\
 & \qquad = \frac{\Pr[X_{1-a} = x_{1-a} \wedge  X_a = x_a \wedge A=a \wedge \mE]}{\Pr[X_a = x_a \wedge A=a \wedge \mE]} \\
 & \qquad \leq 2^{-\alpha}/2^{-\alpha/2-1} = 2^{-\alpha/2 + 1}\;,
\end{align*}
and hence, $\Hmin(X_{1-C} \mid X_{C},A=a,\mE) \geq \alpha/2 - 1$.
Since $R_{0}$ and $R_1$ are uniformly distributed and independent of the rest, it follows from Lemma~\ref{lem:randExt} that, conditioned on $(A=a) \cap \mE$,
$U_{1-C}$ is $\eps$-close to uniform with respect to $(R_0,R_1,U_{C})$.
\item If $A=2$ and 
$\Pr[A=2 \wedge \mE] > 0$, then $C=1$,  $\Pr[A=2 \wedge \mE] \geq \eps$, $\Pr[X_0 = x_0 \wedge X_1 = x_1 \mid A=2 \wedge \mE] \leq 2^{-\alpha}/\eps$, and $\Pr[X_i = x_i \mid A=2 \wedge \mE] \leq 2^{-\alpha/2}/\eps$, for $i \in \{0,1\}$.
It follows that
\begin{align*}
\Hmin(X_0 \mid A=2 \wedge \mE) &\geq \alpha/2 - \log(1/\eps) \\
\Hmin(X_1 \mid A=2 \wedge \mE) &\geq \alpha/2 - \log(1/\eps) \\
\Hmin(X_0 X_1 \mid A=2 \wedge \mE) &\geq \alpha - \log(1/\eps)\;.
\end{align*}
Since $R_{0}$ and $R_1$ are uniformly distributed and independent of the rest,
it follows from Lemma~\ref{lem:distRandExt} that conditioned on  $(A=2) \cap \mE$,
$(U_0,U_1)$
is $\eps$-close to uniform with respect to $(R_0,R_1)$, from which follows that
$U_{1-C}$ is $\eps$-close to uniform with respect to $(R_0,R_1,U_C)$.
\end{itemize}
Therefore, for all $a \in \{0,1,2\}$, conditioned on $(A=a) \cap \mE$, the distribution of $U_{1-C}$ is
$\eps$-close to uniform with respect to $(R_0,R_1,C,U_{C})$.
Since $\Pr[\mE]\geq 1 - \eps$, it follows from Lemma~\ref{lem:statDistEvent2} that $U_{1-C}$ is $2\eps$-close to uniform with respect
to $(R_0,R_1,C,U_{C})$.  Because this holds for every $P=p$,
it follows that $U_{1-C}$ is $2 \eps$-close to uniform with respect to $(C,U_{C},P,R_0,R_1)$.

We define $\bS_\PlayerB$ as follows. After receiving $p: \mX_0 \times \mX_1 \rightarrow [0,1]$ from $\PlayerB$, it simulates $\UOTT_{\{\PlayerB\}}$ on input $p$, from which it gets
the values $X'_0$ and $X'_1$, distributed according to $p$. It calculates
$C' = \min(g(X'_0,X'_1),1)$ according to (\ref{eq:f2}). Then it choses $R'_0$ and $R'_1$ uniformly at random from $\mR$, sends
$(C',h(X'_{C'},R'_{C'}))$ to $\ROTT_{\{\PlayerB\}}$ and outputs $(R'_0,R'_1)$ on the $\Auth$ interface.
$\ROTT_{\{\PlayerB\}}$ will output $(U'_0,U'_1)$ to $\PlayerA$, where $U'_{C} = h(X'_{C'},R_{C'})$
and $U'_{1-C}$ is chosen uniformly at random and independent
from the rest. Since $U_{1-C}$ is $2\eps$-close to
uniform with respect to $(C,P,U_{C},R_0,R_1)$, it is easy to see that
\[(U'_0,U'_1,C',P,R'_0,R'_1) \equiv_{2\eps}(U_0,U_1,C,P,R_0,R_1)\;,\]
from which follows that
\[\ROTfromUOT_\PlayerA(\UOTT_{\{\PlayerB\}}) \equiv_{2\eps} \bS_\PlayerB(\ROTT_{\{\PlayerB\}})\;.\]
\end{proof}

\section{Applications}

The definition of UOT emerged as a generalization of the protocol
presented in \cite{BraCre97,BrCrWo03} to implement string OT out of bit OT.
Therefore it is not surprising that the reduction we presented in this
chapter can be
used to implement string OT from bit OT. Asymptotically, our protocol also
achieves the same bound as the protocol of \cite{BraCre97,BrCrWo03} for this
task. Our protocol can also be used to implement OT from GOT, which leads to better bounds
than the ones presented in \cite{BraCre97,BrCrWo03} or \cite{DFSS06}.

Recently, another very interesting application of UOT has been presented: in
\cite{DFRSS06}, it was shown that in the \emph{bounded quantum-storage model}, it
is possible to implement a simple protocol that achieves a quantum version of
UOT. Whereas it is not clear how the results of \cite{DFSS06} can be used
in that setting to implement OT, they showed that a simplified version of our proof
(only requiring the normal leftover hash lemma)
can directly be applied, using
a quantum version of the leftover hash lemma, 
called \emph{privacy amplification against quantum adversaries} \cite{RenKoe05,Renner05}.
It is also possible to generalize our distributed
leftover hash lemma to the quantum setting, and therefore
the proof we present in this chapter can also
be used in the setting of \cite{DFRSS06} to improve the efficiency of their reduction.

\chapter{Weak Oblivious Transfer} \label{chap:wot}

\begin{figure}
\begin{center} { \input{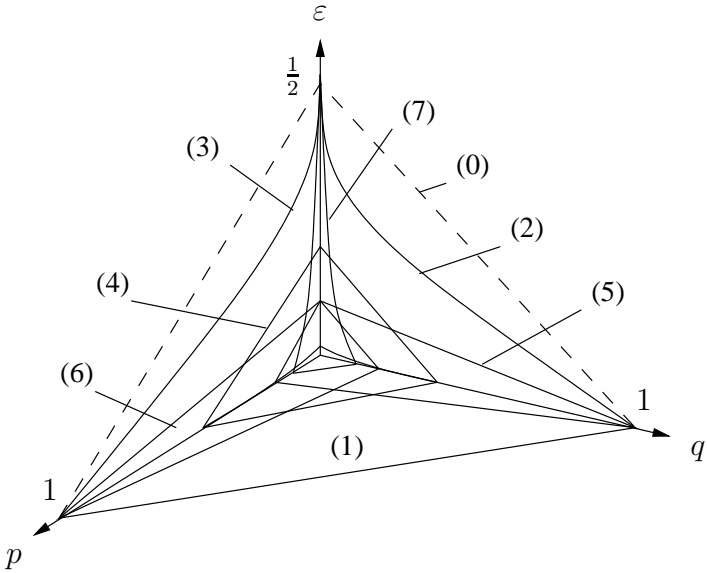} } \end{center}
\caption{The bounds on the parameters $p$, $q$, and $\eps$ for WOT. (0): Impossibility,  Theorem~\ref{thm:wot-imposs-reduction}. (1): Special case where
$\eps = 0$ or $\eps$ is small, Theorem~\ref{thm:otAmp-e0} and Corollary~\ref{cor:generalBound5}. (2-3): Special cases where $p=0$ or $q=0$, Theorem~\ref{thm:wotbound-p0} and Corollary~\ref{cor:wotbound-q0}. (4-7): General case where $p,q,\eps > 0$, Theorem~\ref{thm:generalBound}. 
}
\end{figure}

\emph{Weak oblivious transfer} (WOT), introduced in \cite{DaKiSa99}, is a weak variant of ROT 
where \emph{both} players may obtain additional information about the other player's input, and where the output may have some errors.
In \cite{DaKiSa99} it was
used as a tool to construct OT from \emph{unfair primitives}, i.e., primitives where
the adversary is more powerful than the honest participants, such as the \emph{unfair noisy channel}.
WOT is parameterized by three parameters, $p$, $q$, and $\eps$, where $p$ measures the amount
of side information that the sender gets about the receiver's choice bit, $q$ the amount of side information the receiver gets about the sender's second input bit, and $\eps$
is the maximal probability that an error occurs.

While the definition of WOT is very informal in \cite{DaKiSa99}, the definition used in
\cite{DFMS04} (which gives an ideal functionality of WOT) made implicitly a quite strong assumption, namely that the event that an adversary gains information is independent of the error. Unfortunately, the protocol used in \cite{DaKiSa99,DFMS04} based on unfair noisy channels does not achieve these strong requirements.
We propose two new, weaker definitions of WOT, one for the semi-honest (Definition~\ref{def:wot-semi}) and one for the malicious model (Definition~\ref{def:wot-mal}), that do not have these assumptions.
Also, our definitions make the use of \emph{generalized weak oblivious transfer} \cite{DFMS04}, at least for the protocols we have at the moment, unnecessary.

In Theorem~\ref{thm:wot-imposs-reduction} we restate the impossibility result from \cite{DaKiSa99}
that there does not exist a
protocol which implements OT from WOT if $p + q + 2 \eps \geq 1$.
Then, we give several protocols that implement ROT from WOT.
In Theorem~\ref{thm:otAmp-e0}, we show that the bound of $p+q < 1$ and $\eps=0$ presented in \cite{DaKiSa99} can also be achieved using our definition, both in the semi-honest and the malicious model.
Furthermore, we give a more detailed analysis of the protocols's efficiency.
For the case where $\eps > 0$, our new definition makes it necessary to use a different protocol to reduce the error $\eps$, which implies that we are not able to achieve the same bound as \cite{DaKiSa99}. In Theorems~\ref{thm:wotbound-p0} and Corollary~\ref{cor:wotbound-q0},
we show that for the special case where either $p=0$ or $q=0$ holds,
ROT can securely be implemented from
WOT in the semi-honest model if
\[\left(p=0 \ \wedge \ \sqrt{q} + 2 \eps <1 \right)
 \qquad \vee \qquad 
 \left(q=0 \ \wedge \ \sqrt{p} + 2 \eps <1 \right)\;.\]
We achieve these bounds very easily by using an interesting connection to 
key agreement protocols \cite{HolRen05,Holens06} and the statistical distance polarization problem
\cite{SahVad99,Vadhan99}.
For the general case where $p$, $q$, and $\eps$ may be larger than $0$, we show in Theorem~\ref{thm:generalBound} that if 
\[p+q+2\eps \leq 0.24\]
or
\[(p + 22q + 44\eps < 1) \quad \vee \quad (22p + q + 44\eps < 1) \quad \vee \quad  (7 \sqrt{p+q} + 2\eps < 1)\;,\]
ROT can efficiently be implemented from
WOT secure in the semi-honest model. These bounds do not achieve the bound of
$p + q + 2\eps < 0.45$ from \cite{DaKiSa99} for all values $p$, $q$, and $\eps$, but they are
better for the cases where two parameters are small and one is large.
Finally, we show in Corollary~\ref{cor:generalBound5} that we can also implement ROT from WOT in the semi-honest model if
\[(1 - p - q)^4 < - 178 \cdot \log(1 - 2\eps)\;,\]
which means that if $\eps$ is small enough, then we can achieve OT for all values $p+q <1$.

\section{Definition of WOT} \label{sec:wot-def}

In this section we give formal definitions of WOT.
Because our protocols will reduce the information of the adversary
by using the XOR of several values, the maximum bit-prediction advantage ($\predadv$) turns out to be
a good measure for the adversary's side information. Furthermore, it has the advantage that we can easily find a computational version of this measure, which will be very useful in  Chapter~\ref{chap:compWOT}.
Our definition of WOT is inspired by the definition of \emph{weak bit agreement} in \cite{Holens05,Holens06}.

\subsection{In the Semi-Honest Model}

We start with the definition of WOT in the semi-honest model.
Since the adversary is not able to choose which information he would like to obtain in the semi-honest model, he may only obtain whatever
information the functionality provides him with. But we do not want to fix this information,
as we want to cover a wide range of possibilities --- we might not even
know what information the functionality will provide to the adversary. Therefore, we cannot define
an ideal functionality. Instead, we will define a set of ideal functionalities,
and assume that one instance of this set is provided to us, but we may not know which instance.
We will define this set of ideal functionalities by a list of properties that the ideal functionality
must satisfy.

\begin{definition}[Weak oblivious transfer, semi-honest model] \label{def:wot-semi}
  Let \[\bF = (\bF_{\emptyset},\bF_{\{\widehat \PlayerA\}},\bF_{\{\widehat \PlayerB\}})\] be a collection of systems in the semi-honest model. Let $\bF$ output
  $(X_0,X_1)$ to $\PlayerA$ and $(C,Y)$ to $\PlayerB$. Let $U$ be the auxiliary output to $\PlayerA$ by $\bF_{\{\PlayerA\}}$ and $V$ be the auxiliary output to $\PlayerB$ by $\bF_{\{\PlayerB\}}$.
  Let $E := X_C \oplus Y$. $\bF$ implements $\WOT{p}{q}{\eps}$ in the semi-honest model, if
\begin{itemize}
\item(Correctness) $\Pr[E=1] \leq \eps$.
\item(Security for $\PlayerA$) $\predadv(X_{1-C}  \mid V,E) \leq q$.
\item(Security for $\PlayerB$) $\predadv(C  \mid U,E) \leq p$.
\end{itemize}
\end{definition}

We also use $\WOTtwo{p}{q}$ for $\WOT{p}{q}{0}$.

It is not immediately clear why we require that $X_{1-C}$ and $C$ are difficult to guess even
when additionally the value $E$ is given. We do this for allowing the adversary to
learn the error during the protocol without getting additional information about
$X_{1-C}$ or $C$. For example, in the protocol $\EReduce$, $\PlayerB$ may get to know $X_C$ during
the protocol, which means that he gets to know $E = Y \oplus X_C$. Therefore, we must make sure that
his side information about $X_{1-C}$ is not increased if he gets to know $E$.
Note, however, that for the protocols we present here, it would be sufficient to only require $\predadv(C \mid U) \leq p$ for the security for $\PlayerB$, because $E$ is never leaked to $\PlayerA$. We do not use this definition in order to
keep WOT symmetric, and to get a stronger Theorem~\ref{thm:compOT} that is simpler to proof. (Otherwise,
Theorem~\ref{thm:compOT} would not work for all protocols, but just for the protocols we present here.)

We will now show that the conditions of $\WOTT$ suffice to implement $\ROTT$ in the semi-honest model.
We need the following lemma.

\begin{lemma} \label{lem:almostUniform}
Let $P_U$ be the uniform distribution over $\{0,1\}$ and let $P_{CX_0X_1}$ be a distri\-bution over $\{0,1\}^3$ for which 
 $P_{CX_1X_0} \equiv_\eps P_U P_{X_1X_0}$ and $P_{X_{1-C}X_CC} \equiv_\eps P_U P_{X_CC}$ holds.
 Then $\Delta(P_{CX_0X_1}, P_U P_U P_U) \leq 4 \eps$.
\end{lemma}

\begin{proof} Let
$a := P_{CX_1X_0}(0,0,0)$, $b:=P_{CX_1X_0}(0,0,1)$, $c:=P_{CX_1X_0}(0,1,0)$, $d:=P_{CX_1X_0}(0,1,1)$,
\dots, and let $h := P_{CX_1X_0}(1,1,1)$.
From $P_{CX_1X_0} \equiv_\eps P_UP_{X_1X_0}$ and Lemma~\ref{lem:redBound4}, we get
\[|a-e| + |b-f| + |c-g| + |d-h| \leq 2\eps\;,\]
and from
$P_{X_{1-C}X_CC} \equiv_\eps P_U P_{X_CC}$ and Lemma~\ref{lem:redBound4}
\[|a-c| + |b-d| + |e-f| + |g-h| \leq 2\eps\;.\]
Adding up the two inequalities, we get
\begin{align} 
& |e-a| + |a-c| + |c-g| + |g-h|  \notag \\
& \qquad \qquad + |h-d| + |d-b| + |b-f| + |f-e| \leq 4\eps\;. \label{eq:all}
\end{align}
It is easy to see that the difference between the minimal and the maximal values in the set $\{a,\dots,h\}$
is at most $2\eps$, and that the statistical distance is
maximized
for (\ref{eq:all}) by distributions where $n \in \{1,\dots,7\}$ values have equal probability $1/8 + 2 \eps - \eps n / 4$, and
$8-n$ values have equal probability $1/8 - \eps n / 4$.
The statistical distance is $\eps n (2 - n/4)$, which is maximized for $n=4$\footnote{Note that such a distribution does not satisfy our original,
stricter requirements. Values that do satisfy them are $a=e=f=1/2+5/4\cdot\eps$ and $b=d=h=g=c=1/2-3/4\cdot\eps$, which gives a statistical distance of $3.75$.}, where it is $4 \eps$.
\end{proof}

\begin{lemma} \label{lem:WOT2ROT}
If a protocol $\bF$ implements $\WOT{\eps}{\eps}{\eps}$, then it implements $\semiROT{1}{2}{1}$ secure
in the semi-honest model, with an error of at most $9 \eps$.
\end{lemma}

\begin{proof}
Let $(X_0,X_1,C,Y)$ be the output of $\bF_{\emptyset}$, let $U$ be the auxiliary output of $\bF_{\{\PlayerA\}}$
to $\PlayerA$ and $V$ the auxiliary output of $\bF_{\{\PlayerB\}}$ to $\PlayerB$. From Lemma~\ref{lem:PredAdvStadDist}
follows that $C$ is $\eps/2$-close to uniform with respect to $(E,X_0,X_1,U)$, and that $X_{1-C}$ is $\eps/2$-close to
uniform with respect to $(E,C,Y,V)$. 
Let $P_{\ol{X_0X_1CY}}$ be the output distribution of $\semiROTT_{\emptyset}$.

Lemma~\ref{lem:almostUniform} implies that $P_{X_0X_1C} \equiv_{2\eps} P_{\ol{X_0X_1C}}$. Since
$\Pr[Y \neq X_C] \leq \eps$, we have
\[P_{X_0X_1CY} \equiv_{\eps} P_{X_0X_1C}P_{\ol Y\mid \ol{X_0X_1C}} \equiv_{2\eps} P_{\ol{X_0X_1CY}}\;.\]
We can now apply Theorem~\ref{thm:passiveSecCondforSROT2}.
\end{proof}

\subsection{In the Malicious Model}

We will now also give a formal definition of WOT in the malicious model. The definition differs from the semi-honest case in two important points. Firstly, since we do not have any protocol that can do error reduction in the malicious model, we will only define the case without any error, i.e., $\eps = 0$. Secondly, for the security of $\PlayerA$, we require that the XOR of the two input bits is difficult to guess, because this is a much easier requirement than the standard approach used in Theorem~\ref{thm:SecCondforROT}. Lemma~\ref{lem:OT-XOR-prop} shows that the two conditions are equivalent.
Notice that the security of the XOR does not suffice in the semi-honest model, and, therefore, this trick cannot be applied there. On the other hand,
since in the malicious model a corrupted $\PlayerB$ may choose $C$ freely, we cannot use Lemma~\ref{lem:almostUniform}, and, therefore, the condition $\Pr[Y \neq X_C] = 0$ would not suffice in the
malicious model.

\begin{definition}[Weak oblivious transfer, malicious model]  \label{def:wot-mal}
Let \[\bF = (\bF_{\emptyset},\bF_{\{\PlayerA\}},\bF_{\{\PlayerB\}})\] be a collection of systems in the malicious model.
The system $\bF$ implements $\WOTtwo{p}{q}$ (or, if $p$ and $q$ are clear from the context, $\WOTT$) in the malicious model, if
\begin{itemize}
\item(Correctness): $\bF_{\emptyset} \equiv \ROTT$.
\item(Security for \PlayerA): The system $\bF_{\{\PlayerB\}}$ interacts over the interfaces
belonging to $\PlayerB$ (which produces a transcript $V$), and after the last
input is received over these interfaces, it outputs $(X_0,X_1) \in \{0,1\}^{2}$ to $\PlayerA$  where
$\predadv(X_0 \oplus X_1 \mid V) \leq q$. 
\item(Security for \PlayerB) The system $\bF_{\{\PlayerA\}}$ interacts over the interfaces belonging to $\PlayerA$ (which produces a transcript $U$), and after the last
input is received over these interfaces, it outputs $(C,Y) \in \{0,1\}^2$ to $\PlayerB$ where
$\predadv(C \mid U) \leq p$.
\end{itemize}
\end{definition}

Notice that since we are now in the malicious model, the adversary is able to choose what information he would like to receive, and we could define an ideal functionality in a similar way as we did in Definition~\ref{def:uot} for UOT. We did not do this in order to be closer to Definition~\ref{def:wot-semi} and
Theorem~\ref{thm:SecCondforROT}.

Again, we will first show that $\WOTT$ suffices to implement $\ROTT$ in the malicious model.
We will need the following lemma, which has already been proved in \cite{DFSS06}.

\begin{lemma} \label{lem:OT-XOR-prop}
Let $P_{X_0X_1}$ be given. There exists
a random variable $C$ distributed according to a conditional distribution
$P_{C\mid X_0,X_1}$ such that
$X_{1-C}$ is uniform with respect to $(C,X_C)$, if and only if $X_0 \oplus X_1$ is uniformly distributed.
\end{lemma}

\begin{proof}
Let $P_{X_0X_1C}$ be a distribution such that $X_{1-C}$ is uniform with respect to $(C,X_C)$.
We have
\begin{align*}
\Pr[X_0 \oplus X_1 = 0] 
& = P_{X_0X_1C}(0,0,0) + P_{X_0X_1C}(1,1,0) \\
& \qquad + P_{X_0X_1C}(0,0,1) + P_{X_0X_1C}(1,1,1) \\
& = P_{X_0X_1C}(0,1,0) + P_{X_0X_1C}(1,0,0) \\
& \qquad + P_{X_0X_1C}(1,0,1) + P_{X_0X_1C}(0,1,1) \\
& = \Pr[X_0 \oplus X_1 = 1]\;.
\end{align*}
Hence, $X_0 \oplus X_1$ is uniformly distributed.

The other direction is slightly more complicated.
Let $X_0 \oplus X_1$ be uniformly distributed. We choose
\[ P_{C \mid X_0,X_1}(0 \mid x_0, x_1) := \frac{\min(P_{X_0X_1}(x_0,0),P_{X_0X_1}(x_0,1))}
{P_{X_0X_1}(x_0,x_1)}\;.\]
For $C=0$ and $x_0 \in \{0,1\}$, we have
\begin{align*}
P_{X_0X_1C}(x_0,0,0)
&= P_{X_0X_1}(x_0,0) \cdot P_{C \mid X_0X_1}(0 \mid x_0,0) \\
&= \min(P_{X_0X_1}(x_0,0),P_{X_0X_1}(x_0,1))  \\
&= P_{X_0X_1C}(x_0,1,0)\;.
\end{align*}
Since $X_0 \oplus X_1$ is uniformly distributed, we have
\[ P_{X_0X_1}(0,0) - P_{X_0X_1}(0,1) = P_{X_0X_1}(1,0) - P_{X_0X_1}(1,1)\;,\]
which implies that for $C=1$ and $x_1 \in \{0,1\}$,
\begin{align*}
P_{X_0X_1C}(0,x_1,1)
&= P_{X_0X_1}(0,x_1) \cdot (1 - P_{C \mid X_0X_1}(0 \mid 0,x_1)) \\
& = P_{X_0X_1}(0,x_1) - \min(P_{X_0X_1}(0,0),P_{X_0X_1}(0,1)) \\
& = \max(0,P_{X_0X_1}(0,x_1) - P_{X_0X_1}(0,1 - x_1) \\
& = \max(0,P_{X_0X_1}(1,x_1) - P_{X_0X_1}(1,1 - x_1) \\
& = P_{X_0X_1C}(1,x_1,1)\;.
\end{align*}
Hence, for $c \in \{0,1\}$ and $x_c \in \{0,1\}$, 
\begin{align*}
P_{X_{1-C} X_C C}(0,x_c,c) = P_{X_{1-C} X_C C}(1,x_c,c) = \frac 1 2 P_{X_C C}(x_c,c)\;.
\end{align*}
Therefore, $X_{1-C}$ is uniform with respect to $(C,X_C)$.
\end{proof}

\begin{lemma}
If a protocol $\bF$ implements $\WOTtwo{\eps}{\eps}$ in the malicious model, then it implements
$\ROT{1}{2}{1}$ secure in the malicious model, with an error of at most $\eps/2$.
\end{lemma}

\begin{proof}
Let $\mA = \{\PlayerB\}$. From Lemma~\ref{lem:PredAdvStadDist} follows that there
exists $(X'_0,X'_1)$, such that
\[\Delta((X_0,X_1,V),(X'_0,X'_1,V)) \leq \eps/2\]
and
$X'_0 \oplus X'_1$ is uniform with respect to $V$. We choose $P_{C \mid X'_0,X'_1,V}$ as
proposed in Lemma~\ref{lem:OT-XOR-prop}. $X'_{1-C}$ is uniform with respect to $(C,X'_C,V)$,
and, therefore, $X_{1-C}$ is $\eps/2$-close to uniform with respect to $(C,X_C,V)$.

Let $\mA = \{\PlayerA\}$. From Lemma~\ref{lem:PredAdvStadDist} follows that $C$ is $\eps/2$-close to
uniform with respect to $U$.

The lemma follows now from Theorem~\ref{thm:SecCondforROT}.
\end{proof}

\subsection{Relation to Previous Definitions}

\paragraph{Difference to WOT from \cite{DaKiSa99, DFMS04}.}
Besides the fact that we only consider a randomized version of WOT,
the difference of our definition of $\WOT{p}{q}{\eps}$ to the definitions
used in \cite{DaKiSa99, DFMS04} is that we do
not specify exactly what a malicious player may receive, 
but we only require that his output should not give too much information about the
bits $X_{1-{C}}$ and $C$. This means that a malicious player may, for
example, always receive whether an error occurred in the
transmission or not, if that information is independent of the inputs.
The most important difference is, however, that 
our definitions do not require
that the error must occur independently of the event
that a player gets side information, which is very important when we want to apply it.

Lemmas~\ref{lem:Hol22} and~\ref{lem:Hol22-converse} imply that
our definitions still are quite close to the definitions from \cite{DaKiSa99, DFMS04}, because
there exist events with probability $1-p$ and $1-q$, such that, if they occur, then
the adversary does not get any side information.

\paragraph{Connection to GWOT from \cite{DFMS04}.}
In \cite{DFMS04}, \emph{Generalized WOT} (GWOT) was introduced 
to improve the achievable range of the reductions. It was shown
in Lemma~3 in \cite{DFMS04} that in the reductions they used,
WOT can be replaced by a GWOT, if the probability
to guess the bits $X_{1-{C}}$ and $C$, respectively, remain the same for the adversary.
Since we defined WOT over the advantage to guess these values, Lemma~3 in \cite{DFMS04}
is not needed anymore, and therefore, at least for the moment, the use of GWOT does
not give any advantage over WOT.

\section{Impossibility Results} \label{sec:wot-imposs}

In this section we prove the impossibility result stated in \cite{DaKiSa99}, that WOT cannot be amplified if $p + q + 2 \eps \geq 1$. 
Note that the proof does not work for the definition of WOT used in \cite{DaKiSa99,DFMS04}.
We start with the protocol $\SimWOT_{(p,q)}(\Auth)$ that implements $\WOT{p}{q}{\eps}$ for $p + q + 2\eps = 1$ in the semi-honest model.

\begin{protocol}
$\SimWOT_\PlayerA$:
\begin{enumerate}
\item Choose $(x'_0,x'_1) \in \{0,1\}^2$ uniformly at random.
\item With probability $q$, send $a=(x'_0,x'_1)$ to $\Auth$. Otherwise, send $a=\bot$ to $\Auth$.
\item Receive $b$.
\item If $b = \bot$ then output $(x_0,x_1) := (x'_0,x'_1)$. Otherwise,
$b = (c,y) \in \{0,1\}^2$. Output $(x_0,x_1)$, where $x_c := y$ and $x_{1-c} := x'_{1-c}$.
\end{enumerate}

$\SimWOT_\PlayerB$:
\begin{enumerate}
\item Choose $(c',y') \in \{0,1\}^2$ uniformly at random.
\item Receive $a$.
\item If $a = \bot$ then output
$(c,y) := (c',y')$ and send with probability $p / (1-q)$ the value $(c',y')$ to $\Auth$, and $\bot$ otherwise. Otherwise, $a = (x'_0,x'_1) \in \{0,1\}^2$. Send $\bot$ to $\Auth$ and outputs $(c,y) := (c', x'_{c'})$.
\end{enumerate}
\end{protocol}

\begin{lemma} \label{lem:simWOT}
Protocol $\SimWOT_{(p,q)}(\Auth)$ securely implements $\WOT{p}{q}{(1-p-q)/2}$ in the semi-honest model.
\end{lemma}

\begin{proof}
Let $E := Y \oplus X_C$.
With probability $q$, $\PlayerB$ will adjust his output such that $Y=X_C$,
and with probability $(1-q) \cdot p / (1-q) = p$, $\PlayerA$ will adjust her output such
that $X_C = Y$. With probability $1-p-q$, the values $(X_0,X_1)$ and $(C,Y)$ will be
chosen uniformly at random. Therefore, we have 
\[\Pr[Y \neq X_C] = (1-p-q)/2\;.\] 
When $\SimWOT_\PlayerA$ sends $\bot$ to $\Auth$, then the value $X_{1-C}$ is uniform with respect $(V,E)$.
From Lemma~\ref{lem:Hol22-converse} follows that 
\[\predadv(X_{1-C}  \mid V,E) \leq q\;.\]
When $\SimWOT_\PlayerB$ sends $\bot$ to $\Auth$, then the value $C$ is uniform with respect $(U,E)$.
From Lemma~\ref{lem:Hol22-converse} follows that \[\predadv(C \mid U,E) \leq p\;.\]
\end{proof}

We need the following well-known fact.

\begin{lemma} \label{lem:otImposs}
There cannot exist a protocol
$\bP(\Auth)$ that securely implements $\OT{1}{2}{1}$ in the weak semi-honest model.
\end{lemma}

\begin{theorem} \label{thm:wot-imposs-reduction}
For any $p$, $q$, und $\eps$ with $p + q + 2 \eps \geq 1$ and for any $n$, there cannot exist a protocol
$\bP(\WOT{p}{q}{\eps}^{\|n}\|\Auth)$ that securely implements $\OT{1}{2}{1}$ in the semi-honest or the malicious model.
\end{theorem}

\begin{proof}
From Lemma~\ref{lem:malIsWeakSemiHonest} follows that $\bP$ is secure in the weak semi-honest model.
Therefore, it would follow from Lemma~\ref{lem:simWOT} and Theorem~\ref{thm:compWeakSemiHonest} that the protocol
\[\bP(\SimWOT_{(p,q)}(\Auth)^{\|n}\|\Auth)\]
would implement $\OT{1}{2}{1}$ from scratch in the weak semi-honest model,
which contradicts Lemma~\ref{lem:otImposs}.
\end{proof}

\section{Basic Protocols for WOT Amplification} \label{sec:wot-basic}

We now present the three basic protocols that we use to implement $\ROTT$ from $\WOTT$.
The protocol $\RReduce$  allows for reducing the parameter $p$, and
the protocol $\SReduce$ is used to reduce the parameter $q$.
Both reductions were already used in
\cite{CreKil88,DaKiSa99,DFMS04,Haitne04}, as well as in \cite{HKNRR05,MePrWu07} for building OT combiners.
The protocol $\EReduce$  is used to reduce the parameter $\eps$. Whereas the other two protocols are secure
in both models, $\EReduce$ is merely secure in the semi-honest model. The same protocol
was also used in \cite{Haitne04} and is the one-way variant of the protocol $\EReduce$ presented in
\cite{DaKiSa99}.
Notice that since we defined $\WOTT$ to be a \emph{randomized} primitive, we are not
able to choose the input, which makes the protocols slightly more complicated.

We first present all protocols in the semi-honest model, and later give the proofs for the malicious model.

\subsection{In the Semi-Honest Model}

The protocol $\RReduce(\WOTT^{\|n} \| \Auth)$ is defined as follows.

\begin{protocol}
$\RReduce_\PlayerA$:
\begin{enumerate}
\item Receive $(x_{0,i},x_{1,i})$ from the $i$th $\WOTT$, for all $i \in \{0,\dots,n-1\}$.
\item Receive $d^{n-1} = (d_0,\dots,d_{n-2})$ from $\Auth$. Set $d_{n-1} := 0$.
\item Output $(x_0,x_1):=(\bigoplus_{i=0}^{n-1} x_{d_i,i},\bigoplus_{i=0}^{n-1} x_{d_i \oplus 1,i})$.
\end{enumerate}

$\RReduce_\PlayerB$:
\begin{enumerate}
\item Receive $(c_i,y_i)$ from the $i$th $\WOTT$, for all $i \in \{0,\dots,n-1\}$.
\item Send $d^{n-1} = (d_0,\dots,d_{n-2})$ to $\Auth$, where $d_i := c_{n-1} \oplus c_i$.
\item Output $(c,y) := (c_{n-1},\bigoplus_{i=0}^{n-1} y_i)$.
\end{enumerate}
\end{protocol}

\begin{center} { \input{ps/RReduce.pstex_t} } \end{center}

\begin{lemma} \label{lem:RReduce}
The protocol $\RReduce(\WOT{p}{q}{\eps}^{\|n} \| \Auth)$ securely imple\-ments 
$\WOT{p'}{q'}{\eps'}$ in the semi-honest model, where
$p' = 1 - (1-p)^n \leq np$, $q' = q^n \leq e^{-n(1-q)}$, and
$\eps' = (1 - (1 - 2\eps)^n)/2 \leq n\eps$.
\end{lemma}

\begin{proof} Let $E_i := Y_i \oplus X_{C_i,i}$, and $E := Y \oplus X_C$.
We have
\[E = Y \oplus X_C 
= \bigoplus_{i=0}^{n-1} Y_i \oplus \bigoplus_{i=0}^{n-1} X_{C \oplus D_i,i}
= \bigoplus_{i=0}^{n-1} (Y_i \oplus X_{C_i,i}) = \bigoplus_{i=0}^{n-1}  E_i\;.\]

Let $\mA = \emptyset$. Since $\Pr[E_i=1] \leq \eps$, it follows from Lemma~\ref{lem:corrn} that 
\[\Pr[E=1] \leq  \frac{1 - (1 - 2\eps)^n} {2} \leq n\eps\;.\] 

Let $\mA=\{\PlayerB\}$, and let $V_i$ be the auxiliary output to $\PlayerB$ from the
$i$th instance of $\WOTT_{\{\PlayerB\}}$. The auxiliary output of the protocol to $\PlayerB$ is
$V := V^n$. Since
\[X_{1-C} := \bigoplus_{i=0}^{n-1} X_{1 - D_i \oplus C,i} = \bigoplus_{i=0}^{n-1} X_{1 - C_i,i}\;,\]
and because $E$ is a function of $E^n$, it follows from Lemmas~\ref{lem:predAdv-dataprocessing}, \ref{lem:predXOR} and \ref{lem:redBound2} that
\begin{align*}
\predadv(X_{1-C} \mid V,E)
&\leq \predadv(X_{1-C} \mid V^n,E^n) \\
&= \prod_{i=0}^{n-1} \predadv(X_{1-C_i,i} \mid V_i,E_i) \\
&\leq q^n \leq e^{-n(1-q)}\;.
\end{align*}

Let $\mA=\{\PlayerA\}$, and let $U_i$ be the auxiliary output to $\PlayerA$ from the
$i$th instance of $\WOTT_{\{\PlayerA\}}$. The auxiliary output of the protocol to $\PlayerA$ is
$U := (U^n,D^{n-1})$. Because $E$ is a function of $E^n$, 
Lemmas~\ref{lem:predAdv-dataprocessing}, \ref{lem:predComm} and \ref{lem:redBound1} imply that
\begin{align*}
\predadv(C  \mid U,E) &\leq \predadv(C \mid U^n,E^n,D^{n-1}) \\
&\leq 1 - \prod_{i=0}^{n-1}(1 - \predadv(C_i  \mid U_i,E_i)) \\
&\leq 1 - (1-p)^n \leq np\;.
\end{align*}
\end{proof}

We will also need a protocol $\SReduce$ that reduces the error $p$. To achieve this, we can simply use the protocol $\RReduce$ in the opposite direction, together with the protocol $\ROTOR$. We need the fact that Protocol $\ROTOR(\WOTT)$ implements $\WOTT$ in the inverse direction.

\begin{lemma} \label{lem:rotorWOT}
Protocol $\ROTOR(\WOT{p}{q}{\eps})$ implements $\WOT{q}{p}{\eps}$
in the opposite direction, secure in the semi-honest model.
\end{lemma}

\begin{proof}
Let $(X'_0, X'_1, C', Y')$ be the output of $\ROTT_{\emptyset}$, and
let $(X_0, X_1, C, Y)$ be the output of $\ROTOR$. Let $U'$ be the auxiliary
output to $\PlayerA$ by $\ROTT_{\{\PlayerA\}}$, and let $V'$ be the auxiliary
output to $\PlayerB$ by $\ROTT_{\{\PlayerB\}}$. The auxiliary output
output to $\PlayerA$ by $\ROTOR$ is $V = U'$, and the auxiliary
output to $\PlayerB$ by $\ROTOR$ is $U = V'$.

Let $E := Y \oplus X_C$. It is easy to verify that $E' = Y' \oplus X'_{C'} = E$, and therefore
that the correctness condition is satisfied. 
From Lemma~\ref{lem:predXOR2} follows that
\begin{align*}
\predadv(X_{1-C} \mid V,E)
& = \predadv(X_{1-C} \oplus (E \oplus Y) \mid V,E) \\
& = \predadv(X_{1-C} \oplus X_C \mid V,E') \\
& = \predadv(C'  \mid U',E') \leq p
\end{align*}
and
\begin{align*}
\predadv(C  \mid U,E)
& = \predadv(X'_0 \oplus X'_1 \mid V',E') \\
& = \predadv(X'_{1-C'} \oplus X'_{C'} \mid V',E') \\
& = \predadv(X'_{1-C'} \mid V',E') \leq q\;.
\end{align*}
\end{proof}

We can, therefore, implement $\SReduce$ in the following way: We apply $\ROTOR$
to all $n$ instances of $\WOTT$, then use $\RReduce$ in the opposite direction, and finally apply $\ROTOR$ to the resulting $\WOTT$. We get

\begin{lemma} \label{lem:SReduce}
The protocol $\SReduce(\WOT{p}{q}{\eps}^{\|n} \| \Auth)$ securely imple\-ments $\WOT{p'}{q'}{\eps'}$ in the semi-honest model,
where
 $q' = 1 - (1-q)^n \leq n q$,
$p' = p^n \leq e^{-n(1-p)}$, and
$\eps' = (1 - (1 - 2\eps)^n)/2 \leq n\eps$.
\end{lemma}

Protocol $\EReduce(\WOTT^{\|n} \| \Auth)$ reduces the error $\eps$, and is defined as follows.

\begin{protocol}
$\EReduce_\PlayerA$:
\begin{enumerate}
\item Receive $(x_{0,i},x_{1,i})$ from the $i$th $\WOTT$, for all $i \in \{0,\dots,n-1\}$.
\item Receive $d^{n-1} = (d_0,\dots,d_{n-2})$ from $\Auth$.
\item Send $(s_0^{n-1},s_1^{n-1}) = ((s_{0,0},\dots,s_{0,n-2}),(s_{1,0},\dots,s_{1,n-2}))$
to $\Auth$, where
$s_{j,i} := x_{d_i \oplus j,i} \oplus x_{j,n-1}$.
\item Output $x_0 := x_{0,n-1}$ and $x_1 := x_{1,n-1}$.
\end{enumerate}

$\EReduce_\PlayerB$:
\begin{enumerate}
\item Receive $(c_i,y_i)$ from the $i$th $\WOTT$, for all $i \in \{0,\dots,n-1\}$.
\item Send $d^{n-1} = (d_0,\dots,d_{n-2})$ to $\Auth$, where $d_i := c_{n-1} \oplus c_i$.
\item Receive $(s_0^{n-1},s_1^{n-1})$
from $\Auth$.
\item Output $(c,y) := (c_{n-1}, \maj(\{\ol y_i\}))$ where $\ol y_i := y_i \oplus s_{c_{n-1},i}$ for
$i \in \{0,\dots,n-2\}$ and $\ol y_{n-1} := y_{n-1}$.
\end{enumerate}
\end{protocol}

\begin{center} { \input{ps/EReduce.pstex_t} } \end{center}

\begin{lemma} \label{lem:ered}
Protocol $\EReduce(\WOT{p}{q}{\eps}^{\|n} \| \Auth)$ securely implements $\WOT{p'}{q'}{\eps'}$ in the semi-honest model, where
$p' = 1 - (1-p)^n \leq np$,
$q' = 1 - (1-q)^n \leq nq$ and
\[ \eps' = \sum_{i=\lceil n/2 \rceil}^{n} \binom{n}{i} \eps^{i} (1 - \eps)^{n-i} \leq e^{-2 n (1/2 - \eps)^2}\;.\]
\end{lemma}

\begin{proof}  Let $E_i := Y_i \oplus X_{C_i,i}$, and $E := Y \oplus X_C$.

Let $\mA = \emptyset$. We have $\Pr[E_{i}=1] \leq \eps$.
Since for $i \in \{0,\dots,n-2\}$
\begin{align*}
\ol Y_i
 &= Y_i \oplus S_{C_{n-1},i}
 = Y_i \oplus X_{D_i \oplus C_{n-1},i} \oplus X_{C_{n-1},n-1} \\
 &= Y_i \oplus X_{C_i,i} \oplus X_{C}
 = E_i \oplus X_{C}\;,
 \end{align*}
it follows from Lemma~\ref{lem:errRed} that the protocol satisfies correctness with an error of at most
\[ \eps' = \sum_{i=\lceil n/2 \rceil}^{n} \binom{n}{i} \eps^{i} (1 - \eps)^{n-i} \leq e^{-2 n (1/2 - \eps)^2}\;.\]

Let $\mA=\{\PlayerB\}$. Let $V_i$ be the auxiliary output to $\PlayerB$ from the
$i$th instance of $\WOTT_{\{\PlayerB\}}$.
The auxiliary output
 to $\PlayerB$ is $V = (D^{n-1},S^{n-1}_0,S^{n-1}_{1},V^n)$.
Note that 
$D^{n-1}$ is a function of $V^n$. 
Furthermore, $S^{n-1}_{C}$ is a function of $(V^n,E^n)$, because
\begin{align*}
 S_{C,i}
& = X_{D_i \oplus C_{n-1},i} \oplus X_{C_{n-1},n-1}
 = X_{C_{i},i} \oplus X_{C_{n-1},n-1}  \\
& = Y_i \oplus E_i \oplus Y_{n-1} \oplus E_{n-1}\;,
\end{align*}
for all $i$. Since
\[X_{1-C} = X_{1-C,n-1} = S_{1-C,i} \oplus X_{1-D_i \oplus C,i} = S_{1-C,i} \oplus X_{1-C_i,i}\;, \]
Lemmas~\ref{lem:predAdv-dataprocessing}, \ref{lem:predComm} and \ref{lem:redBound1} imply
\begin{align*}
\predadv(X_{1-C} \mid V,E)
&\leq \predadv(X_{1-C} \mid  V^n,E^n,S^{n-1}_{1 - C}) \\
&\leq 1 - \prod_{i=0}^{n-1}(1 - \predadv(X_{1-C_i,i}  \mid V_i,E_i)) \\
&\leq 1 - (1-q)^n \leq nq\;.
\end{align*}

Let $\mA=\{\PlayerA\}$, and let $U_i$ be the auxiliary output to $\PlayerA$ from the
$i$th instance of $\WOTT_{\{\PlayerA\}}$. The auxiliary output of the protocol to $\PlayerA$ is
$U := (U^n,D^{n-1})$. Because $E$ is a function of $E^n$, 
it follows from Lemmas~\ref{lem:predAdv-dataprocessing}, \ref{lem:predComm} and \ref{lem:redBound1} that
\begin{align*}
\predadv(C  \mid U,E) &\leq \predadv(C \mid U^n,E^n,D^{n-1}) \\
&\leq 1 - \prod_{i=0}^{n-1}(1 - \predadv(C_i  \mid U_i,E_i)) \\
&\leq 1 - (1-p)^n \leq np\;.
\end{align*}
\end{proof}

\subsection{In the Malicious Model}

We will now show that the protocols $\RReduce$ and $\SReduce$ are also secure in the
malicious model, for the same parameters as in the semi-honest model.

\begin{lemma} \label{lem:RReduce-mal}
Protocol $\RReduce(\WOTtwo{p}{q}^{\|n} \| \Auth)$ securely implements $\WOTtwo{p'}{q'}$ in the malicious model,
where
 $p' = 1 - (1-p)^n \leq np$ and
$q' = q^n \leq e^{-n(1-q)}$.
\end{lemma}

\begin{proof}
Let $\mA = \emptyset$. It is easy to verify that $X_0$, $X_1$, and $C$ are uniformly distributed.
Further, we have 
\begin{align*}
Y = \bigoplus_{i=0}^{n-1} Y_i 
 = \bigoplus_{i=0}^{n-1} X_{C_i,i}
 = \bigoplus_{i=0}^{n-1} X_{D_i \oplus C,i} 
 = X_C\;.
\end{align*}
Hence, the protocol achieves correctness.

Let $\mA=\{\PlayerA\}$, and let $U_i$ be the transcript of the interaction with player $\PlayerA$ by the
$i$th instance of $\WOTT$. The transcript of the interaction with $\PlayerA$ of the protocol is
$U := (U^n,D^{n-1})$.
From Lemma~\ref{lem:predComm} and \ref{lem:redBound1} follows that
\begin{align*}
\predadv(C  \mid U) &\leq \predadv(C \mid U^n,D^{n-1}) \\
& \leq 1 - \prod_{i=0}^{n-1}(1 - \predadv(C_i  \mid U_i)) \\
& \leq 1 - (1-p)^n \leq np\;.
\end{align*}

Let $\mA=\{\PlayerB\}$, and let $V_i$ be the transcript of the interaction with player $\PlayerB$ by the
$i$th instance of $\WOTT$. The transcript of the interaction with $\PlayerB$ of the protocol is
$V := (V^n,D^{n-1})$.
Note that since $D^{n-1}$ is a probabilistic function of $V$ it can be ignored.
It follows from Lemmas \ref{lem:predXOR} and \ref{lem:redBound2} that
\begin{align*}
\predadv(X_0 \oplus X_1 \mid V)
&\leq \predadv \left (\bigoplus_{i=0}^{n-1} X_{D_i \oplus 0,i} \oplus \bigoplus_{i=0}^{n-1} X_{D_i \oplus 1,i}
 \mid V^n \right ) \\
&\leq \predadv \left (\bigoplus_{i=0}^{n-1} (X_{0,i} \oplus X_{1,i} ) \mid V^n \right ) \\
&= \prod_{i=0}^{n-1} \predadv(X_{0,i} \oplus X_{1,i} \mid V_i) \\
&\leq q^n \leq e^{-n(1-q)}\;.
\end{align*}
\end{proof}

The proof that $\ROTOR(\WOTT)$ implements $\WOTT$ in the opposite direction is very simple.

\begin{lemma} \label{lem:rotorWOT-mal}
$\ROTOR(\WOT{p}{q}{\eps})$ implements $\WOT{q}{p}{\eps}$
in the opposite direction, secure in the malicious model.
\end{lemma}

\begin{proof}
It is easy to verify that the correctness condition is satisfied. 
Furthermore, we have
\begin{align*}
\predadv(X_0 \oplus X_1 \mid U)
& = \predadv(C' \mid U) \leq p \\
\predadv(C  \mid V)
& = \predadv(X'_0 \oplus X'_1 \mid V)
 \leq q\;.
\end{align*}
\end{proof}

In the same way as in the passive case, we can implement $\SReduce$ by first applying $\ROTOR$ to all $n$ instances of $\WOTT$, then use $\RReduce$ in the opposite direction, and by finally applying $\ROTOR$. We get

\begin{lemma} \label{lem:SReduce-mal}
Protocol $\SReduce(\WOTtwo{p}{q}^{\|n} \| \Auth)$ securely implements $\WOTtwo{p'}{q'}$ in the semi-honest model,
where
 $q' = 1 - (1-q)^n \leq nq$ and
$p' = p^n \leq e^{-n(1-p)}$.
\end{lemma}

\section{WOT Amplification if $\eps=0$} \label{sec:wot-amp-e0}

We will now present several protocols that implement ROT from WOT.
We start with the special case where $p,q >0$, but $\eps = 0$.
In \cite{DaKiSa99}, a protocol for this case is presented that works for all values $p$ and $q$
if $p + q < 1$, which is optimal.
We present a slightly simplified protocol and give a more detailed analysis of its efficiency.

The main part of the reduction is the following lemma, which shows that we can implement a $\WOTtwo{p'}{q'}$ out
of 4 instances of $\WOTtwo{p}{q}$, where the value $1 - (1 - p - q)^2$ is squared.

\begin{lemma} \label{lem:132}
Let $f(p,q) := 1 - (1 - p - q)^2$, and let $p + q < 1$.
We can securely implement
$\WOTtwo{p'}{q'}$ out of 4 instances of $\WOTtwo{p}{q}$
with 
\[ f(p',q') \geq f^2(p,q)\;,\]
secure in the semi-honest and the malicious model.
\end{lemma}

\begin{proof}
It suffices to show that
\[ 1-p'-q' \geq \sqrt{2 - (1-p-q)^2} \cdot (1-p-q)\;,\]
since then
\begin{align*}
 f(p',q')
 &= 1 - (1-p' -q')^2 \\
 &\leq 1 - (2 - (1 - p - q)^2)(1-p-q)^2 \\
 & = f^2(p,q)\;.
\end{align*}

Twice, we apply either the protocol $\RReduce(\WOTT^{\|2} \| \Auth)$ or protocol
$\SReduce(\WOTT^{\|2} \| \Auth)$, such that each time the larger of the two parameter gets
reduced.

Since the protocols are symmetric, we can assume that $p > q$. Therefore, the
first protocol that will be applied is $\SReduce$.
We have to distinguish between two cases. If $p^2 \geq 1 - (1-q)^2 = 2q - q^2$, then also
the second protocol is $\SReduce$, and, therefore,
\begin{align*}
p' = p^4, \qquad
q' = 1 - (1-q)^4\;.
\end{align*}
Let
\begin{align*}
f_1(p,q)
&:= \frac{1-p'-q'}{1-p-q} = \frac{(1-q)^4 - p^4}{1-p-q} \\
&= p^3 + p^2(1-q) + (1-q)^2p + (1-q)^3
\end{align*}
and
\[g_1(p,q) := f_1(p,q) - (1 + p - q) = (p^2 - 2q + q^2)(1+p-q)\;.\]
We will now show that $f_1(p,q) \geq \sqrt{2 - (1-p-q)^2} $ if $p^2 \geq 2q - q^2$.
Since for $0 < p< 1$
and $0 <q <1$, we have $1 + p - q > 0$.
It follows that $g_1(p,q) \geq 0$
for all $p$ and $q$ that satisfy
$p^2 \geq 2q - q^2$ and, therefore, also
\[f_1(p,q) \geq 1 + p - q\]
for all these values. Hence, it suffices to show that 
\[1 + p - q \geq \sqrt{2 - (1-p-q)^2}\] for $p^2 \geq 2q - q^2$.

Let us fix the value $d := 1 - p - q$. We have $1 + p - q = 2p+d$, and thus $1+p-q$
is minimal for $p^2 = 2q - q^2$.
It is taken on by the values $q_0$ and
$p_0 = \sqrt{2q_0 - q_0^2}$, which can be calculated by solving the equation
$\sqrt{2q_0 - q_0^2} + q_0 = 1 - d$, which is equal to
$2q_0^2 - (4-2d)q_0 + (1-2d + d^2) = 0$. We get
\[ q_0 = \frac{(4-2d) - \sqrt{(4-2d)^2 - 4\cdot 2 \cdot (1-2d + d^2)}}{4}
=\frac{2 - d - \sqrt{2 - d^2}}{2}\;.\]
So, for $p + q = 1 - d$, we have
\[ f_1(p,q) \geq 1 + (1-d-q_0) - q_0 = 2 - d - (2 - d - \sqrt{2 - d^2}) = \sqrt{2 - d^2}\;.\]

If $p^2 < 2q - q^2$, the second protocol will be $\RReduce$, and, therefore,
\begin{align*}
p' &= 1 - (1-p^2)^2 = 2p^2 - p^4\;, \\
q' &= (1 - (1-q)^2)^2 = 4q^2 - 4q^3 + q^4\;.
\end{align*}
Let
\begin{align*}
f_2(p,q) := & \; \frac{1-p'-q'}{1-p-q}
= \frac{1-2p^2 + p^4- 4q^2 + 4q^3 - q^4}{1-p-q} \\
= & \; q^3-3q^2-q^2p+q+2qp+q p^2+1+p-p^2-p^3\;.
\end{align*}
We will now show that $f_2(p,q) \geq \sqrt{2 - d^2}$ for $p^2 \leq 2q + q^2$.
Let
\[g_2(p,q) := f_2(p,q) - (1 + p - q) = (p^2 - 2q + q^2)(q-p-1)\;, \]
which is equal to $0$ if $p^2 = 2q + q^2$. Therefore, we have
\[ f_2(p,q) = 1 + p - q = f_1(p,q)\]
for all $p$ and $q$ that satisfy $p^2 = 2q + q^2$. Again, let us fix $d := 1 - p - q$ and let 
\begin{align*}
h_2(q)
:=  & \; f_2(1 - d - q,q) \\
=  & \; 4q^3 - (12-6d)q^2 + (8 - 12d + 4d^2)q + 4d-4d^2+d^3\;.
\end{align*}
We differentiate $h_2(q)$ twice, and get
\begin{align*}
h'_2(q) & = 12q^2 - (24-12d)q + 8 - 12d + 4d^2\;, \\
h''_2(q) & = 24q - 24-12d\;.
\end{align*}
Since $h''_2(q) \leq 24q - 24 < 0$ for $q < 1$, $h_2(q)$ is concave for $0 \leq q \leq 1$ and
$p^2 \leq 2q + q^2$.
It will therefore take on its minimum on a point on the bound. One one side, we have
$p^2 = 2q - q^2$, and therefore 
$q_0$ (see above) is the value on the bound, for which we have $h_2(q_0) = \sqrt{2 - d^2}$.
On the other side, $q_1 = (1-d)/2$ is the value on the bound, for which we have
\[h_2((1-d)/2) = \frac{3 -d^2} {2}\;.\]
For all $d$ we have
\[\frac{3 -d^2} {2} = \sqrt{ \frac { (d^2 - 1)^2}{4} + (2 - d^2)} \geq \sqrt{2 - d^2}\;,\]
so the minimum is always in $q_0$. Therefore, both $f_1(p,q)$ and $f_2(p,q)$
take on their minimum in $(1-d-q_0,q_0)$,
and are always larger than $\sqrt{2 - d^2}$. The statement follows.
\end{proof}

\begin{theorem} \label{thm:otAmp-e0}
Let $p(k)$ and $q(k)$ be functions computable in time $\poly(k)$ such that $p(k) + q(k) < 1$ for all $k$.
$\WOTtwo{2^{-k}}{2^{-k}}$ can efficiently be implemented using
\[\frac{2 \cdot k^{2}}{(1-p(k)-q(k))^{4}}\]
instances of $\WOTtwo{p}{q}$, secure in the semi-honest and the malicious model.
\end{theorem}

\begin{proof}
We apply $t$ times Lemma~\ref{lem:132}, which gives us a $\WOTtwo{p'}{q'}$ with
 $f(p',q') \leq f^{(2^t)}(p,q)$. Using Lemmas~\ref{lem:redBound2} and \ref{lem:redBound3}, we get
\begin{align*}
p' + q' &= 1 - \sqrt{1 - f(p',q')} 
\leq 1 - \sqrt{1 - f^{(2^t)}(p,q)} \leq f^{(2^t)}(p,q) \\
& \leq \exp(-2^t(1-f(p,q)))
= \exp(-2^t(1-p-q)^2)\;.
\end{align*}
To satisfy $p' + q' \leq 2^{-k}$, we choose
\[ t := \left \lceil \log \left ( \frac {-\ln(2^{-k})} {(1 - p - q)^2} \right ) \right \rceil
\leq \log \left ( \frac {\ln(2) \cdot k } {(1 - p - q)^2} \right )  + 1\;.\]
Our protocol requires 
\[ 4^t
\leq \frac {4 \cdot \ln^2(2) \cdot k^2} {(1 - p - q)^4}
\leq \frac {2 \cdot k^2} {(1 - p - q)^4}\]
instances of $\WOTtwo{p}{q}$.
\end{proof}

\paragraph{OT-Combiners.}
As shown in \cite{HKNRR05,MePrWu07}, Theorem~\ref{thm:otAmp-e0} can be used to implement an efficient \emph{$(\alpha,\beta;n)$-robust oblivious transfer combiner}.
We have $n$ different implementations of OT, out of which $\alpha$ are secure for the sender, and $\beta$
are secure for the receiver, where $\alpha + \beta > n$. Choosing randomly one of these $n$ different implementations of OT and using random inputs
implements a $\WOTtwo{p}{q}$ for $p = (n-\beta)/n$, and $q = (n-\alpha)/n$. Since $1 - p - q \geq 1/n$,
we can implement a $\WOTtwo{2^{-k}}{2^{-k}}$ using
$2 k^{2}n^{4}$
instances of the weak implementations of OT, and common randomness.

\section{WOT Amplification if $p=0$ or $q=0$} \label{sec:wot-amp-p0}

We will now look at the special case where $\eps > 0$, but either $p=0$ or $q=0$. This special
case has not been considered in \cite{DaKiSa99}. 
There is a strong connection of this problem to the one-way key-agreement problem studied
in \cite{HolRen05,Holens06}, as well as to the statistical-distance polarization problem
studied in \cite{SahVad99,Vadhan99}.

We will make the amplification in two steps. First, in Lemma~\ref{lem:holae-sahvad} (which is related to Lemma~4.13 in \cite{Holens06}), we implement a $\WOTT$ with constant errors. In Lemma~\ref{lem:sahvad} (related to Lemma~4.1 in \cite{SahVad99}),
we show how the error can be made arbitrarily small.

\begin{lemma} \label{lem:holae-sahvad}
Let $q(k)$ and $\eps(k)$ be functions computable in time $\poly(k)$ such that $\sqrt{q(k)} + 2 \eps(k) <1$ for all $k$.
Let 
\[ \lambda := \max \left (1,\frac{1}{ \log\left((1 - 2\eps)^2/q\right)} \right )\;.\]
Then $\WOT{0}{1/3}{1/50}$ can efficiently be implemented using at most
\[\frac {128 \lambda}{(1 - 2\eps)^{(12 \lambda)}}\] instances of
$\WOT{0}{q}{\eps}$ secure in the semi-honest model.
\end{lemma}

\begin{proof}
Let $\alpha = 1 - 2\eps$ and $\beta = \max(q,\alpha^2/2)$. Note that $\lambda = 1/ \log(\alpha^2/\beta)$.
We use
\begin{align*}
\bG & = \RReduce(\bF^{\|s}\|\Auth)\;,\\
\bH & = \EReduce(\bG^{\|r}\|\Auth)
\end{align*}
for $s := \lceil 5 \lambda \rceil$ and $r := \lceil 1/(4 \beta^s) \rceil$.
Notice that $s < 5 \lambda + 1 \leq 6 \lambda$. Further, since
$s > 5/ \log(\alpha^2 / \beta) > 5/\log(1/\beta) = \log_{\beta}(1/32)$,
we get
\[r < \frac{1}{4\beta^s} + 1 = \frac{1 + 4\beta^s}{4\beta^s} 
< \frac{1 + 4/32}{4 \beta^s} = \frac{9}{32 \beta^s}
< \frac {1}{3\beta^s}\;.\]
Using Lemmas \ref{lem:RReduce} and \ref{lem:ered}, we get that $\bG$ is a $\WOT{0}{\beta'}{(1-\alpha')/2}$ with
$\beta' = \beta^s$ and $\alpha' = \alpha^s$, and $\bH$ is a $\WOT{0}{q''}{\eps''}$ with
\begin{align*}
\eps''
&\leq \exp \left ( -2 r \left (\frac 1 2 - \frac {1-\alpha'} 2 \right )^2 \right )
\leq \exp \left ( -r \frac{\alpha^{2s}}{2} \right ) \\
& \leq \exp \left ( -\frac {\alpha^{2s}}{8\beta^s} \right )
= \exp \left ( -\frac {1}{8} \left ( \frac {\alpha^{2}}{\beta}\right)^s \right ) \\
&\leq \exp \left ( -\frac {1}{8} 2^{\log(\alpha^2/\beta)\frac{5}{\log(\alpha^{2}/\beta)}} \right )
= \exp \left( -32/8 \right ) < 1/50
\end{align*}
and, using that $r < 1/(3\beta^s)$, we get $q '' \leq r \beta' \leq r \beta^s < 1/3$.

Finally, the number of instances used is $s \cdot r$, which is at most
\[
6 \lambda \cdot \frac 1 {3 \beta^{6\lambda}}
=  \frac {2 \lambda} {\beta^{6\lambda}} = \frac{128\lambda}{\alpha^{12\lambda}}
\;,\]
since $2^{1/\lambda} = \alpha^2/\beta$ and thus $\beta^{6\lambda} = \alpha^{12\lambda}/64$.
\end{proof}

\begin{lemma} \label{lem:sahvad}
$\WOT{0}{2^{-k}}{2^{-k}}$ can efficiently be implemented using
\[116 \cdot \log(20 k) \cdot  k^{\log 3+1} = O\left( k^{2.6}\right)\] instances of
$\WOT{0}{1/3}{1/11}$ secure in the semi-honest model.
\end{lemma}

\begin{proof}
Let $\beta = 1/3$, and $\alpha = 1 - 2 \cdot 1/11 = 9/11$.
Let $\ell = \lceil \log(4 k + 4\log k) \rceil$ and $m  = 3^{\ell} / 2$.
We use the reductions
\begin{align*}
\bG & = \RReduce(\bF^{\|\ell}\|\Auth)\;, \\
\bH & = \EReduce(\bG^{\|m}\|\Auth)\;, \\
\bI & = \RReduce(\bH^{\|k}\|\Auth)\;.
\end{align*}
Using Lemmas \ref{lem:RReduce} and \ref{lem:ered} and since $\bF$ is a $\WOT{0}{\beta}{(1-\alpha)/2}$,
$\bG$ is a $\WOT{0}{\beta'}{(1-\alpha')/2}$, where
$\beta' = \beta^\ell$ and $\alpha' = \alpha^\ell$. $\bH$ is a $\WOT{0}{\beta''}{\eps''}$ with
\[\beta'' \leq m \beta' = 3^\ell/2 \cdot (1/3)^\ell = 1/2\]
and, since $3 \cdot \alpha^2 > 2$,
\begin{align*}
\eps''
&\leq \exp \left ( -2m \left ( \frac 1 2 - \frac{1-\alpha'} 2 \right ) ^2 \right ) =
\exp \left ( - 3^{\ell} \cdot \frac {(\alpha^\ell)^2}{4} \right )
\\
&= \exp \left ( - \frac {(3 \cdot \alpha^2)^{\ell}} 4 \right )
\leq \exp \left ( - \frac {2^{\ell}} 4 \right )
\leq \exp \left ( - k - \log k \right )
< 2^{ - k - \log k}
\;.
\end{align*}
 Finally, $\bI$ is a $\WOT{0}{\beta'''}{\eps'''}$ with 
$\eps''' \leq k 2^{-k - \log k} = 2^{-k}$ and $\beta''' \leq 2^{-k}$.

From Lemma~\ref{lem:redBound2} follows that 
\[4 k + 4\log k = 4k + 4 \ln(k)/\ln(2) \leq 4k + (4k - 1)/\ln(2) \leq 10k\;.\]
The number of instances used is, using Lemma~\ref{lem:redBound2},
\begin{align*}
\ell \cdot m \cdot k
&\leq (\log(4 k + 4\log k)+1) \cdot 3^{\log(4 k + 4 \log k)+1} \cdot k \\
&\leq (\log(10 k)+1) \cdot 3 \cdot (10 k)^{\log 3} \cdot k \\
&\leq 116 \cdot \log(20 k) \cdot  k^{\log 3+1} = O\left( k^{2.6}\right)\;.
\end{align*}
\end{proof}

Combining Lemma~\ref{lem:holae-sahvad} and Lemma~\ref{lem:sahvad}, we get the following theorem.

\begin{theorem} \label{thm:wotbound-p0}
Let $q(k)$ and $\eps(k)$ be functions computable in time $\poly(k)$ such that $\sqrt{q(k)} + 2 \eps(k) <1$ for all $k$. Let 
\[ \lambda := \max \left (1,\frac{1}{ \log\left((1 - 2\eps)^2/q\right)} \right )\;.\]
$\WOT{0}{2^{-k}}{2^{-k}}$ can efficiently be implemented using at most
\[O\left( \frac {k^{2.6} \lambda}{(1 - 2\eps)^{(12 \lambda)}}\right)\] instances of
$\WOT{0}{q}{\eps}$ secure in the semi-honest model.
\end{theorem}

Since $\RReduce$ and $\SReduce$ are symmetrical, we immediately get

\begin{corollary} \label{cor:wotbound-q0}
Let $p(k)$ and $\eps(k)$ be functions computable in time $\poly(k)$ such that $\sqrt{p(k)} + 2 \eps(k) <1$ for all $k$. Let 
\[ \lambda := \max \left (1,\frac{1}{ \log\left((1 - 2\eps)^2/p\right)} \right )\;.\]
$\WOT{2^{-k}}{0}{2^{-k}}$ can efficiently be implemented using at most
\[O\left( \frac {k^{2.6} \lambda}{(1 - 2\eps)^{(12 \lambda)}}\right)\] instances of
$\WOT{p}{0}{\eps}$ secure in the semi-honest model.
\end{corollary}

Since any protocol (using our basic protocols) for the special cases where either $p=0$ or $q=0$
can directly be translated into a one-way key-agreement protocol for distributions studied in
\cite{HolRen05}, it follows from Theorem~4 in \cite{HolRen05} that
using our basic protocols, this is the best bound that we can achieve. 
However, it is not clear whether other reductions, 
would be able to achieve a better bound.

\section{WOT Amplification if $p,q,\eps >0$.} \label{sec:wot-amp}

To find an optimal protocol for the general case where all three parameters are non-zero
turns out to be much harder than the other three special cases. It is still unknown
what the exact bound is in this case. In this section we present some
partial results.

We start with the case where all values are non-zero, but smaller than $1/50$.

\begin{lemma} \label{lem:genLowBound1}
$\WOT{2^{-k}}{2^{-k}}{2^{-k}}$ can efficiently be implemented using
\[175 \cdot k^{2+\log(3)} \leq 175 \cdot  k^{3.6} \] instances of
$\WOT{1/50}{1/50}{1/50}$ secure in the semi-honest model.
\end{lemma}

\begin{proof}
We set $\bF_0 := \WOT{p}{q}{\eps}$ and iterate the reduction 
\[\bF_{i+1} := \SReduce(\RReduce(\EReduce(\bF_{i}^{\|3}\|\Auth)^{\|2}\|\Auth)^{\|2}\|\Auth)\;,\]
until $\bF_j$ is a $\WOT{p_j}{q_j}{\eps_j}$ with $\max(p_j,q_j,\eps_j) \leq  2^{-k}$.
In every iteration, we have
$p_{i+1} \leq (2 \cdot (3 p_i))^2 = 36 p^2_i$, $q_{i+1} \leq 2 \cdot ((3 q_i)^2) = 18 q^2_i$, and
$\eps_{i+1} \leq 2 \cdot 2 \cdot (3 \eps^2 - 2 \eps^3) \leq 12 \eps^2$, from which follows that
\[ \max(p_j,q_j,\eps_j) \leq 36^{2^j-1} \cdot \frac 1 {50^{2^j}} =  \frac 1 {36} \left ( \frac{ 36}{50} \right )^{2^j} \leq \left ( \frac{ 36}{50} \right )^{2^j}\;.\]
To achieve $\max(p_j,q_j,\eps_j) \leq 2^{-k}$,
we choose
\[j := \left \lceil \log \frac k {\log(50/36)} \right \rceil \leq \log(2.1101 \cdot k) + 1 = \log(4.2202 \cdot k)\;.\]
To implement one instance of $\bF_j$, we need at most 
\[12^j \leq (4.2202 \cdot k)^{\log(12)} \leq 175 \cdot k^{2+\log(3)} \leq 175 \cdot k^{3.6}\]
instances of $\bF_0$.
\end{proof}

We will now give a similar bound as in Lemma~5 in \cite{DaKiSa99}, which was
$p + q + 2\eps \leq 0.45$.
But since our protocol $\EReduce$ is different, we are only able to achieve a smaller bound.
As in \cite{DaKiSa99}, we are only able to obtain our bound using a simulation. Our simulation
works as follows:
Let $l_i(p,q)$ be a function such that for all $p$, $q$ and $\eps < l_i(p,q)$,
$\WOT{1/50}{1/50}{1/50}$ can be implemented using $\WOT{p}{q}{\eps}$. Using $l_i(p,q)$, we define
\begin{align*}
 l_{i+1}(p,q) & := \max(S^{-1}_{\eps}(l_i(S_p(p),S_q(q))), R^{-1}_{\eps}(l_i(R_p(p),R_q(q))),\\
 & \qquad \qquad E^{-1}_{\eps}(l_i(E_p(p),E_q(q))) )\;,
\end{align*}
where 
\begin{align*}
S_p(p) := p^2\;, \quad
S_q(q) := 1 - (1-q)^2\;, \quad
S^{-1}_{\eps}(\eps) := (1 - \sqrt{1-2\eps})/2\;,
\end{align*}
\begin{align*}
R_p(p) := 1 - (1-p)^2\;, \quad
R_q(q) := q^2\;, \quad
R^{-1}_{\eps}(\eps) := (1 - \sqrt{1-2\eps})/2\;,
\end{align*}
\begin{align*}
E_p(p) := 1 - (1-p)^3\;,  \quad \quad E_q(q) := 1 - (1-q)^3\;,
\end{align*}
and $E^{-1}_{\eps}(\eps)$ is the inverse of $E_{\eps}(\eps) := 3\eps^2 - 2 \eps^3$.

Now, for all $p$, $q$ and $\eps < l_{i+1}(p,q)$,
$\WOT{1/50}{1/50}{1/50}$ can be implemented using $\WOT{p}{q}{\eps}$, since applying one of the three protocols $\SReduce(\WOT{p}{q}{\eps}^{\|2}\|\Auth)$, $\RReduce(\WOT{p}{q}{\eps}^{\|2}\|\Auth)$,
or $\EReduce(\WOT{p}{q}{\eps}^{\|3}\|\Auth))$ gives us an instance of $\WOT{p'}{q'}{\eps'}$ with
$\eps' < l_i(p',q')$, from which $\WOT{1/50}{1/50}{1/50}$ can be implemented.

\begin{figure} 
\begin{center}
\input{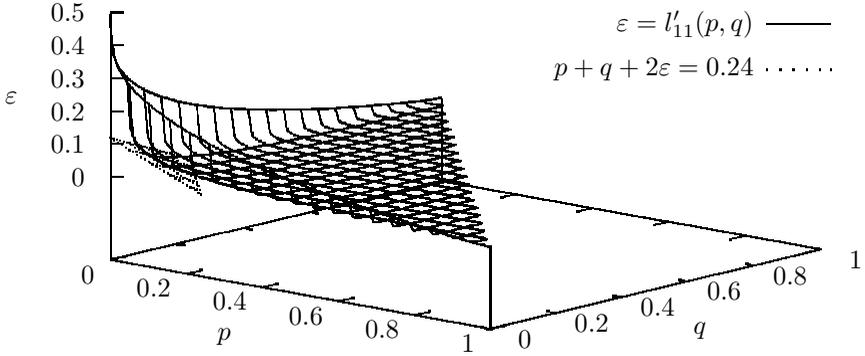}
\end{center}
\vspace{-1.5cm}
\caption{Plot of the bounds $\eps = l'_{11}(p,q)$ and $p+q+2\eps=0.24$. \label{fig:wot-bound2}
}
\end{figure}

Obviously, $l_0(p,q) := (0.02 - p - q)/2$ satisfies our condition.
Iterating $8$ times, we get $l_8(p,q)$, where for all $p,q$ we have $l_8(p,q) \geq (0.15 - p - q)/2$.
Using $l'_0(p,q) := (0.15 - p - q)/2$ and iterating
$11$ times, we get $l'_{11}(p,q)$, were for all $p,q$ we have $l'_{11}(p,q) \geq (0.24 - p - q)/2$
(See also Figure~\ref{fig:wot-bound2}).

\begin{lemma} \label{lem:generalBound}
If $p+q+2\eps \leq 0.24$, then $\WOT{1/50}{1/50}{1/50}$ can efficiently be implemented using 
$O(1)$ instances of
$\WOT{p}{q}{\eps}$, secure in the semi-honest model.
\end{lemma}

We will now further extend this result and give bounds for the cases where one of the
three values is large, while the others are small.

\begin{lemma} \label{lem:generalBound2}
If $p + 22q + 44\eps < 1$, then $\WOT{p'}{q'}{\eps'}$ with $p'+q'+2\eps' \leq 0.24$ can efficiently be implemented using $4/(1-p)$ instances of
$\WOT{p}{q}{\eps}$, secure in the semi-honest model.
\end{lemma}

\begin{proof}
We apply \[\bF = \SReduce(\WOT{p}{q}{\eps}^{\|n}\|\Auth)\] for an $n > 0$ such that $\bF$ is a $\WOT{p'}{q'}{\eps'}$ with $p'+q'+2\eps' \leq 0.24$. 
Using Lemma~\ref{lem:SReduce}, we need to find a value $n$ and constants $\alpha$ and $\beta$ with $\alpha + \beta \leq 0.24$, such that $e^{-n(1-p)} \leq \alpha$
and $n q + 2 n \eps \leq \beta$, which is equivalent to $n (1-p) \geq \ln (1/\alpha)$ and $q + 2\eps \leq \beta / n$. We can choose
\[n := \left \lceil \frac{\ln (1/\alpha)}{1-p} \right \rceil \leq \frac{\ln (1/\alpha)}{1-p} + 1 \leq \frac{\ln (1/\alpha)+1}{1-p}\;.\]
The first inequality is satisfied by definition of $n$, and the second if
\[ q + 2\eps \leq \frac {\beta (1-p)} {\ln (1/\alpha) + 1}\;,\]
which is equivalent to
\[ \frac{\ln (1/\alpha) + 1}{\beta}(q + 2\eps) + p \leq 1\;. \]
Choosing $\alpha = 0.05$, and $\beta = 0.19$, we get $(\ln (1/\alpha) + 1)/\beta \leq 22$.
Our protocol needs $n \leq 4/(1-p)$ instances.
\end{proof}

In the same way, we get

\begin{lemma} \label{lem:generalBound3}
If $22p + q + 44\eps < 1$, then $\WOT{p'}{q'}{\eps'}$ with $p'+q'+2\eps' \leq 0.24$ can efficiently be implemented using $4/(1-q)$ instances of
$\WOT{p}{q}{\eps}$, secure in the semi-honest model.
\end{lemma}

The proof of Lemma~\ref{lem:generalBound3} is omitted, as it can be done in the same way as the proof of Lemma~\ref{lem:generalBound2}.

\begin{lemma} \label{lem:generalBound4}
If $7 \sqrt{p+q} + 2\eps < 1$, then $\WOT{p'}{q'}{\eps'}$ with $p'+q'+2\eps' \leq 0.24$ can efficiently be implemented using $3(1/2-\eps)^{-2}$ instances of
$\WOT{p}{q}{\eps}$, secure in the semi-honest model.
\end{lemma}

\begin{proof}
We apply \[\bF = \EReduce(\WOT{p}{q}{\eps}^{\|n}\|\Auth)\]
 for an $n > 0$ such that $\bF$ is a $\WOT{p'}{q'}{\eps'}$ with $p'+q'+2\eps' \leq 0.24$. 
Using Lemma~\ref{lem:ered}, we need to find a value $n$ and constants $\alpha$ and $\beta$ with $2\alpha + \beta \leq 0.24$, such that $e^{-2n(1/2-\eps)^2} \leq \alpha$ and $n p + n q \leq \beta$, which is equivalent to $2n (1/2-\eps)^2 \geq \ln (1/\alpha)$ and $p+q \leq \beta / n$. Furthermore, we need $\eps < \frac12$. 
We choose
\[n := \left \lceil \frac {\ln (1/\alpha)}{2(1/2-\eps)^2} \right \rceil
\leq \frac {\ln (1/\alpha) }{ 2(1/2-\eps)^2} + 1 \leq \frac {\ln (1/\alpha) + 1/2 }{ 2(1/2-\eps)^2}\;.\]
The last inequality follows from the fact that $2(1/2-\eps)^2 \leq 1/2$.
The first inequality is satisfied by definition of $n$, and the second if
\[ p+q \leq  \frac {2 \beta (1/2-\eps)^2} {\ln (1/\alpha) + 1/2}\;,\]
which is equivalent to
\[ \sqrt{\frac{2\ln (1/\alpha) + 1}{\beta}}\sqrt{p+q} + 2\eps < 1\;. \]
Choosing $\alpha = 0.02$ and $\beta = 0.20$, we get
\[ \sqrt{\frac{2\ln (1/\alpha) + 1}{\beta}} \leq 7\;.\]
 Our protocol needs $n \leq 3(1/2-\eps)^{-2}$ instances.
\end{proof}

Theorem~\ref{thm:generalBound} summarizes all the partial results we obtained in this section.

\begin{theorem} \label{thm:generalBound}
Let $p(k)$, $q(k)$ and $\eps(k)$ be functions computable in time $\poly(k)$ 
such that
 \[p+q+2\eps \leq 0.24\;,\] or  
\[\min( p + 22q + 44\eps,22p + q + 44\eps, 7 \sqrt{p+q} + 2\eps) < 1\]
for all $k$. Then
$\WOT{2^{-k}}{2^{-k}}{2^{-k}}$ can efficiently be implemented using
\[O \left ( \frac{k^{3.6}}{(1-p) (1-q)(1/2-\eps)^{2}} \right ) \]
instances of
$\WOT{p}{q}{\eps}$ secure in the semi-honest model.
\end{theorem}

\begin{proof}
Follows directly from Lemmas \ref{lem:generalBound}, \ref{lem:generalBound2}, \ref{lem:generalBound3}, and \ref{lem:generalBound4}.
\end{proof}

Since Theorem~\ref{thm:otAmp-e0} gives us a bound on the number of instances used, we can also bound
the error probability, and therefore, we can extend the result of Theorem~\ref{thm:otAmp-e0} to allow
for a (small) error.

\begin{corollary} \label{cor:generalBound5}
Let $p(k)$, $q(k)$ and $\eps(k)$ be functions computable in time $\poly(k)$
 such that
\[(1 - p - q)^4 < - 178 \cdot \log(1 - 2\eps)\]
for all $k$. Then $\WOT{2^{-k}}{2^{-k}}{2^{-k}}$ can efficiently be implemented using 
\[O \left ( \frac{k^{3.6}}{(1-p-q)^{4}} \right )\]
instances of $\WOT{p}{q}{\eps}$, secure in the semi-honest model.
\end{corollary}

\begin{proof}
We apply the reduction used in Theorem~\ref{thm:otAmp-e0} for $k = 5$. We get $p' \leq 2^{-5}$, $q' \leq 2^{-5}$,
and $\eps' \leq (1 - (1- 2\eps)^n)/2$, for $n = 50 \cdot (1-p-q)^{-4}$. We have
\begin{align*}
\log(1 - 2\eps')
& = n \log(1- 2\eps) = 50 \cdot (1-p-q)^{-4} \cdot (1 - p - q)^4 / {-178} \\
& = - 50/178
\end{align*}
and therefore
\[p' + q' + 2 \eps' \leq 2 \cdot 2^{-5} + (1 - 2^{- 50/178}) < 0.24\;.\]
The statement follows now by applying Lemmas~\ref{lem:genLowBound1} and \ref{lem:generalBound}.
\end{proof}

\section{Discussion and Open Problems}

We have presented several protocols that implement ROT from many instances of WOT.
For
the special case where $\eps = 0$, we were able to achieve the optimal bound,
and when either $p=0$ or $q=0$, we were at least able to give protocols which
 achieve the optimal bound for the basic protocols that we use.

However, for the general case, we still do not have very satisfactory results.
One of the main difficulties is that we do not know exactly which of the basic
protocols needs to be applied in which
situation. To be able to do that, we would need a better understanding of
how these protocols work together.

There are still many open problems concerning WOT amplification. Here are some of them:

\begin{itemize}
\item
Can we improve the impossibility bound?
\item
For what parameters of WOT can we implement ROT with our basic protocols?
How many instances do we need?
\item
Are there other basic protocols that give better bounds?
Is it possible to use (a modified version of) the protocol $\EReduce$ from \cite{DaKiSa99}?
Is it possible to reduce two parameters at the same time?
\item Is there a (simple) way to make $\EReduce$ secure in the malicious model?
\item
Can GWOT be used to improve WOT amplification?
\item
Is it possible to define WOT in another, more general way?
\item
How do we have to define WOT in a multi-party setting?
\end{itemize}

\chapter{Computational Weak Oblivious Transfer} \label{chap:compWOT}

In this chapter we show how an OT which may contain errors and
which is only \emph{mildly computationally secure} for the
two players can be amplified to a computationally-secure OT. In particular, we show in
Theorem~\ref{thm:compOT} --- using Holenstein's uniform hard-core lemma \cite{Holens05,Holens06},
which is a uniform variant of Impagliazzo's hard-core lemma \cite{Impagl95} ---
that if WOT can be amplified to
ROT in the information-theoretic setting, then also the corresponding computational version of WOT
 can be amplified to a computationally-secure ROT, \emph{using the same protocol}.
 
Our results
generalize the results presented in \cite{Haitne04}, as we cover a much larger
region for the values $p$, $q$ and $\eps$, and in our case the security for both players may be computational.

\section{Preliminaries}

In the following, $k \in \bbN$ is always the the security parameter.
We say that a function $f: \bbN \rightarrow \bbN$ is \emph{polynomial in $k$}, denoted by $\poly(k)$, if there exist constants $c > 0$ and $k_0$, such that $f(k) \leq k^{c}$ for all $k \geq k_0$.
A function $f: \bbN \rightarrow [0,1]$ is \emph{negligible in $k$}, denoted by $\negl(k)$, if for all constant $c > 0$ there exists a constant $k_0$, such that $f(k) \leq k^{-c}$ for all $k \geq k_0$. A function $f: \bbN \rightarrow [0,1]$ is \emph{noticeable} if there exit constants $c>0$ and $k_0$ such that $f(k) \geq k^{-c}$ for all $k \geq k_0$. An algorithm $B$ which has oracle access to an algorithm $A$ will be denoted by $B^A$.

We will need the following lemmas, which are, when put together, the computational version of Lemma~\ref{lem:PredAdvStadDist}. 

\begin{lemma} \label{lem:comp-dist2pred}
Let functions $f: \{0,1\}^k \rightarrow \{0,1\}^\ell$, $P: \{0,1\}^k \rightarrow \{0,1\}$, and a
distribution $P_W$ over $\{0,1\}^k$ be given. There is an oracle algorithm $B^{(\cdot)}$ such that, for any algorithm $A$ where
\[\Pr[A(f(W),P(W)) = 1] - \Pr[A(f(W),U) = 1] = \eps\;,\]
where $W$ is distributed according to $P_W$  and $U$ is uniformly distributed, algorithm $B^A$ satisfies
\[\Pr[ B^A(f(W)) = P(W)] = \frac 1 2 + \eps\;,\]
does one oracle call to $A$, and computes one XOR.
\end{lemma}

\begin{proof}
On input $f(w)$, let algorithm $B^A$ choose a bit $u$ uniformly at random and output $A(f(w),u) \oplus u \oplus 1$. Let
\[g(w,u) := \Pr[ A(f(w),u) = 1]\;.\]
The output of $B^A$ is correct either if $U=P(W)$ and the output of $A$ is $1$, or $U \neq P(W)$
and the output of $A$ is $0$. We get
\begin{align*}
& \Pr[ B^A(f(W)) = P(W)] \\
& \qquad = \sum_w P_W(w) \left (\frac{g(w,P(w))}{2} + \frac{1- g(w,1-P(w))}{2} \right ) \\
& \qquad = \frac 1 2 + \sum_w P_W(w) \frac{g(w,P(w))- g(w,1-P(w))}{2} \\
& \qquad = \frac 1 2 + \sum_w P_W(w) \left ( g(w,P(w)) - \frac{g(w,P(w))- g(w,1-P(w))}{2} \right ) \\
& \qquad = \frac 1 2 + \Pr[A(f(W),P(W)) = 1] - \Pr[A(f(W),U) = 1] \\
& \qquad = \frac 1 2 + \adv^A(f(W),P(W)),(f(W),U) \;.
\end{align*}
\end{proof}

\begin{lemma}  \label{lem:comp-pred2dist}
Let functions $f: \{0,1\}^k \rightarrow \{0,1\}^\ell$, $P: \{0,1\}^k \rightarrow \{0,1\}$, and a
distribution $P_W$ over $\{0,1\}^k$ be given. There is an oracle algorithm $A^{(\cdot)}$ such that, for any algorithm $B$
where \[\Pr[ B(f(W)) = P(W)] = \frac 1 2 + \eps\;,\]
we have
\[\adv^{A^B}(f(W),P(W)),(f(W),U) = \eps\;,\]
where $W$ is distributed according to $P_W$  and $U$ is uniformly distributed, does one oracle call to $A$, and computes one XOR.
\end{lemma}

\begin{proof}
On input $(f(w),b)$, let Algorithm $A$ output $B(f(w)) \oplus b \oplus 1$.
If $b$ is a uniform random bit, than we have $\Pr[A^B(f(w),b)=1]=1/2$, and if $b = P(w)$, then
$\Pr[A^B(f(w),b)=1]=1/2 + \eps$. Therefore, we have $\adv^{A^B}((f(W),P(W)),(f(W),U)) = \eps$.
\end{proof}

\section{Pseudo-Randomness Extraction} \label{sec:PseudoRandExt}

In this section we state a \emph{pseudo-randomness extraction theorem}, Theorem \ref{thm:holae}, that we need later to prove our main theorem of
this chapter, Theorem~\ref{thm:compOT}. Theorem \ref{thm:holae} is based on the \emph{uniform hard-core lemma}
\cite{Holens05,Holens06}, which is a uniform variant of the hard-core lemma from \cite{Impagl95}.

\begin{lemma}[Uniform hard-core lemma \cite{Holens05,Holens06}] \label{lem:hardcore}
Let the functions $f:\{0,1\}^k \rightarrow \{0,1\}^\ell$, $P:\{0,1\}^k \rightarrow \{0,1\}$, 
$\delta: \bbN \rightarrow [0,1]$ and $\gamma: \bbN \rightarrow [0,1]$ computable in time $\poly(k)$ be given, such that $\gamma$ and $\delta$ are noticeable.
Assume that there is no polynomial time algorithm $B$ such that
\[\Pr[B(f(W))=P(W)] \geq 1 - \frac \delta 2 + \frac {\gamma^2 \delta^5}{8192}\;,\]
where $W$ is chosen uniformly at random from $\{0,1\}^k$, for infinitely many $k$.
Then, there is no polynomial time oracle algorithm $A^{(\cdot)}$\footnote{$A^{(\cdot)}$ has oracle access to the \emph{characteristic function} $\chi_\mS$ of the set $\mS$, which is defined as $\chi_{\mS}(w) := 1$ if $w \in \mS$ and $\chi_{\mS}(w) := 0$ otherwise.} such that for infinitely many $k$ the following holds: For any set $\mS \subseteq \{0,1\}^k$ with $|\mS| \geq \delta 2^k$,
\[ \Pr[ A^{\chi_\mS}(f(W)) = P(W)] \geq \frac{1+\gamma} 2\;,\]
where $W$ is chosen uniformly at random from $\mS$ and the queries of $A$ to $\chi_\mS$ are computed independently of the input $f(W)$.
\end{lemma}

Theorem \ref{thm:holae} is a modified version of Theorem 7.3 in \cite{Holens06} and differs from it in two points. First, we simplified it by omitting
the function $q(w)$ that indicates whether $w$ is valid, because in our setting all $w$ are valid. Second, we allow the functions $\extr$ and $\leak$ to
depend on the value $Z^n$, and not only on $X^n$.
The proof of Theorem~\ref{thm:holae} is basically the same as the proof of Theorem 7.3 in \cite{Holens06}. Notice that in the proof of Theorem 7.3 in \cite{Holens06} there is a step
missing before equation (7.8), which is fixed in our proof.

The main difference of Theorem 7.3 in \cite{Holens06} and our Theorem~\ref{thm:holae} compared to the (implicit) extraction lemma in \cite{Hastad90,HILL99} and
the extraction lemma in \cite{HaHaRe05} is that it allows the adversary to gain some additional knowledge during the extraction, expressed by the function $\leak$.

\begin{theorem}[Pseudo-randomness extraction theorem, \cite{Holens06}] \label{thm:holae}
Let the functions
$f : \{0,1\}^k \rightarrow \{0,1\}^\ell$,
$P : \{0,1\}^k \rightarrow \{0,1\}$, and
$\beta : \mathbb N \rightarrow [0,1]$, all computable in time $\poly(k)$, be given, and let
$1 - \beta(k)$ be noticeable.
Assume that every polynomial time algorithm $B$ satisfies
\[ \Pr [ B(f(W)) = P(W) ] \leq \frac {1+\beta(k)} 2 \]
for all but finitely many k, for a uniform random $W \in \{0,1\}^k$. Further, let also functions $n(k)$, $s(k)$, 
\begin{align*}
\extr &: \{0,1\}^{\ell \cdot n} \times \{0,1\}^n \times \{0,1\}^s \rightarrow \{0,1\}^t\;,\\
\leak &: \{0,1\}^{\ell \cdot n} \times  \{0,1\}^n \times \{0,1\}^s \rightarrow \{0,1\}^{t'}\;,
\end{align*}
be given which are computable in time $\poly(k)$, and satisfy the following: For any distribution $P_{XZ}$ over
$\{0,1\} \times \{0,1\}^\ell$ where $\predadv(X \mid Z) \leq \beta(k)$,
$\extr(Z^n,X^n,R)$ is $\eps(k)$-close to uniform with
respect to $\leak(Z^n,X^n,R)$, for $R \in \{0,1\}^s$ chosen uniformly at random.
Then, no polynomial time algorithm $A$, which gets as input 
\[\leak((f(W_0), \dots, f(W_{n-1})),(P(W_0), \dots, P(W_{n-1})),R)\;,\]
(where $(W_1, \dots,W_n)$ is chosen uniformly at random) distinguishes
\[\extr((f(W_0), \dots, f(W_{n-1})),(P(W_0), \dots, P(W_{n-1})),R)\;\]
from a uniform random string of length $t$ with advantage $\eps(k) + \gamma(k)$,
for any non-negligible function $\gamma(k)$.
\end{theorem}

\begin{proof}
Let us assume there exists an algorithm $A$ that contradicts our assumption. We will use $A$ to construct
an oracle algorithm $\ol A^{\chi_\mS}$ for which the following holds for infinitely many $k$ for a noticeable function $\gamma'$. For any set $\mS \subseteq \{0,1\}^k$ with $|\mS| \geq (1-\beta(k))2^k$, we have
\[ \Pr[\ol A^{\chi_\mS}(f(W)) = P(W)] \geq \frac{1+\gamma'} 2\;,\]
where the probability is over the randomness of $\ol A^{\chi_\mS}$, $W$ is chosen uniformly at random from $\{0,1\}^k$, and $\ol A^{\chi_\mS}$ calls $\chi_\mS$ only with queries which are computed independently of the input.

Since $\gamma(k)$ is non-negligible, there exists a constant $c$, such that $\gamma(k) \geq k^{-c}$ for infinitely many $k$. Let $\gamma^*(k) := k^{-c}$. $\gamma^*(k)$ is a noticeable function with
$\gamma^*(k) \leq \gamma(k)$ for infinitely many $k$.

For any fixed $j \in \{0,\dots,n\}$ and any fixed set $\mS \subseteq \{0,1\}^k$ with $|\mS| \geq (1-\beta)2^k$,
we define the following values. For all $i \in \{0,\dots,n-1\}$, we choose $w_i \in \{0,1\}^k$
and $u_i \in \{0,1\}$ uniformly at random. Then we compute
\begin{align} 
y_i &:= \left \{
\begin{array}{ll}
P(w_i) & \textrm{if $i \geq j$ or $w_i \not \in \mS$\;,} \label{eq:yi} \\
u_i & \textrm{otherwise\;,}
\end{array}
\right. \\
e_j &:= \extr((f(w_1),\dots,f(w_n)), y^n, r)\;, \quad \textrm{and} \label{eq:ej} \\ 
\ell_j &:= \leak((f(w_1),\dots,f(w_n)), y^n, r)\;, \label{eq:lj}
\end{align} 
where $r \in \{0,1\}^s$ is chosen uniformly at random.

Let $P_{E_jL_j}$ be the distribution of $(e_j,\ell_j)$. From our assumption follows that 
\[\adv^A((E_0,L_0),(U,L_0)) \geq \eps + \gamma^*\]
for infinitely many $k$,
where $U \in \{0,1\}^t$ is chosen uniformly at random.
On the other hand, for $j=n$, with probability $1-\beta$ (over the choice of $w_i$) we have $y_i = u_i$, and
therefore, by Lemma~\ref{lem:Hol22}, $\predadv(Y_i \mid f(W_i)) \leq \beta$. 
The information-theoretic requirement on the functions $\extr$ and $\leak$ imply that $E_n$ is $\eps$-close to uniform with respect to
$L_n$ and therefore
\[\adv^A((E_n,L_n),(U,L_n)) \leq \eps \;.\]
The triangle inequality implies
\[\adv^A((E_0,L_0),(E_n,L_n)) +  \adv^A((U,L_0),(U,L_n)) \geq \gamma^*\]
for infinitely many $k$.
It follows that at least one of the four inequalities
$\Pr[A(E_0,L_0)=1] - \Pr[A(E_n,L_n)=1]  \geq \gamma^*/2$,
$\Pr[A(E_n,L_n)=1] - \Pr[A(E_0,L_0)=1]  \geq \gamma^*/2$,
$\Pr[A(U,L_0)=1] - \Pr[A(U,L_n)=1]  \geq \gamma^*/2$, or
$\Pr[A(U,L_n)=1] - \Pr[A(U,L_0)=1]  \geq \gamma^*/2$ holds for infinitely many $k$,
from which follows that there exists an algorithm $A'$ such that
\[\Pr[A'(E_0,L_0)=1] - \Pr[A'(E_n,L_n)=1] \geq \frac {\gamma^*} 2\]
for infinitely many $k$.
For a $J \in \{0,\dots,n-1\}$ chosen uniformly at random, we have
\[\Pr[A'(E_J,L_J)=1] - \Pr[A'(E_{J+1},L_{J+1})=1] \geq \frac{\gamma^*}{2n}\]
for infinitely many $k$.
We can now give an implementation of a distinguisher which distinguishes $(f(W),P(W))$ from $(f(W),U)$ with
advantage $\gamma^*/(2n)$ for infinitely many $k$, if $W$ is chosen uniformly from $\mS$ and $U$ is a uniform random bit, as long as oracle access to $\chi_{\mS}$ is given. Let $(f(w),b)$ be the input to the distinguisher.
It chooses $j \in \{0,\dots, n-1\}$, and for all $i \in \{0,\dots,n-1\}$ the values $w_i \in \{0,1\}^k$ and $u_i \in \{0,1\}$ uniformly at random. Then, for all $i \in \{0,\dots,n-1\}$, it computes the values $f(w_i)$, $P(w_i)$ and $y_i$ as in (\ref{eq:yi}). If $w_j \in \mS$, it replaces $f(w_j)$ with $f(w)$ and $y_i$ with $b$. Then, it computes $e_j$ and $\ell_j$ as in (\ref{eq:ej}) and (\ref{eq:lj}). 
If $b$ is a uniform bit, then this process gives random variables $(E_j,L_j)$ distributed according to $P_{E_{j+1}L_{j+1}}$, otherwise it gives random variables distributed according to
$P_{E_{j}L_{j}}$. Therefore, $A'$ distinguishes $(f(W),P(W))$ from $(f(W),U)$ with
advantage $\gamma^*/(2n)$ for infinitely many $k$, if $W$ is chosen uniformly at random from $\mS$. From  Lemma~\ref{lem:comp-dist2pred} follows that there exists a polynomial time
algorithm that predicts $P(W)$ from $f(W)$, where $W$ is chosen uniformly at random from $\mS$, with probability at least $1/2 + \gamma^*/(2n)$ for infinitely many $k$. We can now apply Lemma~\ref{lem:hardcore} for $\gamma := \gamma^*/n$ and $\delta := 1-\beta$ to obtain the statement.
\end{proof}

\section{Definition of Computational WOT} \label{sec:compOTDef}

In order to define security in the computational setting, i.e., where the running time of the adversary is bounded by a polynomial, we need to introduce a security parameter $k$ on which the players agree beforehand. We consider the \emph{uniform} model, that is, we require the same protocols to run on all
security parameters, which they get as a separate input. Additionally, we require the security parameter to be larger than the sum of the length of all the inputs and outputs of the protocol. The security in the \emph{computational
semi-honest model} is very similar to the (information-theoretic) semi-honest model
(Definition \ref{def:passiveSec}). The only differences are that
we require the distinguishers to be efficient, i.e., to run in time $\poly(k)$,
and we require the advantage of these distinguishers to be negligible in $k$. Furthermore,
we require that the simulator is efficient, i.e., runs in time $\poly(k)$.

We say that $X(k)$ and $Y(k)$ are \emph{computationally indistinguishable}, denoted by
$X \compIndist Y$, if $\adv^\mD(X,Y) \leq \negl(k)$, where 
$\mD$ is the set of all distinguishers that run in time $\poly(k)$.

\begin{definition} \label{def:compPassiveSec}
A protocol $\bP(\bF) = (\bP_\PlayerA \| \bP_\PlayerB)(\bF)$ \emph{securely implements $\bG$ in the computational
semi-honest model}, if
\begin{itemize}
\item(Correctness) $\bP(\bF_{\emptyset}) \compIndist \bG_{\emptyset} $\;.
\item(Security for \PlayerA) There exists a system $\bS_\PlayerB$ (called \emph{the simulator for $\PlayerB$}), that runs in time $\poly(k)$ and only modifies the auxiliary interfaces, such that
\[(\bP_\PlayerA \| \underline \bP_\PlayerB)(\bF_{\{\widehat \PlayerB\}}) \compIndist \bS_\PlayerB(\bG_{\{\widehat \PlayerB\}}) \;.\]
\item(Security for \PlayerB) There exists a system $\bS_\PlayerA$ (called \emph{the simulator for $\PlayerA$}), that runs in time $\poly(k)$ and only modifies the auxiliary interfaces, such that
\[(\underline \bP_\PlayerA \| \bP_\PlayerB)(\bF_{\{\widehat \PlayerA\}}) \compIndist \bS_\PlayerA(\bG_{\{\widehat \PlayerA\}}) \;.\]
\end{itemize}
\end{definition}

The primitive $\compWOT{p}{q}{\eps}$ denotes the computational version of $\WOT{p}{q}{\eps}$.
The difference to the definition of $\WOTT$ is that we require the algorithm that
guesses $X_{1-C}$ or $C$ to be efficient.

\begin{definition}[Computational WOT, semi-honest model]
Let functions $\eps: \mathbb N \rightarrow [0,1/2]$,
$p: \mathbb N \rightarrow [0,1]$, and $q: \mathbb N \rightarrow [0,1]$ computable in time $\poly(k)$ be given. Let $\bF = (\bF_{\emptyset},\bF_{\{\widehat \PlayerA\}},\bF_{\{\widehat \PlayerB\}})$ be a collection of systems in the computational semi-honest model. On input $k$, $\bF$ outputs
  $(X_0,X_1)$ to $\PlayerA$ and $(C,Y)$ to $\PlayerB$. Let $U$ be the auxiliary output to $\PlayerA$ by $\bF_{\{\widehat \PlayerA\}}$ and $V$ be the auxiliary output to $\PlayerB$  by $\bF_{\{\widehat \PlayerB\}}$.
  Let $E := X_C \oplus Y$. $\bF$ implements $\compWOT{p(k)}{q(k)}{\eps(k)}$ in the computational semi-honest model, if
\begin{itemize}
\item(Efficiency) $\bF$ can be executed in time $\poly(k)$.
\item(Correctness) $\Pr[E=1] \leq \eps(k)$ for all $k$.
\item(Security for \PlayerA) 
All polynomial time algorithms $A$ satisfy
\[ \Pr[A(V,E)=X_{1-{C}}] \leq \frac {1 + q(k)} 2\]
for all but finitely many $k$.
\item(Security for \PlayerB)
All polynomial time algorithms $A$ satisfy
\[ \Pr[A(U,E)=C] \leq \frac {1 + p(k)} 2\]
for all but finitely many $k$.
\end{itemize}
\end{definition}

Lemma~\ref{lem:compSecCond} is the computational version of Lemma~\ref{lem:WOT2ROT}.

\begin{lemma} \label{lem:compSecCond}
A collection of systems $\bF$ that securely implements 
\[\compWOT{\negl(k)}{\negl(k)}{\negl(k)}\] also securely implements $\ROT{1}{2}{1}$ in the
computational semi-honest model.
\end{lemma}

\begin{proof}
From the (computational) security conditions for $\PlayerA$ follows
that $C$ is (statistically) $\negl(k)$-close to uniform with respect to $(X_0,X_1)$.
Otherwise, it could easily and efficiently be distinguished from uniform. Similarly, it
follows from the security condition for $\PlayerB$ that $X_{1-C}$ is $\negl(k)$-close
to uniform with respect to $(C,X_C)$. From Lemma~\ref{lem:almostUniform} 
follows that $(C,X_0,X_1)$ is $\negl(k)$-close to uniform. Together with the
correctness condition, we get
\[ \bF_{\emptyset} \equiv_{\negl(k)} \ROTT_{\emptyset}\;.\]

Let $\bF_{\{\PlayerB\}}$ produce the output distribution $P_{X_0X_1CYV}$,
and let $P_{\ol {X_0X_1CY}}$ be the output distribution of $\ROTT$. We define $\bS_\PlayerB$ as follows. After receiving $(c,y)$ from 
$\ROTT$, it simulates $\bF_{\{\PlayerB\}}$, which outputs $(c',y',v')$,
until $c'=b$ and $y'=y$. It outputs $v'$.

From the correctness condition follows
that $(C',Y')$ is $\negl(k)$-close to uniform, and, therefore, the probability $C'=c$ and $Y'=y$ is at least $1/4 - \negl(k)$. The expected number of iterations\footnote{If we want the algorithm to be worst-case polynomial, we simply abort after a polynomial amount of simulations.} is therefore constant and the simulator is efficient since the system $\bF$ is efficient.

Let us assume that there exists an algorithm $A$ with
\[ \adv^A( {X_0X_1CYV}, \ol {X_0X_1CY}V' ) \geq \gamma(k)\;,\]
for a non-negligible function $\gamma(k)$. There exists a constant $c$, such that $\gamma(k) \geq k^{-c}$ for infinitely many $k$. Let $\gamma^*(k) := k^{-c}$. $\gamma^*(k)$ is a noticeable function with
$\gamma^*(k) \leq \gamma(k)$ for infinitely many $k$.

Since $(C,X_C,Y,V)$ is $\negl(k)$-close
to $(\ol C,\ol X_{\ol C},\ol Y,V')$, and
$\ol X_{1-\ol C}$ is uniform
with respect to $(\ol C,\ol X_{\ol C},\ol Y,V')$, we have
\[ \adv^A( {R C X_{C}YV}, \ol X_{1-\ol C} \ol C \ol X_{\ol C}\ol YV') \leq \negl(k)\;,\]
where $R$ is chosen uniformly at random. It follows that
\[ \adv^A( R C X_{C}YV, {X_{1-C} C X_{C}YV}) \geq \gamma^*(k) - \negl(k)\]
for infinitely many $k$, and therefore either
\[ \Pr[A(R C X_{C}YV) = 1] - \Pr[A({X_{1-C} C X_{C}YV}) = 1] \geq \gamma^*(k) - \negl(k) \]
for infinitely many $k$, or
\[\Pr[A({X_{1-C} C X_{C}YV}) = 1] - \Pr[A(R C X_{C}YV) = 1]  \geq \gamma^*(k) - \negl(k) \]
for infinitely many $k$. Note that $(C,Y)$ is a function of $V$ and $E = X_{C} \oplus Y$.
In both cases, it follows from Lemma~\ref{lem:comp-dist2pred} that
there exists an algorithm that can predict $X_{1-C}$ with probability $1/2 + \gamma^*(k) - \negl(k)$
for infinitely many $k$, which contradicts our assumption that no such algorithm exists.

The proof for the security of $\PlayerB$ can be done the same way.
\end{proof}

\section{Computational-WOT Amplification} \label{sec:compWOTamp}

In \cite{Holens05}, Lemma~\ref{lem:hardcore} was used to show that any information-theoretic key-agreement protocol can also be used in the computational
setting. We will use a very similar proof to show that any protocol that efficiently implements
$\ROTT$ out of many instances of $\WOTT$ in the semi-honest model can be used to implement $\ROTT$ out of many instances of $\compWOTT$ in the computational semi-honest model.

\begin{theorem} \label{thm:compOT}
Let the functions $\eps(k)$, $p(k)$, $q(k)$ and $n(k)$ computable in time $\poly(k)$ be given. Let a protocol $\bP(\Auth)$
achieve $\compWOT{p}{q}{\eps}$. Further, let an efficient protocol $\bQ(\WOT{p}{q}{\eps}^{\|n} \| \Auth)$ be given which takes $k$ as input and
 securely implements $\WOT{\negl(k)}{\negl(k)}{\negl(k)}$ in the semi-honest model. Then the protocol\footnote{This is an execution of $\bQ$, where all calls to $\WOTT$ are replaced by independent executions of $\bP$.} $\bQ(\bP(\Auth)^{\|n} \| \Auth)$ implements $\compWOT{\negl(k)}{\negl(k)}{\negl(k)}$
in the computational semi-honest model.
\end{theorem}

\begin{proof}
Let $W = (W_\PlayerA,W_\PlayerB)$ be the randomness used in $\bP(\Auth)$ by the sender and the receiver, and let
$Z$ be the communication. The honest protocols $\bP_\PlayerA$ and $\bP_\PlayerB$ output $(X_0,X_1)$ and $(C,Y)$, respectively,  while the semi-honest protocols $\ul \bP_\PlayerA$ and $\ul \bP_\PlayerB$ additionally have the auxiliary outputs $U = (X_0,X_1,Z,W_\PlayerA)$ and $V = (C,Y,Z,W_\PlayerB)$, respectively. Let $E := Y \oplus X_C$.
All these values are functions of $W$.

$\bQ_{\PlayerA}$ receives $(X_0^n,X_1^n)$ from $\bP(\Auth)^{\|n}$
and
outputs $(X_0^*,X_1^*)$. $\bQ_{\PlayerB}$ receives 
$(C^n,Y^n)$ from $\bP(\Auth)^{\|n}$
and  outputs $(C^*,Y^*)$.
Let $R = (R_{\PlayerA},R_{\PlayerB})$ be the randomness used in $\bQ$ by both players, and let $Z'$ be the communication sent
over $\Auth$ in $\bQ$. Let $E^* := Y^* \oplus X^*_{C^*}$.
The values $E^*$, $X_0^*$, $X_1^*$, $C^*$, $Y^*$ and $Z'$ are functions of $(X_0^n,X_1^n,C^n,Y^n,R)$.

First of all, the resulting protocol $\bQ(\bP(\Auth)^{\|n} \| \Auth)$ will be correct and efficient, as every outcome of $\bP(\Auth)$ satisfies $\Pr[Y \neq X_C] \leq \eps$.

For the security for $\PlayerA$, we define the following functions: let $f(W) := (V,E)$ and
$P(W) := X_{1-C}$. Since $X_C = E \oplus Y$, it is possible to simulate the protocol $\bQ$ using the values
$(V,E)^n$, $(X_{1-C})^n$, and $R$. Therefore, we can define
 \[\extr((V,E)^n,(X_{1-C})^n,R) := X^*_{1-C^*}\] and \[\leak((V,E)^n,(X_{1-C})^n,R) := (E^*,C^*,Y^*,V^n,Z',R_{\PlayerB})\;.\]
$\bQ$ implements $\WOT{\negl(k)}{\negl(k)}{\negl(k)}$. It follows from
Lemma~\ref{lem:comp-pred2dist} that the functions $\extr$ and $\leak$ satisfy the extraction requirements from
Theorem~\ref{thm:holae} with $\eps(k) = \negl(k)$. Furthermore, $\extr$ and $\leak$ can be computed efficiently, since the protocol $\bQ$ is efficient. 
From the security condition of $\compWOTT$ follows that every polynomial-time algorithm $B$ satisfies
\[  \Pr [ B(f(W)) = P(W) ] = \Pr [ B(V,E) = X_{1-C} ] \leq \frac {1+q(k)} 2 \]
for all but finitely many $k$, for $W$ chosen uniformly at random. Theorem~\ref{thm:holae} tells us that
no polynomial time algorithm $A$, which gets as input $\leak((V,E)^n,(X_{1-C})^n,R)$ distinguishes $\extr((V,E)^n,(X_{1-C})^n,R)$
from a uniform random bit with advantage $\negl(k) + \gamma(k)$,
for any non-negligible function $\gamma(k)$. The security for $\PlayerA$ follows now from Lemma~\ref{lem:comp-pred2dist}.

For the security for $\PlayerB$, we define the following functions: let $f(W) := (U,E)$ and
$P(W) := C$. Since $X_C = E \oplus Y$, it is possible to simulate the protocol $\bQ$ using the values
$(U,E)^n$, $C^n$, and $R$. Therefore, we can define
\[\extr((U,E)^n,C^n,R) := C^*\;,\] and 
\[\leak((U,E)^n,C^n,R) := (E^*,X_0^*,X_1^*,U^n,Z',R_{\PlayerA})\;.\] $\bQ$ implements $\WOT{\negl(k)}{\negl(k)}{\negl(k)}$. It follows from
Lemma~\ref{lem:comp-pred2dist} that the functions $\extr$ and $\leak$ satisfy the extraction requirements from
Theorem~\ref{thm:holae} with $\eps(k) = \negl(k)$.
Furthermore, $\extr$ and $\leak$ can be computed efficiently, since the protocol $\bQ$ is efficient.
From the security condition of $\compWOTT$ follows that every polynomial time algorithm $A$ satisfies
\[  \Pr [ A(f(W)) = P(W) ] = \Pr [ A(U,E) = C ] \leq \frac {1+p(k)} 2 \]
for all but finitely many k, for $W$ chosen uniformly at random. Theorem~\ref{thm:holae} tells us that
no polynomial time algorithm $B$, which gets as input $\leak((U,E)^n,C^n,R)$ distinguishes $\extr((U,E)^n,C^n,R)$
from a uniform random bit with advantage $\negl(k) + \gamma(k)$,
for any non-negligible function $\gamma(k)$. The security for $\PlayerB$ follows now from Lemma~\ref{lem:comp-pred2dist}.
\end{proof}

\noindent
Together with the information-theoretic reductions presented in Chapters \ref{chap:ot} and \ref{chap:wot}, we get a protocol that securely amplifies $\compWOT{p}{q}{\eps}$ to $\OT{2}{1}{1}$  in the computational semi-honest model.

\begin{corollary} \label{cor:comOT}
Let the functions $\eps(k)$, $p(k)$, and $q(k)$, computable in time $\poly(k)$, be given, where either for all $k$
\[\eps=0 \ \wedge \ p+q < 1 - 1/\poly(k)\;,\]
\[p+q+2\eps \leq 0.24\;,\] or 
\[\min( p + 22q + 44\eps,22p + q + 44\eps, 7 \sqrt{p+q} + 2\eps) < 1 - 1/\poly(k)\;,\]
or, for constant functions $p(k)$, $q(k)$ and $\eps(k)$,
\[p=0 \ \wedge \ \sqrt{q} + 2 \eps <1\;,\]
\[q=0 \ \wedge \ \sqrt{p} + 2 \eps <1\;,\] or 
\[(1 - p - q)^4 < - 178 \cdot \log(1 - 2\eps)\;.\]
If there exists a protocol $\bP(\Auth)$ that securely implements $\compWOT{p}{q}{\eps}$ in the computational semi-honest model, then there exists a protocol $\bQ(\Auth)$ that implements $\OT{2}{1}{1}$ in the computational semi-honest model.
\end{corollary}

\section{Discussion and Open Problems}

We have shown that Holenstein's hard-core lemma \cite{Holens05,Holens06} can also be applied
in the setting of two-party computation, and presented a new computational assumption, namely \emph{computational weak oblivious transfer}, under which
oblivious transfer and hence any two-party computation is possible in a computationally secure way.

The \emph{pseudo-randomness extraction theorem} presented in \cite{Holens06} turned out not to be
general enough for our application. It would be interesting to know
whether our generalization is also useful in other applications.

A very interesting open problem is whether our results can be used to improve the results 
from \cite{Haitne04}, i.e., whether it is possible to implement computationally-secure OT
from weaker requirements on trapdoor permutations.

\bibliographystyle{alpha}

\begin{thebibliography}{IMQNW04}

\bibitem[AC93]{AhlCsi93}
R.~Ahlswede and I.~Csisz{\'a}r.
\newblock Common randomness in information theory and cryptography -- part {I}:
  Secret sharing.
\newblock {\em {IEEE} Transactions on Information Theory}, 39(4):1121--1132,
  1993.

\bibitem[AIR01]{AiIsRe01}
W.~Aiello, Y.~Ishai, and O.~Reingold.
\newblock Priced oblivious transfer: How to sell digital goods.
\newblock In {\em Advances in Cryptology --- EUROCRYPT '01}, Lecture Notes in
  Computer Science, pages 119--135. Springer-Verlag, 2001.

\bibitem[BBCM95]{BBCM95}
C.~H. Bennett, G.~Brassard, C.~Cr{\'e}peau, and U.~Maurer.
\newblock Generalized privacy amplification.
\newblock {\em IEEE Transactions on Information Theory}, 41, 1995.

\bibitem[BBCS92]{BBCS92}
C.~H. Bennett, G.~Brassard, C.~Cr{\'e}peau, and H.~Skubiszewska.
\newblock Practical quantum oblivious transfer.
\newblock In {\em Advances in Cryptology --- CRYPTO '91}, volume 576 of {\em
  Lecture Notes in Computer Science}, pages 351--366. Springer, 1992.

\bibitem[BBR88]{BeBrRo88}
C.~H. Bennett, G.~Brassard, and J.-M. Robert.
\newblock Privacy amplification by public discussion.
\newblock {\em SIAM Journal on Computing}, 17(2):210--229, 1988.

\bibitem[BC97]{BraCre97}
G.~Brassard and C.~Cr{\'e}peau.
\newblock Oblivious transfers and privacy amplification.
\newblock In {\em Advances in Cryptology --- EUROCRYPT '97}, volume 1233 of
  {\em Lecture Notes in Computer Science}, pages 334--347. Springer-Verlag,
  1997.

\bibitem[BCR86]{BrCrRo86b}
G.~Brassard, C.~Cr{\'e}peau, and J.-M. Robert.
\newblock Information theoretic reductions among disclosure problems.
\newblock In {\em Proceedings of the 27th Annual IEEE Symposium on Foundations
  of Computer Science (FOCS~'86)}, pages 168--173, 1986.

\bibitem[BCS96]{BrCrSa96}
G.~Brassard, C.~Cr{\'e}peau, and M.~S{\'a}ntha.
\newblock Oblivious transfers and intersecting codes.
\newblock {\em IEEE Transactions on Information Theory, special issue on coding
  and complexity}, 42(6):1769--1780, 1996.

\bibitem[BCW03]{BrCrWo03}
G.~Brassard, C.~Cr{\'e}peau, and S.~Wolf.
\newblock Oblivious transfers and privacy amplification.
\newblock {\em Journal of Cryptology}, 16(4):219--237, 2003.

\bibitem[Bea89]{Beaver89b}
D.~Beaver.
\newblock Multiparty protocols tolerating half faulty processors.
\newblock In {\em Advances in Cryptology --- CRYPTO '89}, volume 435 of {\em
  Lecture Notes in Computer Science}, pages 560--572. Springer-Verlag, 1989.

\bibitem[Bea92]{Beaver91}
D.~Beaver.
\newblock Foundations of secure interactive computing.
\newblock In {\em Advances in Cryptology --- CRYPTO '91}, volume 1233 of {\em
  Lecture Notes in Computer Science}, pages 377--391. Springer-Verlag, 1992.

\bibitem[Bea95]{Beaver95}
D.~Beaver.
\newblock Precomputing oblivious transfer.
\newblock In {\em Advances in Cryptology --- EUROCRYPT '95}, volume 963 of {\em
  Lecture Notes in Computer Science}, pages 97--109. Springer-Verlag, 1995.

\bibitem[BGW88]{BeGoWi88}
M.~{Ben-Or}, S.~Goldwasser, and A.~Wigderson.
\newblock Completeness theorems for non-cryptographic fault-tolerant
  distributed computation.
\newblock In {\em Proceedings of the 21st Annual ACM Symposium on Theory of
  Computing (STOC~'88)}, pages 1--10. ACM Press, 1988.

\bibitem[BM90]{BelMic89}
M.~Bellare and S.~Micali.
\newblock Non-interactive oblivious transfer and applications.
\newblock In {\em Advances in Cryptology --- CRYPTO '89}, volume 435 of {\em
  Lecture Notes in Computer Science}. Springer-Verlag, 1990.

\bibitem[BPW03]{BaPfWa03}
M.~Backes, B.~Pfitzmann, and M.~Waidner.
\newblock A universally composable cryptographic library.
\newblock http://eprint.iacr.org/2003/015, 2003.

\bibitem[Cac98]{Cachin98}
C.~Cachin.
\newblock On the foundations of oblivious transfer.
\newblock In {\em Advances in Cryptology --- EUROCRYPT '98}, volume 1403 of
  {\em Lecture Notes in Computer Science}, pages 361--374. Springer-Verlag,
  1998.

\bibitem[Can96]{Canetti96}
R.~Canetti.
\newblock {\em Studies in Secure Multiparty Computation and Applications}.
\newblock PhD thesis, Weizmann Institiute of Science, Israel, 1996.

\bibitem[Can00]{Canetti00b}
R.~Canetti.
\newblock Security and composition of multiparty cryptographic protocols.
\newblock {\em Journal of Cryptology}, 13(1):143--202, 2000.

\bibitem[Can01]{Canetti00}
R.~Canetti.
\newblock Universally composable security: A new paradigm for cryptographic
  protocols.
\newblock In {\em Proceedings of the 42th Annual IEEE Symposium on Foundations
  of Computer Science (FOCS~'01)}, pages 136--145, 2001.
\newblock Updated Version at http://eprint.iacr.org/2000/067.

\bibitem[CCD88]{ChCrDa88}
D.~Chaum, C.~Cr{\'e}peau, and I.~Damg{\aa}rd.
\newblock Multiparty unconditionally secure protocols (extended abstract).
\newblock In {\em Proceedings of the 21st Annual ACM Symposium on Theory of
  Computing (STOC~'88)}, pages 11--19. ACM Press, 1988.

\bibitem[CDvdG88]{ChDaGr87}
D.~Chaum, I.~Damg{\aa}rd, and J.~van~de Graaf.
\newblock Multiparty computations ensuring privacy of each party's input and
  correctness of the result.
\newblock In {\em Advances in Cryptology --- {CRYPTO} '87}, volume 293 of {\em
  Lecture Notes in Computer Science}, pages 87--119. Springer-Verlag, 1988.

\bibitem[CF01]{CanFis01}
R.~Canetti and M.~Fischlin.
\newblock Universally composable commitments.
\newblock In {\em Advances in Cryptology --- CRYPTO '01}, volume 576 of {\em
  Lecture Notes in Computer Science}, pages 19--40. Springer-Verlag, 2001.

\bibitem[Che52]{Cherno52}
H.~Chernoff.
\newblock A measure of asymptotic efficiency for tests of a hypothesis based on
  the sum of observations.
\newblock {\em Annals of Mathematical Statistics}, 23:493--507, 1952.

\bibitem[CK88]{CreKil88}
C.~Cr{\'e}peau and J.~Kilian.
\newblock Achieving oblivious transfer using weakened security assumptions
  (extended abstract).
\newblock In {\em Proceedings of the 29th Annual IEEE Symposium on Foundations
  of Computer Science (FOCS~'88)}, pages 42--52, 1988.

\bibitem[CLOS02]{CLOS02}
R.~Canetti, Y.~Lindell, R.~Ostrovsky, and A.~Sahai.
\newblock Universally composable two-party and multi-party secure computation.
\newblock In {\em Proceedings of the 34th Annual ACM Symposium on Theory of
  Computing (STOC~'02)}, pages 494--503. ACM Press, 2002.
\newblock Full version available at http://eprint.iacr.org/2002/140.

\bibitem[CMW04]{CrMoWo04}
C.~Cr{\'e}peau, K.~Morozov, and S.~Wolf.
\newblock Efficient unconditional oblivious transfer from almost any noisy
  channel.
\newblock In {\em Proceedings of Fourth Conference on Security in Communication
  Networks (SCN)}, volume 3352 of {\em Lecture Notes in Computer Science},
  pages 47--59. Springer-Verlag, 2004.

\bibitem[Cr{\'e}88]{Crepea87}
C.~Cr{\'e}peau.
\newblock Equivalence between two flavours of oblivious transfers (abstract).
\newblock In {\em Advances in Cryptology --- CRYPTO '87}, volume 293 of {\em
  Lecture Notes in Computer Science}, pages 350--354. Springer-Verlag, 1988.

\bibitem[Cr{\'e}90]{Crepea89}
C.~Cr{\'e}peau.
\newblock Verifiable disclosure of secrets and applications.
\newblock In {\em Advances in Cryptology --- {CRYPTO}~'89}, volume 434 of {\em
  Lecture Notes in Computer Science}, pages 181--191. Springer-Verlag, 1990.

\bibitem[Cr{\'e}97]{Crepea97}
C.~Cr{\'e}peau.
\newblock Efficient cryptographic protocols based on noisy channels.
\newblock In {\em Advances in Cryptology --- CRYPTO '97}, volume 1233 of {\em
  Lecture Notes in Computer Science}, pages 306--317. Springer-Verlag, 1997.

\bibitem[CS91]{CreSan91}
C.~Cr{\'e}peau and M.~S{\'a}ntha.
\newblock On the reversibility of oblivious transfer.
\newblock In {\em Advances in Cryptology --- EUROCRYPT '91}, volume 547 of {\em
  Lecture Notes in Computer Science}, pages 106--113. Springer, 1991.

\bibitem[CS06]{CreSav06}
C.~Cr{\'e}peau and G.~Savvides.
\newblock Optimal reductions between oblivious transfers using interactive
  hashing.
\newblock In {\em Advances in Cryptology --- EUROCRYPT '06}, volume 4004 of
  {\em Lecture Notes in Computer Science}, pages 201--221. Springer-Verlag,
  2006.

\bibitem[CSSW06]{CSSW06}
C.~Cr{\'e}peau, G.~Savvides, C.~Schaffner, and J.~Wullschleger.
\newblock Information-theoretic conditions for two-party secure function
  evaluation.
\newblock In {\em Advances in Cryptology --- EUROCRYPT '06}, volume 4004 of
  {\em Lecture Notes in Computer Science}, pages 538--554. Springer-Verlag,
  2006.
\newblock {Full version available at http://eprint.iacr.org/2006/183}.

\bibitem[CvdGT95]{CrvGTa95}
C.~Cr{\'e}peau, J.~van~de Graaf, and A.~Tapp.
\newblock Committed oblivious transfer and private multi-party computation.
\newblock In {\em Advances in Cryptology --- {CRYPTO}~'95}, Lecture Notes in
  Computer Science, pages 110--123. Springer-Verlag, 1995.

\bibitem[CW79]{CarWeg79}
J.~L. Carter and M.~N. Wegman.
\newblock {Universal classes of hash functions}.
\newblock {\em Journal of Computer and System Sciences}, 18:143--154, 1979.

\bibitem[DFMS04]{DFMS04}
I.~Damg{\aa}rd, S.~Fehr, K.~Morozov, and L.~Salvail.
\newblock Unfair noisy channels and oblivious transfer.
\newblock In {\em Theory of Cryptography Conference --- TCC~'04}, volume 2951
  of {\em Lecture Notes in Computer Science}, pages 355--373. Springer-Verlag,
  2004.

\bibitem[DFR{\etalchar{+}}06]{DFRSS06}
I.~Damg{\aa}rd, S.~Fehr, R.~Renner, L.~Salvail, and C.~Schaffner.
\newblock A tight high-order entropic uncertinty relation with applications in
  the bounded quantum-storage model.
\newblock In preparation, 2006.

\bibitem[DFSS06]{DFSS06}
I.~Damg{\aa}rd, S.~Fehr, L.~Salvail, and C.~Schaffner.
\newblock Oblivious transfer and linear functions.
\newblock In {\em Advances in Cryptology --- CRYPTO '06}, volume 4117 of {\em
  Lecture Notes in Computer Science}. Springer-Verlag, 2006.

\bibitem[DKS99]{DaKiSa99}
I.~Damg{\aa}rd, J.~Kilian, and L.~Salvail.
\newblock On the (im)possibility of basing oblivious transfer and bit
  commitment on weakened security assumptions.
\newblock In {\em Advances in Cryptology --- EUROCRYPT '99}, volume 1592 of
  {\em Lecture Notes in Computer Science}, pages 56--73. Springer-Verlag, 1999.

\bibitem[DM99]{DodMic99}
Y.~Dodis and S.~Micali.
\newblock Lower bounds for oblivious transfer reductions.
\newblock In {\em Advances in Cryptology --- {EUROCRYPT} '99}, volume 1592 of
  {\em Lecture Notes in Computer Science}, pages 42--55. Springer-Verlag, 1999.

\bibitem[EGL85]{EvGoLe85}
S.~Even, O.~Goldreich, and A.~Lempel.
\newblock A randomized protocol for signing contracts.
\newblock {\em Commun. ACM}, 28(6):637--647, 1985.

\bibitem[Fis06]{Fischl06}
M.~Fischlin.
\newblock Universally composable oblivious transfer in the multi-party setting.
\newblock In {\em RSA Security Cryptographer's Track 2006}, volume 3860 of {\em
  Lecture Notes in Computer Science}, pages 332--349. Springer-Verlag, 2006.

\bibitem[GL91]{GolLev90}
S.~Goldwasser and L.~A. Levin.
\newblock Fair computation of general functions in presence of immoral
  majority.
\newblock In {\em Advances in Cryptology --- {CRYPTO}~'90}, Lecture Notes in
  Computer Science, pages 77--93. Springer-Verlag, 1991.

\bibitem[GMR85]{GoMiRa85}
S.~Goldwasser, S.~Micali, and C.~Rackoff.
\newblock The knowledge complexity of interactive proof-systems.
\newblock In {\em Proceedings of the 17th Annual ACM Symposium on Theory of
  Computing (STOC~'85)}, pages 291--304. ACM Press, 1985.

\bibitem[GMW87]{GoMiWi87}
O.~Goldreich, S.~Micali, and A.~Wigderson.
\newblock How to play any mental game.
\newblock In {\em Proceedings of the 21st Annual ACM Symposium on Theory of
  Computing (STOC~'87)}, pages 218--229. ACM Press, 1987.

\bibitem[GMY04]{GaMaYa04}
J.~Garay, P.~MacKenzie, and K.~Yang.
\newblock Efficient and universally composable committed oblivious transfer and
  applications.
\newblock In {\em Theory of Cryptography Conference --- TCC~'04}, volume 2951
  of {\em Lecture Notes in Computer Science}, pages 297--316. Springer-Verlag,
  2004.

\bibitem[Gol04]{Goldreich04}
O.~Goldreich.
\newblock {\em Foundations of Cryptography}, volume II: Basic Applications.
\newblock Cambridge University Press, 2004.

\bibitem[GV88]{GolVai87}
O.~Goldreich and R.~Vainish.
\newblock How to solve any protocol problem - an efficiency improvement.
\newblock In {\em Advances in Cryptology --- {CRYPTO}~'87}, Lecture Notes in
  Computer Science, pages 73--86. Springer-Verlag, 1988.

\bibitem[Hai04]{Haitne04}
I.~Haitner.
\newblock Implementing oblivious transfer using collection of dense trapdoor
  permutations.
\newblock In {\em Theory of Cryptography Conference --- TCC~'04}, volume 2951
  of {\em Lecture Notes in Computer Science}, pages 394--409. Springer-Verlag,
  2004.

\bibitem[H{\aa}s90]{Hastad90}
J.~H{\aa}stad.
\newblock Pseudo-random generators under uniform assumptions.
\newblock In {\em Proceedings of the 22st Annual ACM Symposium on Theory of
  Computing (STOC~'90)}, pages 395--404. ACM Press, 1990.

\bibitem[HHR06]{HaHaRe05}
I.~Haitner, D.~Harnik, and O.~Reingold.
\newblock On the power of the randomized iterate.
\newblock In {\em Advances in Cryptology --- CRYPTO '06}, volume 4117 of {\em
  Lecture Notes in Computer Science}, pages 21--40. Springer-Verlag, 2006.

\bibitem[HILL99]{HILL99}
J.~H{\aa}stad, R.~Impagliazzo, L.~A. Levin, and M.~Luby.
\newblock A pseudorandom generator from any one-way function.
\newblock {\em SIAM J. Comput.}, 28(4):1364--1396, 1999.

\bibitem[HKN{\etalchar{+}}05]{HKNRR05}
D.~Harnik, J.~Kilian, M.~Naor, O.~Reingold, and A.~Rosen.
\newblock On robust combiners for oblivious transfer and other primitives.
\newblock In {\em Advances in Cryptology --- EUROCRYPT '05}, volume 3494 of
  {\em Lecture Notes in Computer Science}, pages 96--113, 2005.

\bibitem[Hoe63]{Hoeffd63}
W.~Hoeffding.
\newblock Probability inequalities for sums of bounded random variables.
\newblock {\em Journal of the American Statistical Association},
  58(301):13--30, 1963.

\bibitem[Hol05]{Holens05}
T.~Holenstein.
\newblock Key agreement from weak bit agreement.
\newblock In {\em Proceedings of the 37th ACM Symposium on Theory of Computing
  (STOC~'05)}, pages 664--673. ACM Press, 2005.

\bibitem[Hol06]{Holens06}
T.~Holenstein.
\newblock {\em Strengthening key agreement using hard-core sets}.
\newblock PhD thesis, {ETH} Zurich, Switzerland, 2006.
\newblock Reprint as vol.~7 of {\em ETH Series in Information Security and
  Cryptography}, {H}artung-{G}orre {V}erlag.

\bibitem[HR05]{HolRen05}
T.~Holenstein and R.~Renner.
\newblock One-way secret-key agreement and applications to circuit polarization
  and immunization of public-key encryption.
\newblock In {\em Advances in Cryptology --- CRYPTO '05}, volume 3621 of {\em
  Lecture Notes in Computer Science}, pages 478--493. Springer-Verlag, 2005.

\bibitem[ILL89]{ILL89}
R.~Impagliazzo, L.~A. Levin, and M.~Luby.
\newblock Pseudo-random generation from one-way functions.
\newblock In {\em Proceedings of the 21st Annual ACM Symposium on Theory of
  Computing (STOC~'89)}, pages 12--24. ACM Press, 1989.

\bibitem[IMN06]{ImMoNa06}
H.~Imai, K.~Morozov, and A.~Nascimento.
\newblock On the oblivious transfer capacity of the erasure channel.
\newblock In {\em Proceedings of 2006 IEEE International Symposium on
  Information Theory (ISIT~'06)}, pages 1428--1431, 2006.

\bibitem[Imp95]{Impagl95}
R.~Impagliazzo.
\newblock Hard-core distributions for somewhat hard problems.
\newblock In {\em Proceedings of the 36th Annual IEEE Symposium on Foundations
  of Computer Science (FOCS~'95)}, pages 538--545. IEEE Computer Society, 1995.

\bibitem[IMQNW04]{IMNW04}
H.~Imai, J.~M{\"u}ller-Quade, A.~Nascimento, and A.~Winter.
\newblock Rates for bit commitment and coin tossing from noisy correlation.
\newblock In {\em Proceedings of the IEEE International Symposium on
  Information Theory (ISIT~'04)}, 2004.

\bibitem[IR89]{ImpRud89}
R.~Impagliazzo and S.~Rudich.
\newblock Limits on the provable consequences of one-way permutations.
\newblock In {\em Proceedings of the 21st Annual ACM Symposium on Theory of
  Computing (STOC~'89)}, pages 186--208. ACM Press, 1989.

\bibitem[Kil88]{Kilian88}
J.~Kilian.
\newblock Founding cryptography on oblivious transfer.
\newblock In {\em Proceedings of the 20th Annual ACM Symposium on Theory of
  Computing (STOC~'88)}, pages 20--31. ACM Press, 1988.

\bibitem[KM01]{KorMor01}
V.~Korjik and K.~Morozov.
\newblock Generalized oblivious transfer protocols based on noisy channels.
\newblock In {\em Proceedings of the International Workshop MMM ACNS}, volume
  2052 of {\em Lecture Notes in Computer Science}, pages 219--229.
  Springer-Verlag, 2001.

\bibitem[LC97]{LoChau97}
H.~K. Lo and H.~F. Chau.
\newblock Is quantum bit commitment really possible?
\newblock {\em Physical Review Letters}, 78:3410--3413, 1997.

\bibitem[Mau93]{Maurer93}
U.~Maurer.
\newblock Secret key agreement by public discussion.
\newblock {\em IEEE Transaction on Information Theory}, 39(3):733--742, 1993.

\bibitem[Mau06]{Maurer06}
U.~Maurer.
\newblock Lecture notes information security, 2006.

\bibitem[May97]{Mayers97}
D.~Mayers.
\newblock Unconditionally secure quantum bit commitment is impossible.
\newblock {\em Physical Review Letters}, 78:3414--3417, 1997.

\bibitem[Mor05]{Morozo05}
K.~Morozov.
\newblock {\em On Cryptographic Primitives Based on Noisy Channels}.
\newblock PhD thesis, University of Aarhus, Denmark, 2005.

\bibitem[MPW07]{MePrWu07}
R.~Meier, B.~Przydatek, and J.~Wullschleger.
\newblock Robuster combiners for oblivious transfer.
\newblock In {\em Theory of Cryptography Conference --- TCC '07}, Lecture Notes
  in Computer Science. Springer-Verlag, 2007.

\bibitem[MR92]{MicRog91}
S.~Micali and P.~Rogaway.
\newblock Secure computation (abstract).
\newblock In {\em Advances in Cryptology --- CRYPTO '91}, volume 576 of {\em
  Lecture Notes in Computer Science}, pages 392--404. Springer-Verlag, 1992.

\bibitem[MT98]{MolTie98}
B.\ Moldovanu and M.\ Tietzel.
\newblock Goethe's second-price auction.
\newblock {\em The Journal of Political Economy}, 106(4):854--859, 1998.

\bibitem[NP01]{NaoPin01}
M.~Naor and B.~Pinkas.
\newblock Efficient oblivious transfer protocols.
\newblock In {\em Proceedings of the 12th annual ACM-SIAM symposium on Discrete
  algorithms (SODA~'01)}, pages 448--457. Society for Industrial and Applied
  Mathematics, 2001.

\bibitem[NW06]{NaWi06}
A.~Nascimento and A.~Winter.
\newblock On the oblivious transfer capacity of noisy correlations.
\newblock In {\em Proceedings of the IEEE International Symposium on
  Information Theory (ISIT~'06)}, 2006.

\bibitem[OVY93]{OsVeYu91}
R.~Ostrovsky, R.~Venkatesan, and M.~Yung.
\newblock Fair games against an all-powerful adversary.
\newblock In {\em Advances in Computational Complexity Theory}, volume~13 of
  {\em AMS DIMACS Series in Discrete Mathematics and Theoretical Computer
  Science}, pages 155--169. AMS, 1993.

\bibitem[PW01]{PfiWai00}
B.~Pfitzmann and M.~Waidner.
\newblock A model for asynchronous reactive systems and its application to
  secure message transmission.
\newblock In {\em Proceedings of the 2001 IEEE Symposium on Security and
  Privacy (SP~'01)}, page 184, 2001.
\newblock Also available at http://eprint.iacr.org/2000/066.

\bibitem[Rab81]{Rabin81}
M.~O. Rabin.
\newblock How to exchange secrets by oblivious transfer.
\newblock Technical Report TR-81, Harvard Aiken Computation Laboratory, 1981.

\bibitem[RB89]{RabBen89}
T.~Rabin and M.~{Ben-Or}.
\newblock Verifiable secret sharing and multiparty protocols with honest
  majority.
\newblock In {\em Proceedings of the 21st Annual ACM Symposium on Theory of
  Computing (STOC~'89)}, pages 73--85. ACM Press, 1989.

\bibitem[R{\'e}n61]{Renyi61}
A.~R{\'e}nyi.
\newblock On measures of information and entropy.
\newblock In {\em Proceedings of the 4th Berkeley Symposium on Mathematics,
  Statistics and Probability}, pages 547--561, 1961.

\bibitem[Ren05]{Renner05}
R.~Renner.
\newblock {\em Security of Quantum Key Distribution}.
\newblock PhD thesis, {ETH} Zurich, Switzerland, 2005.
\newblock Available at http://arxiv.org/abs/quant-ph/0512258.

\bibitem[RK05]{RenKoe05}
R.~Renner and R.~K{\"o}nig.
\newblock Universally composable privacy amplification against quantum
  adversaries.
\newblock In {\em Theory of Cryptography Conference --- TCC '05}, volume 3378
  of {\em Lecture Notes in Computer Science}, pages 407--425. Springer-Verlag,
  2005.
\newblock Also available at http://arxiv.org/abs/quant-ph/0403133.

\bibitem[RW05]{RenWol05}
R.~Renner and S.~Wolf.
\newblock Simple and tight bounds for information reconciliation and privacy
  amplification.
\newblock In {\em Advances in Cryptology --- ASIACRYPT 2005}, volume 3788 of
  {\em Lecture Notes in Computer Science}, pages 199--216. Springer-Verlag,
  2005.

\bibitem[Sho94]{Shor94}
P.~Shor.
\newblock Algorithms for quantum computation: discrete logarithms and
  factoring.
\newblock In {\em Proceedings of the 35th Annual IEEE Symposium on Foundations
  of Computer Science (FOCS~'94)}, pages 124--134, 1994.

\bibitem[SV99]{SahVad99}
A.~Sahai and S.~Vadhan.
\newblock Manipulating statistical difference.
\newblock In {\em Randomization Methods in Algorithm Design ({DIMACS}
  Workshop~'97)}, volume~43 of {\em {DIMACS} Series in Discrete Mathematics and
  Theoretical Computer Science}, pages 251--270. American Mathematical Society,
  1999.

\bibitem[Vad99]{Vadhan99}
S.~Vadhan.
\newblock {\em A study of statistical zero-knowledge proofs}.
\newblock PhD thesis, Massachusets Institute of Technology, USA, 1999.

\bibitem[Wie83]{Wiesner70}
S.~Wiesner.
\newblock Conjugate coding.
\newblock {\em SIGACT News}, 15(1):78--88, 1983.

\bibitem[Wul07]{Wullsc07}
J.~Wullschleger.
\newblock Oblivious-transfer amplification.
\newblock In {\em Advances in Cryptology --- {EUROCRYPT}~'07}, Lecture Notes in
  Computer Science. Springer-Verlag, 2007.

\bibitem[WW04]{WolWul04}
S.~Wolf and J.~Wullschleger.
\newblock Zero-error information and applications in cryptography.
\newblock In {\em Proceedings of 2004 IEEE Information Theory Workshop
  (ITW~'04)}, 2004.

\bibitem[WW06]{WolWul06}
S.~Wolf and J.~Wullschleger.
\newblock Oblivious transfer is symmetric.
\newblock In {\em Advances in Cryptology --- EUROCRYPT '06}, volume 4004 of
  {\em Lecture Notes in Computer Science}, pages 222--232. Springer-Verlag,
  2006.

\bibitem[Yao82]{Yao82}
A.~C. Yao.
\newblock Protocols for secure computations.
\newblock In {\em Proceedings of the 23rd Annual IEEE Symposium on Foundations
  of Computer Science (FOCS~'82)}, pages 160--164, 1982.

\end{thebibliography}

\newcommand{\etalchar}[1]{$^{#1}$}

\appendix

\chapter{Appendix}

\section{Formal Technicalities}

\begin{lemma}[Chernoff/Hoeffding Bound \cite{Cherno52,Hoeffd63}] \label{lem:chernoff1}
Let $P_{X_0\dots X_n} = P_{X}^n$ be a product distribution
with $X_i \in [0,1]$.  Let
$X := \frac 1 n \sum_{i=0}^{n-1} X_i$, and $\mu = E[X]$. Then, for any $\eps > 0$, 
\begin{align*}
\Pr\left[ X \geq \mu + \eps \right ] &\leq e^{-2n\eps^2}\;,\\
\Pr\left[ X \leq \mu - \eps \right ] &\leq e^{-2n\eps^2}\;.
\end{align*}
\end{lemma}

\begin{lemma}[Cauchy-Schwartz] \label{lem:cauchySchwartz}
For all $x_0, \dots, x_{n-1}, y_0, \dots, y_{n-1} \in \bbR$, we have
\[\left(\sum_{i=0}^{n-1} x_i y_i\right)^2\leq \left(\sum_{i=0}^{n-1} x_i^2\right) \cdot \left(\sum_{i=0}^{n-1} y_i^2\right)\;.\]
\end{lemma}

\begin{lemma}\label{lem:cauchySchwartz2}
For all $a_0, \dots, a_{n-1} \in \bbR$, we have
\[\left(\sum_{i=0}^{n-1} a_i\right)^2\leq n \cdot \sum_{i=0}^{n-1} a_i^2\;.\]
\end{lemma}

\begin{proof}
The statement follows from Lemma~\ref{lem:cauchySchwartz},
choosing $x_i :=1$ and $y_i := a_i$.
\end{proof}

\begin{lemma} \label{lem:redBound2}
For all $x \in \bbR$, we have  $\ln(x+1) \leq x \leq e^{x-1}$.
\end{lemma}

\begin{proof}[Proof sketch]
The function $\ln(x+1)$ is convex, and goes through the point $(0,0)$ with slope $1$, and the function $e^{x-1}$ is concave, and goes through the point $(1,1)$ with slope $1$. Hence, we have
$\ln(x+1) \leq x \leq e^{x-1}$.
\end{proof}

\begin{lemma} \label{lem:redBound3}
For $0 \leq x \leq 1$, we have $1 - \sqrt{1 - x} \leq x$.
\end{lemma}

\begin{proof}
From $(1-x)^2 \leq 1 - x$ follows that $1 - x \leq \sqrt{1-x}$, and hence $1 - \sqrt{1 - x} \leq x$.
\end{proof}

\begin{lemma} \label{lem:redBound4}
For all $x, y \in \bbR$, we have
\[ \left |x - \frac{x+y} 2 \right | + \left |y - \frac{x+y} 2 \right| = |x-y| \]
\end{lemma}

\begin{proof}
If $x \geq y$, we have
\begin{align*}
\left |x - \frac{x+y} 2 \right | + \left |y - \frac{x+y} 2 \right|
= x - \frac{x+y} 2  + \frac{x+y} 2 - y
= x-y = |x-y|\;.
\end{align*}
The same holds for $y > x$.
\end{proof}

\begin{lemma} \label{lem:corr2}
Let $X_0$ and $X_1$ be two independent binary random variables with $\Pr[X_0=1] \leq (1-\alpha_0)/2$ and
$\Pr[X_1=1] \leq (1-\alpha_1)/2$, where $\alpha_0,\alpha_1 \geq 0$. Then $\Pr[X_0 \oplus X_1=1] \leq (1-\alpha_0\alpha_1)/2$.
\end{lemma}

\begin{proof}
For $\Pr[X_0=1] = (1-\alpha'_0)/2$ and $\Pr[X_1=1] = (1-\alpha'_1)/2$, we have
\begin{align*}
\Pr[X_0 \oplus X_1 =1] 
 = \frac{1+\alpha'_0}{2} \cdot \frac{1-\alpha'_1}{2} + \frac{1-\alpha'_0}{2} \cdot \frac{1+\alpha'_1}{2}
 = \frac{1- \alpha'_0 \alpha'_1} 2\;.
\end{align*}
The lemma follows from the fact that 
\[\frac{1- \alpha'_0 \alpha'_1} 2 \geq \frac{1- \alpha_0 \alpha_1} 2\]
for all $\alpha'_0 \in [\alpha_0,1]$ and $\alpha'_1 \in [\alpha_1,1]$.
\end{proof}

\begin{lemma} \label{lem:corrn}
For $i \in\{0,\dots,n-1\}$, let $X_i$ be independent binary random variables
where $\Pr[X_i = 1] \leq \alpha$, for $\alpha \leq 1/2$. Then
\[\Pr[X_0 \oplus \cdots \oplus X_{n-1} = 1] \leq \frac{1 - (1-2 \alpha)^n}{2} \leq n \alpha\;.\]
\end{lemma}

\begin{proof}
The first inequality follows by induction from Lemma~\ref{lem:corr2}, and the second
by the union bound, since
\[\Pr[X_0 \oplus \cdots \oplus X_{n-1} = 1] \leq \Pr[\exists i: X_i = 1] \leq n \alpha\;.\]
\end{proof}

\begin{lemma} \label{lem:redBound1}
For $i \in \{0,\dots,n-1\}$, let $X_i \in \{0,1\}$ be independently distributed with $\Pr[X_i = 1] \leq \alpha$.
We have
\[ \Pr[ X_0 = 1 \vee \dots \vee X_{n-1} = 1 ] \leq 1 - (1-\alpha)^n \leq n \alpha.\]
\end{lemma}

\begin{proof}
Follows directly from the union bound.
\end{proof}

\begin{lemma} \label{lem:errRed}
For $i \in \{0,\dots,n-1\}$, let $X_i \in \{0,1\}$ be independently distributed with $\Pr[X_i = 1] \leq \alpha$.
We have
\[ \Pr\left[\sum_{i=0}^{n-1} X_i \geq n/2\right ] \leq \sum_{i=\lceil n/2 \rceil}^{n} \binom{n}{i} \alpha^{i} (1 - \alpha)^{n-i} \leq e^{-2 n (1/2 - \alpha)^2}\;.\]
\end{lemma}

\begin{proof}
We apply Lemma~\ref{lem:chernoff1} for $\mu := \alpha$ and $\eps := 1/2 - \alpha$.
\end{proof}

\end{document}